\documentclass[10pt]{article}

\usepackage{booktabs}
\usepackage{amsmath}
\usepackage{amssymb,latexsym,color,dsfont,bm,mathtools,float}
\usepackage{mathrsfs} % makes \mathscr work properly
\usepackage{enumitem}
\usepackage[left=1.5cm,right=1.5cm,
    top=2.5cm,bottom=2.5cm,bindingoffset=0cm]{geometry}\usepackage[auth-sc]{authblk}
\usepackage{tikz}
\usepackage{pgf}

\usepackage{enumitem}
\usepackage{mathtools}
\usepackage{array}
\numberwithin{equation}{section}

\newtheorem{theorem}{Theorem}[section]
\newtheorem{lemma}[theorem]{Lemma}

\newtheorem{proposition}[theorem]{Proposition}

\newtheorem{corollary}[theorem]{Corollary}

\newtheorem{example}{Example}
\newtheorem{remarkx}[theorem]{Remark}
\newenvironment{remark}[1][]%
  {\if\relax\detokenize{#1}\relax\begin{remarkx}\else\begin{remarkx}[#1]\fi
   \normalfont}%
  {\end{remarkx}}
\newenvironment{proof}{\noindent
  \textbf{Proof.}}{\hfill$\Box$\\}

\newtheorem{observation}[theorem]{Observation}%

\def\rmbr#1{{\rm{(}\/}#1\/{\rm{)}}}
\newcommand{\vp}{\varphi}

\newcommand{\param}[1]{\ensuremath{\mathop{\mathit{par}} #1}}
\newcommand{\Forall}           [0]     {{\mathrlap{\forall}\mkern1.3mu\forall}}% informal quantifier (font-independent offset)
\newcommand{\Exists}           [0]     {{\mathrlap{\exists}\mkern1.7mu\exists}}% informal quantifier (font-independent offset)

\newcommand{\N}{\mathbb{N}}
\newcommand{\Npos}{\mathbb{N}^{+}}
 \newcommand{\lo}[1]{\mathbf{#1}} % logics
\newcommand{\cl}[1]{\mathcal{#1}} % classes of frames
\newcommand{\fr}[1]{\mathfrak{#1}} % frames, models
\newcommand{\Dplus}{D^{+}}
\newcommand{\CMLang}{\mathcal{ML}_{\mathrm{C}}} % the language
\newcommand{\CLang}{\mathcal{L}_{\mathrm{C}}} % the language
\newcommand{\MLang}{\mathcal{ML}} % the language
\newcommand{\MLangeq}{\mathcal{ML}_=} % the language
\newcommand{\CMLmon}{\CMLang^{\mathit{mon}}}

\newcommand{\md}{\operatorname{md}} \newcommand{\Typ}{\mathsf{T}}
\newcommand{\tp}{\mathsf{t}} \newcommand{\hist}{\mathsf{h}}

\newcommand{\sat}{\Vdash} \newcommand{\nsat}{\nVdash}
\newcommand{\ex}[1]{\exists^{\geqslant #1}}
\newcommand{\exle}[1]{\exists^{\leqslant #1}}
\newcommand{\exeq}[1]{\exists^{=#1}}
\newcommand{\NP}{\mathrm{NP}}
\newcommand{\coNP}{\mathrm{coNP}}
\newcommand{\PSPACE}{\mathrm{PSPACE}}

\newcommand{\EXP}{\mathrm{EXPTIME}}
\newcommand{\coEXP}{\mathrm{coEXPTIME}}
\newcommand{\NEXP}{\mathrm{NEXPTIME}}
\newcommand{\coNEXP}{\mathrm{coNEXPTIME}}

\newcommand{\cotwoNEXP}{\mathrm{co\text{-}2\text{-}NEXPTIME}}
\newcommand{\TwoEXP}{2\text{-}\mathrm{EXPTIME}}
\newcommand{\coTwoEXP}{\mathrm{co}\text{-}2\text{-}\mathrm{EXPTIME}}
\newcommand{\tower}{\mathrm{TOWER}}

\newcommand{\QCL}{\mathbf{QCL}}
\newcommand{\QCLe}{\mathbf{QCL}_=}
\newcommand{\CQCL}{\mathbf{QCL}_{\mathrm{C}}}

\newcommand{\GFe}{\ensuremath{\mathcal{G}_=}}
\newcommand{\FL}{\mathcal{F}}
\newcommand{\FLc}{\mathcal{F}_{\mathrm{C}}} % fluted modal formulas with counting quantifiers
\newcommand{\cFL}{\mathcal{FL}} % classical fluted formulas with counting quantifiers
\newcommand{\cFLc}{\mathcal{FL}_{\mathrm{C}}}
\newcommand{\ST}{\mathrm{ST}}
\newcommand{\STg}{\mathrm{gST}}
\newcommand{\Ax}{\mathit{Ax}}
\newcommand{\CML}{\mathop{\mathbf{ML}_{\mathrm{C}}^{\mathit{xd}}}}
\newcommand{\CMLcd}{\mathop{\mathbf{ML}_{\mathrm{C}}^{\mathit{cd}} }}
\newcommand{\ML}{\mathop{\mathbf{ML}^{\mathit{xd}}}}
\newcommand{\MLcd}{\mathop{\mathbf{ML}^{\mathit{cd}} }}

\newcommand{\dvalid}{\Vdash^{\mathit{d}}}

\newcommand{\subf}{\mathop{\mathit{sub}}}
\newcommand{\varset}{\mathop{\mathit{var}}}

\providecommand{\rc}{\mathit{c}}

\providecommand{\dom}{\mathrm{dom}\,}
\providecommand{\CMLangrc}{\CMLang^{\rc}}
\providecommand{\MLangeqrc}{\mathcal{ML}_{=}^{\rc}}
\providecommand{\MLangrc}{\mathcal{ML}^{\rc}}
\providecommand{\FLrc}{\FL^{\rc}}
\providecommand{\GFerc}{\mathcal{G}_{=}^{\rc}}
\providecommand{\Const}[1]{C_{#1}}

\newcommand{\Pn}[2]{P_{#1,#2}(\bar x_{#2})}
\newcommand{\Nn}[2]{N_{#1,#2}(\bar x_{#2})}
\newcommand{\Def}{\mathit{Def}}

\newcommand{\Ops}{\mathcal{O}}

\newcommand{\CMLO}[1]{\mathbf{ML}^{#1}_{\mathrm{C},\Ops}}
\newcommand{\MLeqO}[1]{\mathbf{ML}^{#1}_{=,\Ops}}
\newcommand{\MLO}[1]{\mathbf{ML}^{#1}_{\Ops}}
\newcommand{\Until}{\mathbin{\mathcal{U}}}
\newcommand{\Since}{\mathbin{\mathcal{S}}}
\newcommand{\Next}{\bigcirc}
\newcommand{\Prev}{\ominus}

\begin{document}

\title{Complexity transfer from classical fragments to modal predicate logics
  of finite Kripke frames}

\author{Dmitry Shkatov}

\maketitle

\begin{abstract}
  Over a single Kripke frame with finitely many worlds but possibly infinite
  domains, the decidable fragments of classical predicate logic pass to modal
  predicate logic at no cost in complexity.  With equality interpreted as
  identity, the two-variable and monadic fragments with and without counting
  quantifiers, the fluted fragments with or without counting quantifiers, and
  the guarded fragments are, without any monodicity restriction, each
  polynomial-time equivalent to their classical counterparts, also when the
  frame is part of the input.  Wherever the classical fragment has the finite
  model property, the modal fragment has the finite domain property, with
  explicit bounds.  The results extend to rigid constants, except for the fluted
  fragments with counting quantifiers, and, under a mild condition, to other
  operators definable on the frame, such as the universal modality and nominals.
  Over the usual classes of frames, by contrast, decidability typically requires
  monodicity, and can fail even then with counting quantifiers.  Our main tool
  is a translation into classical logic that preserves the arity of predicate
  letters and the number of variables, made polynomial by renaming subformulas
  within each fragment.  The monadic fragments, where renaming fails, are
  treated by histograms counting the elements of each type up to a cap; for a
  fixed frame, this gives $\PSPACE$ for a fixed number of predicate letters and
  polynomial time if the numbers of variables and constants are fixed too, and,
  as a by-product, $\coNEXP$-completeness of the classical monadic fragment with
  counting quantifiers, with a parameterwise optimal bound on model size.
\end{abstract}

%\keywords{modal predicate logic, finite Kripke frames, decidable fragments,
%  two-variable fragment, monadic fragment, fluted fragment, guarded fragment,
%  counting quantifiers, computational complexity, finite model property}

\maketitle

\section{Introduction}

Decidability in modal predicate logic is scarce.  For most common predicate
logics, undecidability obtains already for languages with two individual
variables and a single monadic predicate letter~\cite{RS19,Ryb24,RS25-trick}.
Decidable fragments can be obtained either by restricting the syntactic shape of
formulas so that full two-variable fragments with a single predicate letter are
avoided (this is done in one-variable fragments~\cite{GS98,Hampson2018,SS19} and
various monodic fragments~\cite{WZ01}; the latter allow more than one variable
but restrict the scope of modal operators to formulas with at most one
parameter) or by restricting the class of semantic structures used in the
interpretation of formulas. Unlike the situation in propositional modal logic,
the restriction to Kripke frames with finitely many worlds does not help: in
fact, it increases the complexity~\cite{RS20}.

This paper concerns the setting in which the frame is not only finite but fixed
in advance: formulas are interpreted over a single Kripke frame with finitely
many worlds, whose domains may nonetheless be infinite.  This is the simplest
semantic setting beyond classical predicate logic, and the one closest to it; it
is not trivial since the standard translation~\cite{Benthem} sends many
fragments of interest to undecidable fragments of classical logic.  Our main
result is that, in this setting, modality costs nothing: each of the well-known
decidable fragments of classical predicate logic---the one- and two-variable
fragments, the fluted fragments and the monadic fragments, with or without
counting quantifiers, and the guarded fragments---gives rise to a modal fragment
that has exactly the complexity of its classical counterpart, and the finite
domain property whenever the classical fragment has the finite model property,
with no restriction on the free variables of modalised formulas, for every
finite frame, and also when the frame is part of the input.

Single finite frames arise whenever a fixed finite set of states, contexts or
perspectives is combined with unbounded data.  Examples are first-order temporal
logics of finite-state systems~\cite{HWZ00,DFK06} and formalisms for
verification of data-aware processes, where the control is finite but the data
are not~\cite{DHPV09,BCDDM13,CDMP18}.  In knowledge representation, standpoint logic~\cite{GR21} interprets ontologies
relative to a set of precisifications; its first-order version~\cite{GRS22}
involves rigid constants and uses constant-domain semantics.
%Sentential
%formulas need only boundedly many precisifications and translate into the
%underlying first-order fragment~\cite{GRS22}, and monodic standpoint extensions
%of the two-variable fragment with counting retain its complexity~\cite{GR25};
%non-sentential formulas may need infinitely many precisifications~\cite{GRS22}.
%If standpoints are encoded by proposition letters, reasoning with $n$
%precisifications becomes reasoning over the $n$-element cluster with constant
%domains, where modal operators may apply to formulas with any number of free
%variables.
Contextualised knowledge repositories, finally, place description-logic
knowledge bases in a finite hierarchy of contexts along which knowledge
propagates~\cite{SH12}; in the arising computational problems, the frame is part
of the input.

Unfolding a fixed finite set of worlds into a formula of the underlying
classical fragment is a natural idea, and the translation of sentential
standpoint formulas in~\cite{GRS22} is an instance of it; what this paper adds
is a way of carrying out the unfolding, with polynomial overhead, for formulas
with free variables under modal operators and within each of the classical
fragments in turn, together with a different method for the monadic fragments,
where the unfolding cannot be kept polynomial in this way.

The only prior work on this setting that we are aware of is~\cite{RS24}, on
monadic fragments, where a L\"owenheim-style argument gives decidability and
the finite domain property, but no complexity bounds.  It suggested that the
methods used for the decidable fragments of classical logic can be adapted to
the modal setting one by one; this route seems available, but laborious.  We
take a different route: a tool that transfers results from classical logic
wholesale, and a separate method for the monadic fragments, where the tool
does not give optimal complexity.

\paragraph{Main result.}  Precise definitions are given in
Section~\ref{sec:prelim}.  Our main result is the following transfer theorem;
Table~\ref{tab:summary} gives, fragment by fragment, the complexity classes and
the bounds on the size of domains, with pointers to the proofs, and
Table~\ref{tab:contrast} sets some of these results against their counterparts
over classes of frames.

\begin{theorem}[Transfer]
  \label{thm:main}
  Let $\fr{F}$ be a Kripke frame with finitely many worlds, and let $\Gamma$ be
  one of the fragments of modal predicate logic listed in
  Table~\ref{tab:summary}, in which equality, where present, is interpreted as
  identity and modal operators may be applied to formulas with any number of
  free variables: the one-variable fragment with counting quantifiers; the
  two-variable fragments with and without counting quantifiers; the fluted
  fragments with and without counting quantifiers, with or without a bound on
  the number of variables; the guarded fragments with and without equality,
  with or without a bound $k\geqslant2$ on the number of variables; and the
  monadic fragments with and without counting quantifiers and equality.  Let
  $\Gamma_0$ be the classical fragment consisting of the $\square$-free formulas
  of $\Gamma$.
  \begin{itemize}
  \item[\rmbr{$1$}] Validity on $\fr{F}$, with expanding or with locally
    constant domains, is decidable for $\Gamma$, and is polynomial-time
    equivalent to classical validity for $\Gamma_0$.  This remains true when
    $\fr{F}$, given explicitly by its worlds and accessibility relation, is
    part of the input.
  \item[\rmbr{$2$}] If $\Gamma_0$ has the finite model property, then $\Gamma$
    has the finite domain property on $\fr{F}$, with the bounds on the size of
    domains given in Table~\ref{tab:summary}; hence, validity on $\fr{F}$
    coincides with validity on $\fr{F}$ in models with finite global domains.
    The two-variable fragment with counting quantifiers, whose classical
    counterpart lacks the finite model property, lacks the finite domain
    property.
  \end{itemize}
\end{theorem}

Theorem~\ref{thm:main} collects results proved in Sections~\ref{sec:complexity}
and~\ref{sec:types}, as Table~\ref{tab:summary} indicates.  The decidability
claim in \rmbr{$1$} already follows from the general unfolding of
Section~\ref{sec:transfer}; the complexity and domain bounds for the one- and
two-variable and the fluted fragments come from the polynomial renaming of
Sections~\ref{sec:two-var} and~\ref{sec:fluted}, those for the guarded fragments
from the guarded renaming with positive and negative names of
Section~\ref{sec:guarded}, and those for the monadic fragments from the
histogram method of Section~\ref{sec:types}.  The lower bounds are inherited
from the classical fragments, which embed into the modal ones
(Remark~\ref{rem:conserv}); the content of the theorem lies in the upper bounds,
which match the lower bounds exactly, and in the bounds on domains.  The theorem
continues to hold with rigid individual constants (Section~\ref{sec:constants}),
with the exceptions recorded in the caption of Table~\ref{tab:summary}; the
classical counterpart $\Gamma_0$ then consists of the $\square$-free formulas of
$\Gamma$ with their constants.  The polynomial-time equivalence in \rmbr{$1$} is
supplied by the renaming, which carries constants along unchanged, and the
complexity classes then follow from the known results on the classical
two-variable, one-variable and guarded fragments with constants
(Proposition~\ref{prop:classical-const}).  For the fluted fragments, whose
classical complexity is known only for formulas without constants, a separate
step is needed: we eliminate constants from classical fluted sentences without
increasing the number of variables (Lemma~\ref{lem:fluted-const}), in polynomial
time for each fixed number of variables and in elementary time in general, which
suffices for the $\tower$ bound (note that this does not give a polynomial-time
equivalence with the constant-free fluted fragment).  The theorem also continues
to hold with frame-definable operators in place of the box
(Section~\ref{sec:operators}), again with the exceptions recorded in the caption
of Table~\ref{tab:summary}.

\begin{table}[tp]
  \caption{Validity on a finite Kripke frame $\fr{F}=\langle W,R\rangle$, with
    expanding or with locally constant domains.  Unless the frame is said to be
    fixed or part of the input, every complexity result holds both for a fixed
    frame and for the frame as part of the input, and its lower bound holds
    already for $\square$-free formulas and is that of the classical fragment.
    With rigid individual constants, all results except those for fluted
    formulas with counting quantifiers continue to hold
    (Theorem~\ref{thm:rigid}), but the polynomial-time bound needs the number of
    constants to be fixed as well, the problem being $\coNP$-complete otherwise
    (Proposition~\ref{prop:const-fixed}), and the last column holds with the
    bounds of Theorem~\ref{thm:rigid}\rmbr{$6$}.  All results also hold with
    frame-definable operators in place of the box, under the hypotheses of
    Theorem~\ref{thm:operators}; the lower bound of
    Proposition~\ref{prop:frame-input-mon} then holds if the box is among the
    operators.  In the last column, ``yes'' means that every formula that is not
    valid is refuted in a model with a finite global domain, with the stated
    bound on its size, where one is stated; validity on $\fr{F}$ then coincides
    with validity on $\fr{F}$ in models with finite global domains.  For the
    two-variable fragment with counting quantifiers, which lacks the finite
    domain property, validity in such models is nonetheless $\coNEXP$-complete
    as well (Remark~\ref{rem:finite-validity}).}
  \label{tab:summary}
  \centering
  \footnotesize
  \renewcommand{\arraystretch}{1.2}
  \begin{tabular}{@{}>{\raggedright\arraybackslash}p{0.28\textwidth}>{\raggedright\arraybackslash}p{0.43\textwidth}>{\raggedright\arraybackslash}p{0.23\textwidth}@{}}
    \toprule
    Fragment & Complexity of validity & Finite domain property\\
    \midrule
    one variable, with counting and equality
      & $\coNP$-complete (Theorem~\ref{thm:one-var})
      & yes, $|W|\cdot2^{O(|\vp|)}$ (Theorem~\ref{thm:fdp-one-var})\\
    two variables, with counting and equality
      & $\coNEXP$-complete (Theorem~\ref{thm:two-var})
      & no (Remark~\ref{rem:no-FDP-counting})\\
    two variables, with equality, without counting
      & $\coNEXP$-complete (Theorem~\ref{thm:two-var-plain})
      & yes, $2^{O(|W|\cdot|\vp|)}$ (Theorem~\ref{thm:fdp-two-var})\\
    fluted, $k$ variables, without counting
      & $\coNP$-complete for $k=1$, $\coNEXP$-complete for $k\in\{2,3\}$,
        $\mathrm{co}$-$2$-$\NEXP$-complete for $k=4$, in
        $\mathrm{co}$-$(k{-}2)$-$\NEXP$ and
        $\mathrm{co}$-$\lfloor k/2\rfloor$-$\NEXP$-hard for $k\geqslant5$
        (Theorem~\ref{thm:fluted-complexity})
      & yes, $\max(1,k-2)$-fold exponential (Theorem~\ref{thm:FDP-fluted})\\
    fluted, without counting
      & $\tower$-complete (Theorem~\ref{thm:fluted-complexity})
      & yes (Theorem~\ref{thm:FDP-fluted})\\
    fluted, $k$ variables, with counting
      & $\coNP$-complete for $k=1$, $\coNEXP$-complete for $k=2$, in
        $\mathrm{co}$-$(k{-}1)$-$\NEXP$ and
        $\mathrm{co}$-$\lfloor k/2\rfloor$-$\NEXP$-hard for $k\geqslant3$
        (Theorem~\ref{thm:fluted-counting})
      & yes, $\max(1,k-1)$-fold exponential
        (Theorem~\ref{thm:fluted-FDP-counting})\\
    fluted, with counting
      & $\tower$-complete (Theorem~\ref{thm:fluted-counting})
      & yes (Theorem~\ref{thm:fluted-FDP-counting})\\
    guarded, $k\geqslant2$ variables, with or without equality
      & $\coEXP$-complete (Theorem~\ref{thm:guarded-complexity})
      & yes, $2^{O(|W|\cdot|\vp|)}$ if all arities are at most $k$
        (Theorem~\ref{thm:FDP-guarded})\\
    guarded, with or without equality
      & $\coTwoEXP$-complete (Theorem~\ref{thm:guarded-complexity})
      & yes, $2^{|W|\cdot2^{O(|\vp|\log|\vp|)}}$
        (Theorem~\ref{thm:FDP-guarded})\\
    monadic, with counting and equality
      & $\coNEXP$-complete (Theorem~\ref{thm:nexptime},
        Corollary~\ref{cor:complexity-monadic})
      & yes, $\max\{1,\kappa_\vp\}\cdot|\Typ(\fr{F},s)|$ elements (Theorem~\ref{thm:small});
        $|\Typ(\fr{F},s)|$ without equality and counting
        (Proposition~\ref{prop:fdp-ML})\\
    monadic, $s$ predicate letters fixed
      & $\PSPACE$-complete, for a fixed frame
        (Proposition~\ref{prop:pspace})
      & as above\\
    monadic, $s$ predicate letters and $v$ variables fixed
      & polynomial time, for a fixed frame
        (Proposition~\ref{prop:bounded-var})
      & as above\\
    monadic, one predicate letter
      & $\coNEXP$-complete, for the frame as part of the input, already
        for two variables (Proposition~\ref{prop:frame-input-mon})
      & as above\\
    \bottomrule
  \end{tabular}

\end{table}

\paragraph{Contributions.}  Beyond Theorem~\ref{thm:main} itself, the paper
contributes the following.
\begin{itemize}
\item \emph{A transfer principle.}  We give a translation of modal formulas
  into classical ones that, unlike the standard translation~\cite{Benthem},
  preserves the arity of predicate letters and the number of variables, sends
  fluted formulas to fluted formulas and, slightly amended, guarded formulas
  to guarded formulas.  It reduces validity on a finite frame to classical
  validity (Proposition~\ref{prop:tr}), from which the decidability of all
  the fragments follows at once (Section~\ref{sec:transfer}).
\item \emph{Polynomial renaming.}  The translation is exponential in the modal
  depth of the formula.  By renaming subformulas with fresh predicate letters,
  within each fragment, we bring it down to polynomial size; this gives the
  polynomial-time equivalence of Theorem~\ref{thm:main}\rmbr{$1$} and the finite
  domain property with explicit bounds for the one- and two-variable, fluted and
  guarded fragments (Section~\ref{sec:complexity}).
\item \emph{Histograms for monadic fragments.}  Renaming introduces predicate
  letters of arity greater than one, so it does not apply to monadic
  fragments.  For these we use histograms of Kripke models, which count the
  elements of each type up to a cap computed from the formula.  They give
  $\coNEXP$-completeness over every finite frame, settling a question left
  open in~\cite{RS24}, the finite domain property with a bound whose cap
  cannot be lowered, and, for a fixed frame, $\PSPACE$-completeness for a
  fixed number of predicate letters and polynomial time if the number of
  variables is fixed as well, whereas, with the frame as part of the input,
  one predicate letter and two variables already give $\coNEXP$-hardness.  As
  a by-product, the classical monadic fragment with counting quantifiers is
  $\coNEXP$-complete, with a parameterwise optimal bound on the size of
  models (Section~\ref{sec:types}).
\item \emph{Constants and other operators.}  The results extend to rigid
  individual constants, the classical counterparts then being the fragments
  with constants (Section~\ref{sec:constants}), except for the fluted
  fragments with counting quantifiers.  As the
  complexity of the classical fluted fragments is known only without
  constants, we also show how to eliminate constants from fluted sentences
  without increasing the number of variables, in polynomial time for each
  fixed number of variables and in elementary time in general
  (Lemma~\ref{lem:fluted-const}).  The results also extend to every operator
  whose truth at a world is given by a Boolean combination of truth values at
  the worlds of the frame, the size of these definitions counting towards the
  input---the boxes of a polymodal frame, the universal and
  difference modalities, nominals, graded modalities and the operators of
  linear temporal logic---provided that the operators consult only worlds
  accessible from the current one or that domains are constant on a
  connected frame (Section~\ref{sec:operators}).
\end{itemize}

The paper is structured as follows.  Section~\ref{sec:prelim} introduces the
modal languages and their semantics, and recalls the results on fragments of
classical predicate logic that we rely on.  Section~\ref{sec:transfer} defines
the translation and proves the transfer principle, from which decidability
follows at once.  Section~\ref{sec:complexity} brings the translation down to
polynomial size by renaming subformulas, and establishes the complexity and the
finite domain property of the one- and two-variable fragments
(Section~\ref{sec:two-var}), the fluted fragments (Section~\ref{sec:fluted}) and
the guarded fragments (Section~\ref{sec:guarded}).  Section~\ref{sec:types}
treats the monadic fragments by means of capped histograms.
Section~\ref{sec:constants} adds rigid individual constants, and
Section~\ref{sec:operators} treats frame-definable operators.
Section~\ref{sec:limits} concludes with open problems.  Appendices contain
proofs omitted from Sections~\ref{sec:complexity} through~\ref{sec:operators}.

\section{Preliminaries}
\label{sec:prelim}

\begin{table}[tp]
  \caption{Some results of this paper set against their counterparts over
    classes of frames.  Complexities are those of validity; $\lo{QK}$ is the
    smallest normal modal predicate logic.}
  \label{tab:contrast}
  \centering
  \footnotesize
  \renewcommand{\arraystretch}{1.2}
  \begin{tabular}{@{}>{\raggedright\arraybackslash}p{0.25\textwidth}>{\raggedright\arraybackslash}p{0.38\textwidth}>{\raggedright\arraybackslash}p{0.31\textwidth}@{}}
    \toprule
    Fragment & Over classes of frames & Over a single finite frame (this paper)\\
    \midrule
    one variable, without counting quantifiers
      & $\PSPACE$-complete for $\lo{QK}$~\cite{Hampson2018};
        $\coNEXP$-complete with constant domains~\cite{Marx99}
      & $\coNP$-complete (Theorem~\ref{thm:one-var})\\
    one variable, with counting quantifiers
      & $\coNEXP$-complete for $\lo{QK}$~\cite{Hampson2018}; undecidable for
        linear temporal logics~\cite{HK15} and for some fusions of decidable
        logics~\cite{KShW26}
      & $\coNP$-complete (Theorem~\ref{thm:one-var})\\
    two variables, one monadic predicate letter
      & undecidable for most common logics~\cite{RS19,RS25-trick,RShJLC21c,RSh26-Dedekind,RS26-ARQNL}
      & polynomial time for a fixed frame
        (Proposition~\ref{prop:bounded-var}); $\coNEXP$-complete for the frame
        as part of the input (Proposition~\ref{prop:frame-input-mon})\\
    two variables, with counting quantifiers
      & undecidable (contains the previous row) %; for monodic formulas,
        %$\coNEXP$-complete in standpoint logic~\cite{GR25}
      & $\coNEXP$-complete, with no monodicity restriction
        (Theorem~\ref{thm:two-var})\\
    monodic fragments
      & decidable over the usual classes of frames~\cite{HWZ00,WZ01}, but
        undecidable with counting quantifiers~\cite{HK15} or with non-rigid
        designators~\cite{AHKMW25,KShW26}
      & monodicity not needed; counting quantifiers and rigid constants
        handled (Table~\ref{tab:summary}, Section~\ref{sec:constants})\\
    \bottomrule
  \end{tabular}

\end{table}

\subsection{Syntax and semantics}

%We use the following notation for sets of numbers: $\N := \{0,1,2,\ldots\}$,
%$\Npos := \N-\{0\}$, and $\Nstar := \N\cup\{\aleph_0\}$.

A \emph{modal predicate language with counting quantifiers and equality},
denoted here by~$\CMLang$, contains the following symbols: countably many
(individual) variables $x_1,x_2,x_3,\ldots$ (we assume that this enumeration is
fixed); for every $n\in\N$, countably many $n$-ary predicate letters (nullary
predicate letters are also called \textit{proposition letters}); a designated
binary predicate letter $=$; the propositional constant $\bot$; the binary
Boolean connective $\to$; the unary modal operator~$\square$; for every
$M\in\Npos$, the counting quantifier symbol $\ex{M}$.  \textit{ Atomic
  formulas}, or simply \textit{atoms}, are $\bot$, formulas of the form
$P(x_1,\ldots,x_n)$, where $P$ is an $n$-ary predicate letter and
$x_1,\ldots,x_n$ are variables, and formulas of the form $x=y$, where $x$ and
$y$ are variables. Formulas are built from atoms using $\to$, $\square$ and the
counting quantifiers.  When writing $\CMLang$-formulas, we use the standard
abbreviations $\top$, $\neg$, $\vee$, $\wedge$, $\leftrightarrow$ and
$\Diamond$. In addition, we use the following abbreviations for quantifiers
(here, $M\in\N^+$):
\begin{align*}
\exists x\,\varphi &:= \ex{1}x\,\varphi; & \forall x\,\varphi &:=
                                                                \neg\ex{1}x\,\neg\varphi;
  & \ex{0}x\,\varphi &:=\top;  \\
 \exle{M}x\,\varphi & := \neg\ex{M+1}x\,\varphi; & \exeq{M}x\,\varphi &:= \ex{M}x\,\varphi\wedge\exle{M}x\,\varphi. & & 
\end{align*}
We identify the language $\CMLang$ with the set of its formulas.  We denote by
$\MLangeq$ the subset of $\CMLang$ containing formulas without counting
quantifiers (but with equality) and by $\MLang$ the subset of $\MLangeq$
containing formulas without equality (thus, the only primitive quantifier symbol
of both $\MLangeq$ and $\MLang$ is $\exists$).

A \textit{propositional formula} is a formula built from $\bot$ and proposition
letters using $\to$ and $\Box$; thus, such a formula contains neither variables
nor quantifiers.

The \emph{modal depth} $\md\varphi$ of a formula $\varphi$ is the maximal number
of nested modal operators in $\varphi$. Variables freely occurring in a formula
$\vp$ are called the \emph{parameters} of $\vp$; the set of parameters of $\vp$
is denoted by $\param{\vp}$. If $\param{\vp} = \varnothing$, then $\vp$ is a
\emph{sentence}.  The set of subformulas of a formula $\vp$ is denoted by
$\mathop{\mathit{sub}} \vp$.

The set $\FLc \subseteq \CMLang$ of \textit{fluted formulas with counting
  quantifiers} is defined as follows.  A \emph{fluted $m$-atom} is either
$\bot$, or an atom $P(x_\ell,\ldots,x_m)$, with $\ell \leqslant m+1$, where
arguments form a contiguous final segment of the sequence $x_1,\ldots,x_m$ (if
$\ell = m+1$, then $P$ is a proposition letter), or, in case $m\geqslant2$, the
atom $x_{m-1}=x_m$.  The set $\FLc^{[m]}$, where $m \geqslant 0$, is the least
set containing the fluted $m$-atoms, closed under $\to$ and $\square$, and such
that $\exists^{\geqslant M}x_{m+1}\,\vp$ belongs to $\FLc^{[m]}$ whenever
$\vp\in\FLc^{[m+1]}$ and $M\in\Npos$ (thus, $\exists x_{m+1}\,\vp$,
$\forall x_{m+1}\,\vp$, $\exists^{\leqslant M}x_{m+1}\,\vp$ and
$\exists^{=M}x_{m+1}\,\vp$ belong to $\FLc^{[m]}$, as well).  Lastly,
$\FLc := \bigcup_{m \geqslant 0} \FLc^{[m]}$.  If $\vp \in \FLc^{[m]}$, then the
parameters of $\vp$ form a final segment $x_\ell,\ldots,x_m$ of
$x_1,\ldots,x_m$, for some $\ell\leqslant m+1$, and the bound variables of $\vp$
are among $x_{m+1},x_{m+2},\ldots$\, .

The set of \textit{fluted formulas} is $\FL:=\FLc\cap\MLang$; we also put
$\FL^{[m]}:=\FLc^{[m]}\cap\MLang$. Since all the variables in fluted formulas
can be renamed down, we may assume without a loss of generality that all
\textit{fluted sentences} are members of $\FLc^{[0]}$ or $\FL^{[0]}$.
%the parameters of a formula in $\FLc^{[m]}$ form a final segment
%$x_\ell,\ldots,x_m$ of $x_1,\ldots,x_m$, for some $\ell\leqslant m+1$, and that
%its bound variables are among $x_{m+1},x_{m+2},\ldots$; moreover, by definition,
%a subformula of a formula in $\FLc^{[m]}$ that begins with a quantifier has the
%form $\exists^{\geqslant M}x_{j+1}\,\psi$ with $\psi\in\FLc^{[j+1]}$, for some
%$j\geqslant m$, so a quantifier nested within $d-1$ other quantifiers of a
%sentence in $\FLc^{[0]}$ binds $x_d$.  In particular, the quantifiers of a
%sentence in $\FLc^{[0]}$ whose variables are among $x_1,\ldots,x_k$ are nested
%at most $k$ deep.

The set $\GFe \subseteq \MLangeq$ of \textit{guarded formulas} (with equality)
is defined by the following recursion:
\begin{enumerate}[label=(G\arabic*)]
\item every atomic formula belongs to $\GFe$;
\item $\GFe$ is closed under $\to$ and $\square$;
\item if $\vp \in \GFe$, $\gamma$ is an atomic formula, and $\bar x$ is a
  non-empty tuple of distinct variables such that
  $\param \vp \cup \bar {x} \subseteq \param \gamma$, then
  $\exists\bar x\,(\gamma \wedge\vp) \in \GFe$; %\forall\bar x\,(\gamma \to\vp), 
\item if $\vp \in \GFe$ and $\param \vp \subseteq \{x\}$, then
  $\exists x\,\vp \in \GFe$. %\forall x\,\vp, 
\end{enumerate}
In (G3), the \emph{guard} $\gamma$ may be an atom $P(\bar x)$ or an equality
atom; (G4) admits unguarded quantification of formulas with at most one
parameter, as in the guarded fragment of~\cite{PH23} (see
Remark~\ref{rem:GF-variants}). We will assume that
$\forall\bar x\,(\gamma \to\vp) \in \GFe$ whenever the conditions of (G3) are
satisfied; this is not strictly true, but
$\theta := \forall\bar x\,(\gamma \to\vp)$ is logically equivalent to
$\neg \exists\bar x\,(\gamma \wedge \neg \vp)$, which has the guard $\gamma$ and
has the same variables and modal depth as $\theta$.

We use the following notation for fragments: if $\Gamma \subseteq \CMLang$, then
$$
\begin{array}{lcl}
  \Gamma^k & := & \{ \vp \in \Gamma : \mbox{$\vp$ contains at most $k$
                  variables} \}; \smallskip\\
  \Gamma^{\mathit{mon}} & := & \{ \vp \in \Gamma : \mbox{every predicate letter of
                               $\vp$ other than $=$ is nullary or monadic} \}.
\end{array}
$$
Thus, proposition letters are admitted in the \emph{monadic} fragments
$\Gamma^{\mathit{mon}}$; Remark~\ref{rem:prop-letters} shows how to dispense
with them.

A \emph{Kripke frame} is a pair $\fr{F}=\langle W,R\rangle$ where
$W \ne \varnothing$ and $R \subseteq W \times W$.  Elements of $W$ are called
\textit{worlds} or \textit{points}.  A Kripke frame $\langle W,R\rangle$ is
\emph{finite} if $W$ is a finite set. We say that $\fr{F}$ is \emph{connected}
if every two worlds are joined by a path along $R\cup R^{-1}$.

We use the standard notation for binary relations: if $R\subseteq W\times W$,
$w \in W$, and $X\subseteq W$, then $R^0$ is the identity on $W$;
$R^{n+1} := R\circ R^{n}$, for every $n\in\N$; $R^{\ast} := \bigcup_{n\in\N}R^n$
is the reflexive transitive closure of $R$, and
$R^{+} := \bigcup_{n\geqslant1}R^n$ its transitive closure;
$R(w) := \{v\in W : wRv\}$;
$R(X) := \bigcup\{R(w): w\in X\}$; and $R\restriction X := R\cap(X\times X)$.  A
Kripke frame $\langle W',R'\rangle$ is a \emph{subframe} of a Kripke frame
$\langle W,R\rangle$ if $W'\subseteq W$ and $R'=R\restriction W'$.
%If $n\in\N$, a Kripke frame $\langle W,R\rangle$ is

A \emph{Kripke model with expanding domains} ($\mathit{xd}$-model, for short),
or simply \textit{Kripke model}, over a Kripke frame
$\fr{F} = \langle W, R \rangle$ is a tuple
\mbox{$\fr{M}=\langle W,R,D,I\rangle$} where
\begin{itemize}
\item $D$ is a \emph{domain function} sending each $w\in W$ to a non-empty set
  $D(w)$ (often written as $D_w$ and called \textit{the domain of the world
    $w$}) satisfying the \emph{expanding domains condition}
  (\textit{$xd$-condition}, for short): for all $w,v\in W$,
  \begin{equation}
    \label{eq:expanding}
    wRv \implies D_w\subseteq D_{v};
  \end{equation}
\item $I$ is an \textit{interpretation} function sending a world $w\in W$ and an
  $n$-ary predicate letter~$P$ other than $=$ to an $n$-ary relation $I(w,P)$ on
  $D_w$ (we often write $P^{I,w}$ instead of $I(w,P)$).
\end{itemize}

\begin{remark}
  If $P$ is a proposition letter, then $I(w, P)$ is a subset of
  $D_w^{0}=\{\langle\rangle\}$, so $I(w, P) \in \{ \{\langle \rangle\},
  \varnothing \}$; we identify $\{\langle \rangle\}$ with truth and
  $\varnothing$ with falsity.  Then, we do not need to consider evaluation of
  proposition letters as a special case.
\end{remark}

A \emph{Kripke model with \rmbr{locally} constant domains} ($\mathit{cd}$-model,
for short) is a Kripke model satisfying the local $\mathit{cd}$-condition: for
all $w,v\in W$,
\begin{equation}
\label{eq:constant}
wRv \implies D_w=D_v.
\end{equation}
Note that, if $\fr{F}$ is connected, then all worlds of a $\mathit{cd}$-model
over $\fr{F}$ have the same domain.  

The \textit{global domain} of the model $\fr{M}=\langle W,R,D,I\rangle$ is the
set $\Dplus := \bigcup \{ D_w : w \in W \}$.

An \emph{assignment} in a model is a map $g$ sending each variable to an element
of~$\Dplus$. We write $g'\overset{x}{=}g$ if assignments~$g$ and~$g'$ differ in
at most the value of the variable~$x$.

The truth of a $\CMLang$-formula $\varphi$ at world $w$ of a model
$\fr{M}=\langle W,R,D,I\rangle$ under an assignment $g$ is defined by recursion:
\begin{itemize}
\item $\fr{M},w\sat^{g}P(x_1,\ldots,x_n)$ if $\langle g(x_1),\ldots,g(x_n)\rangle\in P^{I,w}$;
\item $\fr{M},w\sat^{g}x=y$ if $g(x)=g(y)$;
\item $\fr{M},w\not\sat^{g}\bot$;
\item $\fr{M},w\sat^{g}\varphi_1\to\varphi_2$ if $\fr{M},w \not\sat^{g}\varphi_1$ or $\fr{M},w\sat^{g}\varphi_2$;
\item $\fr{M},w\sat^{g}\square\varphi_1$ if $\fr{M},w'\sat^{g}\varphi_1$ whenever $w'\in R(w)$;
\item $\fr{M},w\sat^{g}\ex{M}x\,\varphi_1$ if there exist at least $M$ elements
  $a\in D_w$ such that $\fr{M},w\sat^{g'}\varphi_1$, where $g'\overset{x}{=}g$
  and $g'(x)=a$.
\end{itemize}
Thus, we interpret the equality symbol as identity, and the counting quantifiers
count elements of the domain of the current world.  Note that the truth clauses
for the abbreviations $\exists$ and $\forall$ are the usual ones:
\begin{itemize}
\item $\fr{M},w\sat^{g}\exists x\,\varphi_1$ if there exists $a\in D_w$ such
  that $\fr{M},w\sat^{g'}\varphi_1$, where $g'\overset{x}{=}g$ and $g'(x)=a$;
\item $\fr{M},w\sat^{g}\forall x\,\varphi_1$ if $\fr{M},w\sat^{g'}\varphi_1$
  whenever $g'\overset{x}{=}g$ and $g'(x) \in D_w$.
\end{itemize}

A formula $\vp$ is \emph{true} at world $w$ of a model $\fr{M}$ (notation:
$\fr{M},w\sat\varphi$) if $\fr{M},w\sat^{g}\varphi$ holds for every $g$ such
that $g(x) \in D_w$ for all $x \in \param \vp$; $\vp$ is \emph{true} in $\fr{M}$
(notation: $\fr{M} \sat \vp$) if $\fr{M},w\sat\varphi$ for all $w \in W$. For
$d \in \{\mathit{xd}, \mathit{cd} \}$, a formula $\vp$ is \emph{$d$-valid} at
world $w$ of Kripke frame $\fr{F}$ (notation: $\fr{F}, w \dvalid \vp$) if
$\fr{M}, w \Vdash \vp$ holds for every $d$-model $\fr{M}$ over $\fr{F}$; $\vp$
is \emph{$d$-valid} on Kripke frame $\fr{F}$ (notation: $\fr{F} \dvalid \vp$) if
$\vp$ is true in every $d$-model over $\fr{F}$; $\vp$ is
\emph{$\mathit{d}$-valid} on a class $\cl{C}$ of frames (notation:
$\cl{C} \dvalid \vp$) if it is $\mathit{d}$-valid on every frame from~$\cl{C}$.
A formula $\vp$ is \emph{refuted} at a world $w$ of a model $\fr{M}$ if
$\fr{M},w\nsat\vp$, and \emph{refuted} in $\fr{M}$ if it is refuted at some
world of $\fr{M}$, that is, if $\fr{M}\nsat\vp$.  Thus, $\vp$ is $d$-valid on a
Kripke frame $\fr{F}$ if, and only if, it is refuted in no $d$-model over
$\fr{F}$.  Observe that a formula is valid on a frame if, and only if, its
universal closure is; thus, when considering validity, we restrict our attention
to sentences.

We use the following notation. Let $\fr{M}=\langle W,R,D,I\rangle$ be a Kripke
model, $w\in W$ and $a_1,\ldots,a_n\in D_w$; let also $\varphi(x_1,\ldots,x_n)$
be a formula with
% with no parameters other than
$\param \vp \subseteq \{x_1,\ldots,x_n\}$, and let $g$ be an assignment with
$g(x_1)=a_1,\ldots,g(x_n)=a_n$. Then
\mbox{$\fr{M},w\sat\varphi(a_1,\ldots,a_n)$} means
$\fr{M},w\sat^{g} \varphi(x_1,\ldots,x_n)$. Similarly, if $\fr{A}$ is a
classical model and $a_1,\ldots,a_n$ are elements of its universe, then
$\fr{A}\models\varphi[a_1,\ldots,a_n]$ means that $\varphi$ is true in $\fr{A}$
under an assignment sending $x_i$ to $a_i$, for $i=1,\ldots,n$.

We define logics semantically:
if $\cl{C}$ is a class of Kripke frames, then, for
$d \in \{\mathit{xd}, \mathit{cd} \}$,
$$
\begin{array}{lcl}
  \mathbf{ML}_{\mathrm{C}}^d \cl{C} & := & \{\varphi \in \CMLang : \cl{C} \dvalid \varphi\};
\smallskip\\
  \mathbf{ML}_=^d \cl{C} & := & \{\varphi \in \MLangeq : \cl{C} \dvalid
                                \varphi\}; \smallskip\\
  \mathbf{ML}^d \cl{C} & := & \{\varphi \in \MLang : \cl{C} \dvalid \varphi\}. 
%\Logfin{\cl{C}} &= \{\varphi\in\CML : \AFfin{\cl{C}}\sat\varphi\}; & \Logcfin{\cl{C}} &= \{\varphi\in\CML : \AFcfin{\cl{C}}\sat\varphi\}.
\end{array}
$$
We write $\mathbf{ML}^d_\mathrm{C} \fr{F}$ instead of
$\mathbf{ML}^d_\mathrm{C} \{\fr{F}\}$, for
$d \in \{\mathit{xd}, \mathit{cd} \}$, and similarly for
$\mathbf{ML}^d_= \fr{F}$ and $\mathbf{ML}^d \fr{F}$.  The sets $\CML \cl{C}$ and
$\CMLcd \cl{C}$ are normal modal predicate logics: they contain the smallest
normal modal predicate logic $\mathbf{QK}$ and are closed under Modus Ponens,
Generalisation, Necessitation, and Predicate Substitution.

For $d\in\{\mathit{xd},\mathit{cd}\}$, a set $\Gamma\subseteq\CMLang$ has the
\emph{finite domain property} over a Kripke frame $\fr{F}$ for $d$-models if
every $\vp\in\Gamma\setminus\mathbf{ML}^d_\mathrm{C}\fr{F}$ is refuted in a
$d$-model over $\fr{F}$ with a finite global domain; if $\Gamma=\CMLang$, we
also say that $\mathbf{ML}^d_\mathrm{C}\fr{F}$ has the finite domain property,
and similarly for $\mathbf{ML}^d_= \fr{F}$ and $\mathbf{ML}^d \fr{F}$.

The \emph{size} $|\vp|$ of a formula $\vp$ is the number of occurrences of
symbols in $\vp$, where every predicate letter and every variable counts as a
single symbol, and the number $M$ in a counting quantifier $\ex{M}$ contributes
the length of its binary representation.  The \emph{counting threshold} $M_\vp$
of a formula $\vp$ is the largest $M$ such that $\ex{M}$ occurs in $\vp$ (thus,
$\exle{M}$ and $\exeq{M}$ contribute $M+1$); if $\vp$ contains no quantifiers,
then $M_\vp:=1$.  Since numbers in counting quantifiers are written in binary,
$M_\vp\leqslant 2^{|\vp|}$.

%; this is the measure of~\cite{PH23},
%which we use for classical formulas as well.
Whenever the frame is input to a computational problem, it is given explicitly,
as a list of its worlds and of the pairs of its accessibility relation, so that
its size is $|\fr{F}|=|W|+|R|$.
%When $\fr{F}$ is part of the input, it is given
%in this explicit form, by listing $W$ and $R$; a succinct description of a
%frame, such as the number $n$ written in binary for the chain of length $n$, is
%a different input, of size logarithmic in $|\fr{F}|$, to which our results on
%the frame as part of the input do not apply as stated (see
%Section~\ref{sec:limits}).
When a formula is given as input to an algorithm, it is written as a word in a
finite alphabet, in which the indices of predicate letters and variables are
written out.  As the predicate letters and variables of a formula can be
renamed, in polynomial time and without affecting validity, to be among the
first $|\vp|$ of each kind, this word has length $O(|\vp|\log|\vp|)$; similarly,
$\fr{F}$ can be written as a word of length $O(|\fr{F}|\log|\fr{F}|)$.  All the
complexity classes considered in this paper are closed under such changes of the
size measure, and all formulas constructed in our reductions have predicate
letters and variables with indices polynomial in the size of the input; hence,
the choice between the two measures does not affect any of our results
(cf.~\cite[Section~1.2]{PH23}).

\subsection{Classical fragments}
\label{sec:QCL}

We briefly recall the definitions of classical fragments whose modal
counterparts we shall be considering; to make the paper self-contained, we also
recall the results about these fragments that we shall be relying on.

The sets of validities of classical first-order logic without equality, with
equality, and with counting quantifiers (and equality) are denoted by,
respectively, $\mathbf{QCL}$, $\mathbf{QCL}_=$, and $\CQCL$; the corresponding
languages are denoted by, respectively, $\mathcal{L}$, $\mathcal{L}_=$, and
$\mathcal{L}_{\mathrm{C}}$; we identify these languages with the sets of
$\square$-free formulas of, respectively, $\MLang$, $\MLang_=$ and $\CMLang$.

\begin{proposition}[\cite{Lewis80,Furer84,PH23}]
  \label{prop:lewis}
  The sets $\mathbf{QCL}^{\mathit{mon}}$ and $\mathbf{QCL}_=^{\mathit{mon}}$ are
  $\coNEXP$-complete.  The lower bound holds already for sentences with two
  variables and without equality \rmbr{see the proof of~\cite[Lemma~3.21]{PH23}}.
\end{proposition}

%
%\begin{proposition}
%  \label{prop:QCLmon} The set $\CQCL^{\mathit{mon}}$ is decidable.
%\end{proposition}
%
%\begin{proof}
%  Follows from Proposition~\ref{prop:lewis}, since counting quantifiers can be
%  eliminated preserving monadicity.
%\end{proof}
%
%\begin{remark}
%  Even though $\coNEXP$-completeness of $\CQCL^{\mathit{mon}}$ is a routine
%  consequence of L\"owenheim's identification of elements of the same type,
%  adapted to counting quantifiers (a satisfiable sentence has a model in which
%  every type is realised at most $\kappa_\vp$ times, and such a model can be
%  guessed and checked in exponential time), we have not found it stated in the
%  literature.
%\end{remark}

\begin{proposition}[\cite{GKV97}, Corollary 5.4]
  \label{prop:FO2}
  The sets $\mathbf{QCL}_=^2$ and $\mathbf{QCL}^2$ are $\coNEXP$-complete.  The
  lower bound holds already for sentences without equality whose predicate
  letters are monadic \rmbr{see Proposition~\ref{prop:lewis}}.
\end{proposition}

\begin{proposition}[\cite{PH08}; \cite{PH23}, Theorem 7.18]
  \label{prop:CQCL1}
  The set $\CQCL^1$ is $\coNP$-complete.
\end{proposition}

\begin{proposition}[\cite{PH23}, Theorem 8.10]
  \label{prop:CQCL2}
  The set $\CQCL^2$ is $\coNEXP$-complete.
\end{proposition}

The definition of the sets $\cFLc$ and $\cFL$ of classical fluted
formulas, respectively, with and without counting quantifiers, is obtained from
the definitions of, respectively, the sets $\FLc$ and $\FL$ of modal fluted
formulas by omitting the clause for $\square$. As in the modal case, we will
assume that all classical fluted sentences are members of $\cFLc^{[0]}$
or $\mathcal{FL}^{[0]}$.  The set $\mathcal{GF}_=$ of classical guarded formulas
(with equality) is defined by omitting the $\square$-clause from
the definition of $\mathcal{G}_=$.

\begin{proposition}[\cite{PH23}, Theorem 6.6]
  \label{prop:FL2}
  The set $\QCL^2 \cap \mathcal{FL}$ is $\coNEXP$-complete.
\end{proposition}

Following \cite[Section~6.5]{PH23}, we denote by $\tower$ the class of
problems decidable by a Turing machine in time bounded by a tower of
exponentials whose height is an elementary function of the size of the input
(deterministic and non-deterministic machines give the same class), and we
understand $\tower$-completeness with respect to elementary-time many-one
reductions, as in \cite[Section~6.5]{PH23}; $\tower$ is closed under such
reductions.  The reductions of this paper from modal to classical problems are
polynomial-time, with one exception: the elimination of constants from fluted
sentences with an unbounded number of variables in the proof of
Theorem~\ref{thm:rigid}\rmbr{$2$} is elementary.

\begin{proposition}[\cite{PH23}, Theorems 6.10, 6.13, and 6.14]
  \label{prop:FL}
  \mbox{}
  \begin{itemize}
  \item[\rmbr{$1$}] For every $k \geqslant 3$, the set
    $\QCL^k \cap \mathcal{FL}$ is in $\mathrm{co}$-$(k\!\!-\!\!2)$-$\NEXP$ and is
    $\mathrm{co}$-$\lfloor k/2\rfloor$-$\NEXP$-hard; in particular,
    $\QCL^3 \cap \mathcal{FL}$ is $\coNEXP$-complete and
    $\QCL^4 \cap \mathcal{FL}$ is $\mathrm{co}$-$2$-$\NEXP$-complete.
  \item[\rmbr{$2$}] The set $\mathbf{QCL} \cap \mathcal{FL}$ is $\tower$-complete.
  \end{itemize}
\end{proposition}

\begin{proposition}[\cite{PH21}, Theorem~1 and Lemma~4]
  \label{prop:FLC}
  \mbox{}
  \begin{itemize}
  \item[\rmbr{$1$}] The set $\CQCL^2 \cap \cFLc$ is $\coNEXP$-complete.
  \item[\rmbr{$2$}] For every $k \geqslant 3$, the set
    $\CQCL^k \cap \cFLc$ is in $\mathrm{co}$-$(k\!\!-\!\!1)$-$\NEXP$ and is
    $\mathrm{co}$-$\lfloor k/2\rfloor$-$\NEXP$-hard.
  \item[\rmbr{$3$}] The set $\CQCL \cap \cFLc$ is $\tower$-complete.
  \end{itemize}
\end{proposition}

The lower bounds in Proposition~\ref{prop:FLC} are those of
Propositions~\ref{prop:FL2} and~\ref{prop:FL}, since
$\mathcal{FL}\subseteq\cFLc$, and the upper bound in~(1) also follows from
Proposition~\ref{prop:CQCL2}.

\begin{proposition}[Model size for fluted sentences with counting]
  \label{prop:ffc}
  For every $k\geqslant2$, every satisfiable \mbox{$\cFLc^{k}$-sentence} has a
  model whose size is bounded by a $(k-1)$-fold exponential function of the size
  of the sentence.
\end{proposition}

Proposition~\ref{prop:ffc}, which Theorem~\ref{thm:fluted-FDP-counting}
requires, is not stated in~\cite{PH21}, but follows from the bounds carried
by the lemmas proved there; since one of these lemmas needs a qualification,
we give the derivation in Appendix~\ref{sec:prop:ffc}.

\begin{proposition}[\cite{PH23}, Theorem 4.9]
  \label{prop:GF}
  \mbox{}
  \begin{itemize}
  \item[\rmbr{$1$}] For every $k \geqslant 2$, the set
    $\QCLe^k \cap \mathcal{GF}_=$ is $\coEXP$-complete.
  \item[\rmbr{$2$}] The set $\QCLe \cap \mathcal{GF}_=$ is $\coTwoEXP$-complete.
  \end{itemize}
\end{proposition}

\begin{remark}
  \label{rem:GF-variants}
  The guarded fragment as defined in~\cite{PH23} differs from $\mathcal{GF}_=$
  only in that it allows individual constants, which we add in
  Section~\ref{sec:constants}.  As $\mathcal{GF}_=$ is contained in the guarded
  fragment of~\cite{PH23}, the upper bounds of Proposition~\ref{prop:GF} hold
  for $\mathcal{GF}_=$.  So does the lower bound in~\rmbr{$1$}, already for
  sentences without equality: the proof of~\cite[Lemma~4.7]{PH23} derives it
  from~\cite[Lemma~2.20]{PH23}, which establishes $\EXP$-hardness of
  satisfiability of finite sets of sentences of the forms
  $$
  \exists x\,\alpha(x),\quad \forall x\,\lambda(x),\quad
  \forall x\,\bigl(\alpha(x)\to\exists y\,(r(x,y)\wedge\beta(y))\bigr),\quad
  \forall x\,\bigl(\alpha(x)\to\forall y\,(r(x,y)\to\lambda(y))\bigr),
  $$
  where $r$ is a binary predicate letter, $\alpha$ and $\beta$ are conjunctions
  of monadic literals, and $\lambda$ is a disjunction of monadic literals; these
  sentences contain neither constants nor equality, the quantifiers over $y$ are
  guarded by the atom $r(x,y)$, as in (G3), and the quantifiers over $x$ bind
  the only parameter of their matrix, as in (G4); thus, the conjunction of such
  a set is a sentence of $\mathcal{GF}_=$ with two variables and without
  equality, all of whose predicate letters other than $r$ are monadic.  The
  lower bound in~\rmbr{$2$} is from~\cite{BKP25}: by~\cite[Theorem~3]{BKP25},
  satisfiability of sentences of the equality-free guarded adjacent fragment is
  $\TwoEXP$-hard; formulas of the adjacent fragment of~\cite{BKP25} contain no
  constants, and its guarded sub-fragment is its intersection with the guarded
  fragment; thus, every quantifier in these sentences is guarded by an atom, and
  they are sentences of $\mathcal{GF}_=$ without equality.  
\end{remark}

\begin{remark}[Conservativity]
  \label{rem:conserv}
  The modal logics of this paper are conservative extensions of their classical
  counterparts: for every Kripke frame $\fr{F}$, every
  $d\in\{\mathit{xd},\mathit{cd}\}$ and every $\vp \in \CLang$,
  $$
  \begin{array}{lcl}
    \vp \in \mathbf{ML}^d_{\mathrm{C}} \fr{F} & \iff & \vp \in \CQCL,
  \end{array}
  $$
  and similarly for $\mathbf{ML}^d$ and $\mathbf{ML}^d_=$ with, respectively,
  $\QCL$ and $\QCLe$. Indeed, if $\fr{M}$ is a Kripke model over $\fr{F}$ and
  $w \in W$, then the truth of $\vp$ at $w$ in $\fr{M}$ coincides with its truth
  in the classical model with universe $D_w$ interpreting every predicate letter
  $P$ as $P^{I,w}$; hence, a classically valid formula is true at every world of
  every model over $\fr{F}$.  Conversely, if $\vp$ is refuted in a classical
  model $\fr{A}$ with universe $A$, then $\vp$ is refuted at every world of the
  $\mathit{cd}$-model $\langle W,R,D,I \rangle$ over $\fr{F}$ with $D_w := A$
  and $I(w,P) := P^{\fr{A}}$, for all $w \in W$ and all predicate letters
  $P$. Hence, the lower bounds reported in this section are lower bounds for the
  corresponding fragments of the modal logics of finite Kripke frames.
\end{remark}

\section{A transfer principle for finite frames}
\label{sec:transfer}

In this section, we state our translation of predicate modal formulas evaluated
over a fixed finite Kripke frame $\fr{F} = \langle W, R \rangle$ into classical
formulas evaluated in classical structures that encode Kripke models over
$\fr{F}$.  The main difference of our translation from the well-known standard
translation~\cite{Benthem} is that no predicate symbol is introduced for~$R$;
rather, the evaluation of modalised formulas takes into account the structure of
$\fr{F}$; thus, just like for modal formulas, the structure of the frame is
implicit in the evaluation of classical correspondents; this also allows us to
preserve the arity of atomic formulas.

\subsection{The translation}

Throughout this section, we fix a finite Kripke frame $\fr{F}=\langle W,R\rangle$
and a formula $\vp \in \CMLang$.  We assume that the predicate letters of $\vp$
other than $=$ are $P_1, \ldots, P_s$.  We define the signature
$\sigma_{\fr{F}}(\vp)$ to contain, beside $=$, the following symbols:
\begin{itemize}
\item for every $w\in W$ and every $n$-ary predicate letter $P_i$ ($i = 1, \ldots, s$), an
  $n$-ary predicate letter~$P_i^{w}$;
\item for every $w\in W$, a monadic predicate letter $D_w$.
\end{itemize}
Thus, the signature $\sigma_{\fr{F}}(\vp)$ contains $|W| \cdot (s + 1)$
predicate letters.

We next state the `axiom of $\fr{F}$', a formula describing systems of domains
over $\fr{F}$; it says that domains of worlds are non-empty and expand along
$R$:
$$
\begin{array}{lcl}
  \Ax^{\mathit{xd}}_{\fr{F}} & := & \displaystyle \bigwedge_{w \in W} \Bigl( \exists x\,
                              D_w(x) \wedge \bigwedge_{v \in R(w)}
                    \forall x\,\bigl(D_w(x)\to D_v(x)\bigr) \Bigr).
\end{array}
$$
We also state the $\mathit{cd}$-version of the axiom saying that domains of
$R$-related worlds are equal:
$$
\begin{array}{lcl}
  \Ax^{\mathit{cd}}_{\fr{F}} & := & \displaystyle \Ax^{\mathit{xd}}_{\fr{F}} \wedge
                                    \bigwedge_{w \in W} \bigwedge_{v \in R(w)} \forall x\,\bigl(D_v(x)\to D_w(x)\bigr).
\end{array}
$$

If $\fr{M}=\langle W,R,D,I\rangle$ is a Kripke model over $\fr{F}$, then we
define the classical model $\fr{A}^{\fr{M}}$ in the signature
$\sigma_{\fr{F}}(\vp)$ with universe $D^+$ as follows:
\begin{itemize}
\item $D^{\fr{A}}_w = D(w)$, for every $w \in W$;
\item $(P_i^{w})^{\fr{A}} = I(w,P_i)$, for every $w \in W$ and every
  $i = 1, \ldots, s$.
\end{itemize}
It should be clear that $\fr{A}^{\fr{M}} \models \Ax^{\mathit{xd}}_{\fr{F}}$ and
that $\fr{A}^{\fr{M}} \models \mathit{Ax}^{\mathit{cd}}_{\fr{F}}$ whenever
$\fr{M}$ is a $\mathit{cd}$-model.

Conversely, if $\fr{A}$ is a classical model in the signature
$\sigma_{\fr{F}}(\vp)$ such that $\fr{A} \models \Ax^{\mathit{xd}}_{\fr{F}}$,
then we put
\begin{itemize}
  \item $D^{\fr{A}}(w) := D^{\fr{A}}_w$, for every $w \in W$;
  \item $I^{\fr{A}}(w,P_i) := (P_i^{w})^{\fr{A}}\cap(D^{\fr{A}}_w)^n$, for every $w
    \in W$ and every $n$-ary $P_i$ ($i = 1, \ldots, s$);
  \item \mbox{$\fr{M}^{\fr{A}} := \langle W,R,D^{\fr{A}}, I^{\fr{A}} \rangle$}.
\end{itemize}
Then, $\fr{M}^{\fr{A}}$ is a Kripke model: by the first conjunct of
$\Ax^{\mathit{xd}}_{\fr{F}}$, each $D^{\fr{A}}(w)$ is non-empty, and, by the
second conjunct, domains expand along $R$.  Moreover, if
$\fr{A} \models \mathit{Ax}^{\mathit{cd}}_{\fr{F}}$, then $\fr{M}^{\fr{A}}$ is a
$\mathit{cd}$-model.  The following should be clear:

\begin{observation}
  \label{obs:M}
  If $\fr{M}$ is a Kripke model over $\fr{F}$, then
  $\fr{M}^{\fr{A}^{\fr{M}}}=\fr{M}$.
\end{observation}

Now, we define, for every $w\in W$, the function $\mathrm{ST}^{\mathfrak{F}}_w$
sending $\CMLang$-formulas to classical $\sigma_{\fr{F}}(\vp)$-formulas with
counting quantifiers and equality (below, $i = 1, \ldots, s$, and the empty
conjunction is taken to be $\top$):
\begin{align*}
  \mathrm{ST}^{\mathfrak{F}}_w(P_i(x_1,\ldots,x_n)) &:= P_i^{w}(x_1,\ldots,x_n); & \mathrm{ST}^{\mathfrak{F}}_w(x=y) &:= (x=y);\\
  \mathrm{ST}^{\mathfrak{F}}_w(\bot)&:=\bot; & \mathrm{ST}^{\mathfrak{F}}_w(\psi_1\to\psi_2) &:= \mathrm{ST}^{\mathfrak{F}}_w(\psi_1)\to\mathrm{ST}^{\mathfrak{F}}_w(\psi_2);\\
  \mathrm{ST}^{\mathfrak{F}}_w(\square\psi_1) &:= \displaystyle \bigwedge_{v\in R(w)}\mathrm{ST}^{\mathfrak{F}}_v(\psi_1); & \mathrm{ST}^{\mathfrak{F}}_w(\ex{M}x\,\psi_1) &:= \ex{M}x\,\bigl(D_w(x)\wedge\mathrm{ST}^{\mathfrak{F}}_w(\psi_1)\bigr).
\end{align*}

\begin{remark}
  \label{rem:st}
  Due to the clause for $\square$, the size of $\mathrm{ST}^{\mathfrak{F}}_w(\vp)$
  can be exponential in $|\vp|$.
\end{remark}

\begin{lemma}[Translation lemma]
  \label{lem:st}
  Let $\fr{F}=\langle W,R\rangle$ be a finite Kripke frame, let
  $\vp \in \CMLang$ be a formula whose predicate letters other than $=$ are
  $P_1, \ldots, P_s$, and let $\fr{A}$ be a classical
  $\sigma_{\fr{F}}(\vp)$-model with $\fr{A} \models \Ax^{\mathit{xd}}_{\fr{F}}$. Then,
  for every $\psi(x_1, \ldots, x_n) \in \subf \vp$, every $w\in W$, and every
  $\bar a\in(D^{\fr{A}}_w)^n$,
  \[
    \begin{array}{lcl}
      \fr{M}^{\fr{A}}, w \Vdash \psi(\bar a) & \iff & \fr{A} \models \mathrm{ST}^{\fr{F}}_w(\psi)[\bar a].
    \end{array}
  \]
\end{lemma}

\begin{proof}
  Induction on $\psi$.

  Let $\psi = P_i(\bar{x})$.  Since $\bar{a} \in (D^{\fr{A}}_w)^n$ and
  $I^{\fr{A}}(w,P_i) = (P_i^{w})^{\fr{A}} \cap (D^{\fr{A}}_w)^n$,
  $$
  \begin{array}{lclclclclcl}
    \fr{M}^{\fr{A}}, w \Vdash P_i(\bar a) & \iff & \bar{a} \in I^{\fr{A} }(w,P_i) &
                                                                                \iff &  \bar{a} \in(P_i^{w})^{\fr{A}} 
  & \iff \fr{A} \models P_i^w( \bar{a}).
  \end{array}
  $$

  Let $\psi = \square\psi_1$.  If $v\in R(w)$, then, by the second conjunct of
  $\Ax^{\mathit{xd}}_{\fr{F}}$, we know that $\bar a\in(D^{\fr{A}}_v)^n$.  Thus, due
  to IH and finiteness of $R(w)$,
  $$
  \begin{array}{lclclclclcl}
    \fr{M}^{\fr{A}}, w \Vdash  \square \psi_1 (\bar{a}) & \iff &
    (\Forall v \in R(w))\,  \fr{M}^{\fr{A}}, v \Vdash  \psi_1 (\bar{a}) & \iff &
    \fr{A} \models \displaystyle \bigwedge_{v\in R(w)}\mathrm{ST}^{\mathfrak{F}}_v(\psi_1)[\bar{a}].                                                                             
  \end{array}
  $$
  Let $\psi = \ex{M}x\,\psi_1$, where $x$ may or may not be among
  $x_1,\ldots,x_n$, and let $g$ be an assignment with $g(x_i)=a_i$, for
  $i=1,\ldots,n$.  For $b\in D^{\fr{A}}_w$, let $g_b\overset{x}{=}g$ be the
  assignment with $g_b(x)=b$.  Then, the left-hand side counts the elements
  $b\in D^{\fr{A}}_w$ with \mbox{$\fr{M}^{\fr{A}},w \Vdash^{g_b} \psi_1$}, and
  the right-hand side counts the elements $b\in D^{\fr{A}}_w$ with
  \mbox{$\fr{A} \models^{g_b} \mathrm{ST}^{\fr{F}}_w(\psi_1)$}.  Since every
  parameter of $\psi_1$ is among $x_1,\ldots,x_n,x$, the assignment $g_b$
  sends the parameters of $\psi_1$ to elements of $D^{\fr{A}}_w$; hence, by
  IH, the counts are equal.

  The other cases are straightforward.
\end{proof}

\begin{corollary}
  \label{cor:st}
  Let $\fr{M}$ be a model over a finite Kripke frame $\fr{F}=\langle W,R\rangle$
  and let $\vp \in \CMLang$ be a formula whose predicate letters other than $=$
  are $P_1, \ldots, P_s$. Then, for every
  $\psi(x_1, \ldots, x_n) \in \subf \vp$, every $w\in W$, and every
  $\bar a\in(D_w)^n$,
  \[
    \begin{array}{lcl}
      \fr{M}, w \Vdash \psi(\bar a) & \iff & \fr{A}^{\fr{M}} \models \mathrm{ST}^{\fr{F}}_w(\psi)[\bar a].
    \end{array}
  \]
\end{corollary}

\begin{proof}
  Immediate from Lemma~\ref{lem:st} and Observation~\ref{obs:M}.
\end{proof}

\begin{proposition}[Transfer principle]
  \label{prop:tr}
  Let $\fr{F} = \langle W, R \rangle$ be a finite Kripke frame, let
  $\vp\in\CMLang$ be a sentence, and let $d\in\{\mathit{xd},\mathit{cd}\}$.
  Then,
  $$
  \begin{array}{lcl}
    \fr{F} \dvalid \vp & \iff & \Ax^{d}_{\fr{F}} \to
    \displaystyle\bigwedge_{w\in W} \mathrm{ST}^{\fr{F}}_w(\vp) \in \CQCL.
  \end{array}
  $$
\end{proposition}

\begin{proof}
  If $\vp$ is refuted at a world $w_0$ of a $d$-model $\fr{M}$ over $\fr{F}$,
  then $\fr{A}^{\fr{M}}\models\Ax^{d}_{\fr{F}}$ and, by Corollary~\ref{cor:st},
  $\fr{A}^{\fr{M}}\not\models\ST^{\fr{F}}_{w_0}(\vp)$; thus, the implication is
  not classically valid.  Conversely, if the implication is refuted in a
  classical model $\fr{A}$, then $\fr{A}\models\Ax^{d}_{\fr{F}}$ and
  $\fr{A}\not\models\ST^{\fr{F}}_{w_0}(\vp)$ for some $w_0\in W$; then
  $\fr{M}^{\fr{A}}$ is a $d$-model over $\fr{F}$ and, by Lemma~\ref{lem:st},
  $\fr{M}^{\fr{A}},w_0\nsat\vp$.
\end{proof}

The classical sentence in Proposition~\ref{prop:tr} is computable from $\fr{F}$
and $\vp$.  Hence, the proposition yields the decidability of every modal
fragment that the translation, together with the axiom of $\fr{F}$, keeps inside
a decidable classical fragment: as we shall see, this covers the one- and
two-variable fragments, the monadic fragment, the fluted fragments, and, with
the amended translation of Section~\ref{sec:guarded}, the guarded fragment.  The
complexity of these fragments, however, requires the finer analysis of
Sections~\ref{sec:complexity} and~\ref{sec:types}, as the sentence
$\ST^{\fr{F}}_w (\vp)$ can be exponential in the modal depth of $\vp$
(Remark~\ref{rem:st}).

\begin{observation}
  \label{obs:tr-kvar} For every $k \geqslant 1$, if
  $\vp \in \mathcal{ML}_{\mathrm{x}}^k$, then
  $\ST^{\fr{F}}_w (\vp) \in \mathcal{L}^k_{\mathrm{x}}$, where $\mathrm{x}$ is
  $\mathrm{C}$, $=$, or the empty string.
  %\begin{itemize}
  %\item[\rmbr{$1$}] if $\vp \in \CMLang^k$, then
  %  $\ST^{\fr{F}}_w (\vp) \in \mathcal{L}^k_{\mathrm{C}}$;
  %\item[\rmbr{$2$}] if $\vp \in \MLangeq^k$, then
  %  $\ST^{\fr{F}}_w (\vp) \in \mathcal{L}_=^k$;
  %\item[\rmbr{$3$}] if $\vp \in \MLang^k$, then
  %  $\ST^{\fr{F}}_w (\vp) \in \mathcal{L}^k$.
  %\end{itemize}
\end{observation}

\begin{observation}
  \label{obs:monadicity}
  If $\vp \in \mathcal{ML}^{\mathit{mon}}_{\mathrm{x}}$, then
  $\ST^{\fr{F}}_w (\vp) \in \mathcal{L}^{\mathit{mon}}_{\mathrm{x}}$, where
  $\mathrm{x}$ is $\mathrm{C}$, $=$, or the empty string.
  %\begin{itemize}
  %\item[\rmbr{$1$}] if $\vp \in \CMLang^{\mathit{mon}}$, then
  %  $\ST^{\fr{F}}_w (\vp) \in \mathcal{L}^{\mathit{mon}}_{\mathrm{C}}$;
  %\item[\rmbr{$2$}] if $\vp \in \MLangeq^{\mathit{mon}}$, then
  %  $\ST^{\fr{F}}_w (\vp) \in \mathcal{L}_=^{\mathit{mon}}$;
  %\item[\rmbr{$3$}] if $\vp \in \MLang^{\mathit{mon}}$, then
  %  $\ST^{\fr{F}}_w (\vp) \in \mathcal{L}^{\mathit{mon}}$.
  %\end{itemize}
\end{observation}

\begin{proposition}
  \label{prop:flutedness}
  For all $\vp \in \CMLang$, if $\vp \in \FLc^{[0]}$, then
  $\mathrm{ST}^{\fr{F}}_w(\varphi) \in \cFLc^{[0]}$; in particular, if
  $\vp\in\FL^{[0]}$, then $\mathrm{ST}^{\fr{F}}_w(\varphi) \in \mathcal{FL}^{[0]}$.
\end{proposition}

\begin{proof}
  We show, by induction on the construction of $\psi$, that
  \mbox{$\ST^{\fr{F}}_w(\psi) \in \cFLc^{[m]}$} whenever $\psi \in \FLc^{[m]}$.
  %; for $m=0$, this is the first statement, and
  %the second statement then follows since the translation introduces neither
  %equality nor counting quantifiers.
  
  If $\psi$ is a fluted $m$-atom, then $\ST^{\fr{F}}_w(\psi)$ is a fluted
  $m$-atom: $\ST^{\fr{F}}_w(\bot) = \bot$, $\ST^{\fr{F}}_w(x_{m-1}=x_m)$ is
  $x_{m-1}=x_m$, and $\ST^{\fr{F}}_w(P(x_\ell,\ldots,x_m))$ is
  $P^w(x_\ell,\ldots,x_m)$.  The cases for $\to$ and $\square$ follow from IH and
  closure of sets $\FLc^{[m]}$ and $\cFLc^{[m]}$ under connectives.

  Lastly, if $\psi = \exists^{\geqslant M} x_{m+1}\,\psi_1$, with
  $\psi_1 \in \FLc^{[m+1]}$, then
  $$
  \begin{array}{lcl}
  \ST^{\fr{F}}_w(\psi) & = & \exists^{\geqslant M} x_{m+1}\,\bigl(D_w(x_{m+1}) \wedge
    \ST^{\fr{F}}_w(\psi_1)\bigr),
  \end{array}  
  $$
  where $D_w(x_{m+1})$ is a fluted $(m+1)$-atom and, by IH,
  $\ST^{\fr{F}}_w(\psi_1) \in \cFLc^{[m+1]}$; hence,
  $(D_w(x_{m+1}) \wedge \ST^{\fr{F}}_w(\psi_1)) \in \cFLc^{[m+1]}$, and so
  $\ST^{\fr{F}}_w(\psi) \in \cFLc^{[m]}$.
\end{proof}

\section{Complexity of two-variable, fluted, and guarded fragments}
\label{sec:complexity}

Because of the clause for $\square$, the formula $\ST^{\fr{F}}_w(\vp)$ may have
size exponential in $\md\vp$.  To use the translation to establish
polynomial-time equivalence to well-known classical fragments, we bring it down
to polynomial size by renaming subformulas, as in the Scott normal form for the
two-variable fragment of classical logic.

In the three subsections, the renaming takes the same form.  For every $w\in W$
and every subformula $\psi$ of $\vp$, we introduce a fresh predicate letter, the
\emph{name} of $\psi$ at $w$, with one argument for each parameter of $\psi$,
and we define it by the formula obtained from the translation of $\psi$ at $w$
by replacing the translations of the immediate subformulas of $\psi$ with their
names.  The conjunction of these definitions, of the sentence
$\Ax^{\mathit{xd}}_{\fr{F}}$ or $\Ax^{\mathit{cd}}_{\fr{F}}$ of
Section~\ref{sec:transfer} and of the statement that $\vp$ fails at some world
is a sentence, of size polynomial in $|\fr{F}|+|\vp|$, that is satisfiable if,
and only if, $\vp$ is refuted in a model over $\fr{F}$, with the same bound on
the size of models in both directions; thus, both the upper complexity bounds
and the finite domain property transfer from the classical fragment to the modal
one.  What differs between the subsections is what keeps the definitions inside
the fragment: nothing is needed for the two-variable fragments; for the fluted
fragments, the parameters of $\psi$ are listed, and the definitions closed, in
the order that flutedness prescribes; for the guarded fragments, the translation
is first amended to translate guarded quantifier blocks as blocks, each name
comes in a positive and a negative version, and each definition is an
implication guarded by the name itself.

\begin{remark}
  \label{rem:frame-input}
  The complexity results of this section are stated for a fixed finite Kripke
  frame $\fr{F}$, but they hold in the same form for the problems in which
  $\fr{F}$ is part of the input: given a finite Kripke frame $\fr{F}$, listed
  explicitly as described in Section~\ref{sec:prelim}, and a sentence $\vp$ of
  the fragment in question, decide whether $\vp\in \mathbf{L}$, where
  $\mathbf{L} \in \{ \mathbf{ML}^d_{\mathrm{C}}\fr{F}, \mathbf{ML}^d_=\fr{F},
  \mathbf{ML}^d\fr{F} \}$, with $d \in\{\mathit{xd},\mathit{cd}\}$.  Indeed, the
  upper bounds are obtained by computing from $\fr{F}$ and $\vp$, in time
  polynomial in $|\fr{F}|+|\vp|$, a sentence of the corresponding classical
  fragment, whose satisfiability is then decided; and the lower bounds hold for
  every fixed frame, in particular for the one-world frame.  
\end{remark}

\subsection{Two-variable and one-variable fragments (with counting)}
\label{sec:two-var}

Given a sentence $\vp \in \CMLang^2$, we define sentences
$\Phi^d_{\fr{F}}(\vp) \in \CLang^2$, where
$d \in \{ \mathit{xd}, \mathit{cd} \}$, computable in time polynomial in
$|\fr{F}|+|\vp|$, such that $\Phi^d_{\fr{F}}(\vp)$ is satisfiable iff $\vp$ is
refuted in a $\mathit{d}$-model over $\fr{F}$ (Lemmas~\ref{lem:Phi}
and~\ref{lem:renaming}).

For every $w\in W$ and every $\psi \in \subf \varphi$, introduce a fresh
predicate letter $Q_{w,\psi}$ of arity $|\param \psi|$ (thus, at most $2$, since
$\vp \in \CMLang^2$; if $\psi$ is a sentence, then $Q_{w,\psi}$ is a proposition
letter; when we write $Q_{w,\psi}(\bar x)$, we assume that $\bar x$ lists
$\param{\psi}$ in some fixed order).  The intended meaning of $Q_{w,\psi}$ is
`$\psi$ holds at $w$'.  For every such $w$ and $\psi$, we define, by recursion,
the formula $\delta_{w,\psi}$ whose intended meaning coincides with
$\mathrm{ST}^{\fr{F}}_w (\psi)$ modulo renaming the translations of the
immediate subformulas with the corresponding atoms (below, $\bar{x}_i$ is the
list of $\param{\psi_i}$ in some fixed order):
$$
  \begin{array}{ll}
    \psi\text{ is atomic:} & \delta_{w,\psi}:=\ST^{\fr{F}}_w(\psi); \smallskip\\
    \psi=\psi_1\to\psi_2: & \delta_{w,\psi}:=Q_{w,\psi_1}(\bar x_1)\to
                            Q_{w,\psi_2}(\bar x_2); \smallskip\\
    \psi=\square\psi_1: & \delta_{w,\psi}:=\displaystyle\bigwedge_{v\in R(w)}Q_{v,\psi_1}(\bar{x});\smallskip\\
    \psi=\exists^{\geqslant M} x \,\psi_1 : & \delta_{w,\psi}:=\exists^{\geqslant
                                             M} x \,\bigl(D_w(x)\wedge Q_{w,\psi_1}(\bar{x}_1)\bigr).
  \end{array}
  $$
We then put
$$
\begin{array}{lcl}
  \mathit{Def} & := & \displaystyle \bigwedge_{w \in W} \bigwedge_{\psi \in \subf
                      \vp} \forall\bar x\,(Q_{w,\psi}(\bar
x)\leftrightarrow\delta_{w,\psi}).
\end{array}
$$
Lastly, we put
$$
\begin{array}{lclclcl}
  \Phi^{\mathit{xd}}_{\fr{F}}(\varphi) & := & \displaystyle \Ax^{\mathit{xd}}_{\fr{F}} \wedge \mathit{Def} \wedge\bigvee_{w\in
                                              W}\neg Q_{w,\varphi}, & & 
  \Phi^{\mathit{cd}}_{\fr{F}}(\varphi) & := & \displaystyle \mathit{Ax}^{\mathit{cd}}_{\fr{F}} \wedge \mathit{Def} \wedge\bigvee_{w\in
                                              W}\neg Q_{w,\varphi}.
\end{array}
$$

\begin{remark}
  \label{rem:var-conv}
  We take the variable $x$ of $\Ax^{\mathit{xd}}_{\fr{F}}$ and
  $\Ax^{\mathit{cd}}_{\fr{F}}$ to be one of the variables of $\vp$ whenever
  $\vp$ has any; thus, $\Phi^{\mathit{xd}}_{\fr{F}}(\vp)$ and
  $\Phi^{\mathit{cd}}_{\fr{F}}(\vp)$ contain no more variables than $\vp$ does,
  unless $\vp$ is propositional.
\end{remark}

\begin{observation}
  \label{obs:ML2}
  If $\vp \in \mathcal{ML}_{\mathrm{x}}^k$, then
  $\Phi^{\mathit{d}}_{\fr{F}}(\varphi) \in \mathcal{L}_{\mathrm{x}}^k$, where
  $k \in \Npos$, $\mathrm{x}$ is $\mathrm{C}$, $=$, or the empty string, and
  $d \in \{ \mathit{xd}, \mathit{cd} \}$.
  %Let $d \in \{ \mathit{xd}, \mathit{cd} \}$.  Then,
  %\begin{itemize}
  %\item[\rmbr{$1$}] 
  %\item[\rmbr{$2$}] If $\vp \in \mathcal{ML}^2$, then
  %$\Phi^{\mathit{d}}_{\fr{F}}(\varphi) \in
  %\mathcal{L}^2$.
%\item[\rmbr{$3$}] If $\vp \in \CMLang^1$, then
 % $\Phi^{\mathit{d}}_{\fr{F}}(\varphi) \in \CLang^1$.
 % \end{itemize}
\end{observation}

\begin{lemma}
  \label{lem:Phi}
  The sentences $\Phi^{\mathit{d}}_{\fr{F}}(\vp)$, where
  $d \in \{ \mathit{xd}, \mathit{cd} \}$, have size
  $O\bigl(|W|\cdot|\vp|\cdot(|W|+|\vp|)\bigr)$ and are computable in time
  polynomial in $|\fr{F}|+|\vp|$.
\end{lemma}

\begin{proof}
  The sentences $\Ax^{\mathit{xd}}_{\fr{F}}$ and $\Ax^{\mathit{cd}}_{\fr{F}}$ are conjunctions
  of at most $|W|+2|R|\leqslant3|W|^2$ formulas of constant size, and
  $\bigvee_{w\in W}\neg Q_{w,\vp}$ has $|W|$ disjuncts of constant size.  The
  formula $\mathit{Def}$ has at most $|W|\cdot|\vp|$ conjuncts; each conjunct
  has size $O(|\vp| + |W|$) (the worst case is when $\delta_{w, \psi}$ is
  written for $\psi = \square \psi_1$).
\end{proof}

\begin{lemma}[Renaming for two-variable sentences]
  \label{lem:renaming}
  %\label{rem:FDP}
  Let $\fr{F}=\langle W,R\rangle$ be a finite Kripke frame, let
  $\vp\in\CMLang^2$ be a sentence, and let
  $d \in \{ \mathit{xd}, \mathit{cd} \}$.  Then, for every cardinal
  $\kappa$, the sentence $\vp$ is refuted in a $d$-model over $\fr{F}$ whose
  global domain has at most $\kappa$ elements iff
  $\Phi^{\mathit{d}}_{\fr{F}}(\vp)$ has a model with at most $\kappa$
  elements.  In particular, $\vp$ is refuted in a $d$-model over $\fr{F}$ iff
  $\Phi^{\mathit{d}}_{\fr{F}}(\vp)$ is satisfiable.
\end{lemma}

\begin{proof}
  First, routine induction on $\psi$ shows that, if $\fr{A}$ is a classical
  $\sigma_{\fr{F}}(\vp)$-model and $\fr{B}$ is an expansion of $\fr{A}$ to the
  letters $Q_{w,\psi}$, then,
  $$
  \begin{array}{lcl}
    \fr{B}\models\mathit{Def} & \iff & \fr{B}\models\forall\bar x\,\bigl(Q_{w,\psi}(\bar x)\leftrightarrow\ST^{\fr{F}}_w(\psi)\bigr),
                                      \mbox{for all $w\in W$ and $\psi\in\subf\vp$}. 
  \end{array}
  \eqno{(\ast)}
  $$
  
  Now, suppose that $\fr{M},w_0\nsat\vp$, for some $d$-model
  $\fr{M} = \langle W, R, D, I \rangle$ and some $w_0\in W$.  Then
  $\fr{A}^{\fr{M}} \models \Ax^{d}_{\fr{F}}$ and, by Corollary~\ref{cor:st},
  $\fr{A}^{\fr{M}}\not\models\ST^{\fr{F}}_{w_0}(\vp)$.  Let $\fr{B}$ be the
  expansion of $\fr{A}^{\fr{M}}$ interpreting each $Q_{w,\psi}$ as the relation
  defined by $\ST^{\fr{F}}_w(\psi)$; then, by ($\ast$), $\fr{B} \models \Def$.
  Since $\Ax^d_{\fr{F}}$ contains no letters $Q_{w,\psi}$, surely
  $\fr{B} \models\Ax^d_{\fr{F}}$.  As
  $\fr{A}^{\fr{M}}\not\models\ST^{\fr{F}}_{w_0}(\vp)$, it follows, by the choice
  of $\fr{B}$, that $\fr{B} \models\neg Q_{w_0,\vp}$.  Hence,
  $\fr{B} \models\Phi^d_{\fr{F}}(\vp)$; as the universe of $\fr{B}$ is $D^+$,
  this gives the cardinality claim in one direction.

  Conversely, suppose that $\fr{B}\models\Phi^d_{\fr{F}}(\vp)$.
  % and $|\fr{B}| \leqslant D^+$.
  Let $\fr{A}$ be the $\sigma_{\fr{F}}(\vp)$-reduct of $\fr{B}$. Then,
  $\fr{A} \models \Ax^d_{\fr{F}}$ and
  $\fr{B} \models \bigvee_{w\in W}\neg Q_{w,\vp}$.  Choose $w_0 \in W$ with
  $\fr{B} \models \neg Q_{w_0,\vp}$.  Since $\fr{A}\models\Ax^d_{\fr{F}}$, the
  structure $\fr{M}^{\fr{A}}$ is a $d$-model over $\fr{F}$.  Since
  $\fr{B} \models \neg Q_{w_0,\vp}$ and since $\fr{B}\models\mathit{Def}$, it
  follows by ($\ast$) that $\fr{B}\not \models\ST^{\fr{F}}_{w_0}(\vp)$; since
  $\ST^{\fr{F}}_{w_0}(\vp)$ is a $\sigma_{\fr{F}}(\vp)$-sentence, it follows
  that $\fr{A}\not \models\ST^{\fr{F}}_{w_0}(\vp)$.  Hence, by
  Lemma~\ref{lem:st}, $\fr{M}^{\fr{A}},w_0\nsat\vp$; as the global domain of
  $\fr{M}^{\fr{A}}$ is a subset of the universe of $\fr{B}$, this gives the
  cardinality claim in the other direction.
\end{proof}

\begin{theorem}[Two-variable fragment with counting]
\label{thm:two-var}
Let $\fr{F}=\langle W,R\rangle$ be a finite Kripke frame.  The sets
$(\mathbf{ML}_{\mathrm{C}}^d \fr{F})^2$, for
$d \in \{ \mathit{xd}, \mathit{cd} \}$, are $\coNEXP$-complete.  The lower
bound holds already for $\square$-free monadic sentences without equality and
counting quantifiers.
\end{theorem}

\begin{proof}
  The upper bound is immediate from Lemmas~\ref{lem:Phi}
  and~\ref{lem:renaming}, and Proposition~\ref{prop:CQCL2}.  The lower
  bound, already for the sentences described in the statement, follows from
  Proposition~\ref{prop:FO2} and Remark~\ref{rem:conserv}.
\end{proof}

\begin{theorem}[Two-variable fragments without counting]
\label{thm:two-var-plain}
Let $\fr{F} = \langle W,R\rangle$ be a finite Kripke frame.  The sets
$(\mathbf{ML}^d \fr{F})^2$ and $(\mathbf{ML}^d_= \fr{F})^2$, for $d \in \{
\mathit{xd}, \mathit{cd} \}$,  are $\coNEXP$-complete.  The lower bound holds
already for $\square$-free monadic sentences without equality.
\end{theorem}

\begin{proof}
  The upper bound follows from Theorem~\ref{thm:two-var}, the lower bound,
  already for the sentences described in the statement, from
  Proposition~\ref{prop:FO2} and Remark~\ref{rem:conserv}.
\end{proof}

\begin{theorem}[Finite domain property for two-variable fragments without
  counting]
  \label{thm:fdp-two-var}
  Let $\fr{F}=\langle W,R\rangle$ be a finite Kripke frame and
  $d\in\{\mathit{xd},\mathit{cd}\}$.  If
  $\vp\in\mathcal{ML}_=^2\setminus\mathbf{ML}^d_=\fr{F}$, then $\vp$ is
  refuted in a $d$-model $\langle W,R,D,I\rangle$ with
  $|D^+|\leqslant2^{O(|W|\cdot|\vp|)}$.
\end{theorem}

\begin{proof}
  Follows from an estimate of the size of models of satisfiable
  $\mathcal{L}^2_=$-sentences~\cite[Sections~3.1 and~3.2]{PH23} applied to
  $\Phi^{d}_{\fr{F}}(\vp)$; see Appendix~\ref{sec:thm:fdp-two-var}.
\end{proof}

\begin{remark}[No finite domain property for the two-variable fragment with
  counting]
  \label{rem:no-FDP-counting}
  An analogue of Theorem~\ref{thm:fdp-two-var} does not hold for
  $(\mathbf{ML}^d_{\mathrm{C}}\fr{F})^2$, where
  $d\in\{\mathit{xd},\mathit{cd}\}$: the classical sentence
  $$
  \begin{array}{lcl}
  \chi:=\forall x\,\exeq{1}y\,S(x,y)\wedge\forall y\,\exle{1}x\,S(x,y)\wedge\exists
    y\,\forall x\,\neg S(x,y),
  \end{array}  
  $$
  which says that $S$ is the graph of an injective function that is not
  surjective, is satisfiable, but not finitely satisfiable (see,
  e.g.,~\cite{PH05}); thus, $\CQCL^2$ does not have the finite model property.
  By Remark~\ref{rem:conserv}, this transfers to its modal counterparts.
\end{remark}

\begin{remark}[Finite-domain validity]
  \label{rem:finite-validity}
  Although $(\mathbf{ML}^d_{\mathrm{C}}\fr{F})^2$ lacks the finite domain
  property, validity on $\fr{F}$ in $d$-models with a finite global domain for
  \mbox{$\CMLang^2$-formulas} is, again, $\coNEXP$-complete, also when the frame
  is part of the input: for the upper bound, let $\vp\in\CMLang^2$ be a
  sentence; by Lemma~\ref{lem:renaming}, $\vp$ is refuted in a $d$-model over
  $\fr{F}$ with a finite global domain if, and only if,
  $\Phi^{d}_{\fr{F}}(\vp)$, which is computable in polynomial time by
  Lemma~\ref{lem:Phi}, has a finite model---now the upper bound follows from
  membership of finite satisfiability of $\CLang^2$-sentences in
  $\NEXP$~\cite[Theorem~8.10]{PH23}; for the lower bound, we can use
  Theorem~\ref{thm:two-var}: since the argument of Remark~\ref{rem:conserv}
  preserves the finiteness of domains, for $\square$-free sentences without
  counting quantifiers, validity in finite models over $\fr{F}$ coincides with
  finite classical validity; since $\mathcal{L}^2_=$ has the finite model
  property~\cite[Lemma~3.8]{PH23}, the latter coincides with classical validity.
  For every other fragment in Table~\ref{tab:summary}, validity in models with
  finite global domains coincides with validity, by the finite domain property.
\end{remark}

We next consider one-variable fragments.

\begin{theorem}[One-variable fragment with counting]
\label{thm:one-var}
Let $\fr{F}=\langle W,R\rangle$ be a finite Kripke frame.  The sets
$(\mathbf{ML}^d_{\mathrm{C}}\fr{F})^1$, for $d\in\{\mathit{xd},\mathit{cd}\}$,
are $\coNP$-complete.  The lower bound holds already for propositional
formulas.
\end{theorem}

\begin{proof}
  The upper bound follows from Lemmas~\ref{lem:Phi} and~\ref{lem:renaming},
  Observation~\ref{obs:ML2}, and Proposition~\ref{prop:CQCL1}.  The lower bound
  follows from Remark~\ref{rem:conserv}.
\end{proof}

\begin{remark}
  Theorem~\ref{thm:one-var} shows that, over a single finite Kripke frame,
  adding to the modal propositional logic one individual variable and counting
  quantifiers costs nothing in complexity.  This is in contrast with the
  situation over classes of frames: the one-variable fragment of $\lo{QK}$ is
  $\coNEXP$-complete with counting quantifiers~\cite[Theorem 4.3]{Hampson2018}
  and $\PSPACE$-complete without them~\cite[Corollary 5.5]{Hampson2018}; over
  constant domains, i.e., for $\lo{QK}$ plus the Barcan formula, the
  one-variable fragment is $\coNEXP$-complete both with~\cite[Theorem
  3.1]{Hampson2018} and without counting quantifiers~\cite{Marx99}; counting
  quantifiers may lead to undecidability even with one variable: fusions of
  decidable one-variable logics with counting quantifiers may be
  undecidable~\cite{KShW26}, and one-variable fragments of linear temporal
  logics with counting quantifiers are undecidable~\cite{HK15}.
\end{remark}

\begin{theorem}[Finite domain property for the one-variable fragment with
  counting]
  \label{thm:fdp-one-var}
  Let $\fr{F}=\langle W,R\rangle$ be a finite Kripke frame and
  $d\in\{\mathit{xd},\mathit{cd}\}$.  If
  $\vp\in\CMLang^1\setminus\mathbf{ML}^d_{\mathrm{C}}\fr{F}$, then $\vp$ is
  refuted in a $d$-model $\langle W,R,D,I\rangle$ with
  $|D^+|\leqslant|W|\cdot(1+|\vp|\cdot M_\vp)\leqslant|W|\cdot2^{2|\vp|}$, where
  $M_\vp$ is the counting threshold of $\vp$.
\end{theorem}

%\begin{theorem}[Finite domain property for the one-variable fragment with
%  counting]
%  \label{thm:fdp-one-var}
%  Let $\fr{F}=\langle W,R\rangle$ be a finite Kripke frame and
%  $d\in\{\mathit{xd},\mathit{cd}\}$.  If
%  $\vp\in\CMLang^1\setminus\mathbf{ML}^d_{\mathrm{C}}\fr{F}$, then $\vp$ is
%  refuted in a $d$-model $\langle W,R,D,I\rangle$ with
%  $|D^+|\leqslant2^{O(|W|\cdot|\vp|)}$.
%\end{theorem}

\begin{proof}
  See Appendix~\ref{sec:thm:fdp-one-var}.
\end{proof}

\subsection{Fluted fragments}
\label{sec:fluted}

We now adapt the renaming of Section~\ref{sec:two-var} by allowing any number of
variables and making the definitions of names $Q_{w, \psi}$ fluted; the
construction covers fluted formulas with counting quantifiers at no extra cost.
We shall be concerned with the full fluted fragment as well as with its
subfragments with fixed numbers of variables.

Let $k\geqslant1$ and let $\vp \in \FLc^{k}$ be a sentence; as explained in
Section~\ref{sec:prelim}, we may assume that $\vp \in \FLc^{[0]}$; thus, the
variables of $\vp$ are among $x_1, \ldots, x_k$.  Every occurrence of a
subformula $\psi$ in $\vp$ has a \emph{level}: the number $m$ of quantifiers of
$\vp$ whose scope contains the occurrence (note that $m\leqslant k$).  By the
definition of $\FLc^{[0]}$, an occurrence of level $m$ belongs to $\FLc^{[m]}$,
and its parameters are $x_\ell, \ldots, x_m$, for some $\ell \leqslant m+1$.
The level of an occurrence of $\psi$ depends only on $\psi$, unless $\psi$ is a
propositional formula: if $\psi$ has parameters, the last of them is $x_m$, and
if $\psi$ is a sentence containing a quantifier, its outermost quantifiers bind
$x_{m+1}$.  We may, therefore, speak of the level of $\psi\in\subf\vp$ (rather
than of levels of its occurrences); we take the level of a propositional formula
to be $0$.

For every $w\in W$ and every $\psi\in\subf\vp$ of level $m$, with parameters
$x_\ell,\ldots,x_m$, introduce a fresh predicate letter $Q_{w,\psi}$ of arity
$m-\ell+1\leqslant k$, and use it only in the fluted $m$-atom
$Q_{w,\psi}(x_\ell,\ldots,x_m)$.  Define the formula $\delta_{w,\psi}$ as
follows:
  $$
  \begin{array}{ll}
    \psi\text{ is atomic:} & \delta_{w,\psi}:=\ST^{\fr{F}}_w(\psi);\\
    \psi=\psi_1\to\psi_2: & \delta_{w,\psi}:=Q_{w,\psi_1}(\bar x_1)\to Q_{w,\psi_2}(\bar x_2);\\
    \psi=\square\psi_1: & \delta_{w,\psi}:=\displaystyle\bigwedge_{v\in R(w)}Q_{v,\psi_1}(\bar x);\\
    \psi=\exists^{\geqslant M} x_{m+1}\,\psi' : & \delta_{w,\psi}:=\exists^{\geqslant M} x_{m+1}\,\bigl(D_w(x_{m+1})\wedge Q_{w,\psi'}(\bar x')\bigr),
  \end{array}
  $$
  where $\bar x$ is the list of the parameters of $\psi$, and $\bar x_1$ and
  $\bar x_2$ are the lists of parameters of, respectively, $\psi_1$ and $\psi_2$
  (all of these are final segments of $x_1,\ldots,x_m$); $\bar{x}'$ is the list
  of parameters of $\psi'$, a final segment of $x_1,\ldots,x_{m+1}$.  Then, for
  every $w\in W$,
\begin{equation}
  \label{eq:delta-fluted}
  \delta_{w,\psi}\in \cFLc^{[m]}\quad\text{whenever $\psi\in\subf\vp$ has level $m$;}
\end{equation}
indeed, the atoms of $\delta_{w,\psi}$ are fluted $m$-atoms, the clauses for
$\to$ and $\square$ combine fluted \mbox{$m$-atoms} with $\to$ and $\wedge$, and
the counting quantifier binds $x_{m+1}$ in a formula from $\cFLc^{[m+1]}$ (in
this case, $m+1 \leqslant k$).

\begin{observation}
  \label{rem:fl}
  If $\vp\in\FL$, then $\delta_{w,\psi}\in \cFL^{[m]}$, as, in this case,
  $\delta_{w,\psi}$ contains neither equality nor any quantifiers other than
  $\exists^{\geqslant1}\,(= \exists)$.
\end{observation}

Now, let $\mathit{Def}^{\mathrm{fl}}$ be the conjunction, over all $w\in W$ and
all $\psi\in\subf\vp$, of the sentences
\[
\forall x_1\cdots\forall x_m\,\bigl(Q_{w,\psi}(x_\ell,\ldots,x_m)\leftrightarrow\delta_{w,\psi}\bigr),
\]
where $m$ is the level of $\psi$ and $x_\ell,\ldots,x_m$ are its parameters
(note that the quantifiers $\forall x_1\cdots\forall x_{\ell-1}$ are vacuous).
By~\eqref{eq:delta-fluted}, each conjunct of $\mathit{Def}^{\mathrm{fl}}$ is a
$\cFLc^{k}$-sentence (by Observation~\ref{rem:fl}, it is an $\cFL^{k}$-sentence
whenever $\vp\in\FL$): it is the universal closure, in the prescribed order, of
a formula from $\cFLc^{[m]}$, and all of its variables occur in $\vp$ (these are
$x_1,\ldots,x_m$ and, if $\psi$ begins with a quantifier, $x_{m+1}$).  We take
the variable of $\Ax^{\mathit{xd}}_{\fr{F}}$ and $\Ax^{\mathit{cd}}_{\fr{F}}$ to
be $x_1$; then, since $D_w(x_1)$ is a fluted $1$-atom,
$\Ax^{\mathit{xd}}_{\fr{F}}, \Ax^{\mathit{cd}}_{\fr{F}} \in \cFL^{[0]}$.
Lastly, put
$$
\begin{array}{lclclcl}
  \Psi^{\mathit{xd}}_{\fr{F}}(\varphi) & := & \displaystyle \Ax^{\mathit{xd}}_{\fr{F}} \wedge \mathit{Def}^{\mathrm{fl}} \wedge\bigvee_{w\in
                                  W}\neg Q_{w,\varphi}, & &
  \Psi^{\mathit{cd}}_{\fr{F}}(\varphi) & := & \displaystyle \mathit{Ax}^{\mathit{cd}}_{\fr{F}} \wedge \mathit{Def}^{\mathrm{fl}} \wedge\bigvee_{w\in
                                              W}\neg Q_{w,\varphi}.
\end{array}
$$
These are $\cFLc^{k}$-sentences, and, by Observation~\ref{rem:fl}, they are
$\cFL^{k}$-sentences whenever $\vp\in\FL^k$.

\begin{lemma}
\label{lem:Psi-size}
Let $\fr{F} = \langle W,R \rangle$ be a finite Kripke frame and let
$\vp \in \FLc^{k} \cap \FLc^{[0]}$.  Then, $\Psi^{d}_{\fr{F}}(\vp)$, for
$d \in \{ \mathit{xd}, \mathit{cd} \}$, has size, and is computable in time,
polynomial in $|\fr{F}| + |\vp|$.
\end{lemma}

\begin{proof}
  As in the proof of Lemma~\ref{lem:Phi}, $\Ax^{\mathit{xd}}_{\fr{F}}$ and
  $\Ax^{\mathit{cd}}_{\fr{F}}$ have size $O(|W|^2)$,
  $\bigvee_{w\in W}\neg Q_{w,\vp}$ has size $O(|W|)$, and
  $\mathit{Def}^{\mathrm{fl}}$ has at most $|W|\cdot|\vp|$ conjuncts.  The
  difference from Lemma~\ref{lem:Phi} is that name atoms are no longer of
  constant size: the level $m$ of a subformula of $\vp$ does not exceed the
  number of quantifiers of $\vp$, so $m\leqslant|\vp|$, and hence the prefix
  $\forall x_1\cdots\forall x_m$ of a conjunct of $\mathit{Def}^{\mathrm{fl}}$
  and every name atom $Q_{v,\psi}(x_\ell,\ldots,x_m)$, which has at most $m$
  arguments, have size $O(|\vp|)$.  Now, $\delta_{w,\psi}$ consists of
  $|R(w)|\leqslant|W|$ name atoms if $\psi=\square\psi_1$, and of at most two
  name atoms and $O(|\psi|)$ further symbols otherwise (these are the symbols
  of $\ST^{\fr{F}}_w(\psi)$ if $\psi$ is atomic, and those of the quantifier
  $\exists^{\geqslant M}x_{m+1}$ and the atom $D_w(x_{m+1})$ if
  $\psi=\exists^{\geqslant M}x_{m+1}\,\psi'$).  Hence, each conjunct of
  $\mathit{Def}^{\mathrm{fl}}$ has size $O(|W|\cdot|\vp|)$ (the worst case is
  again when $\delta_{w,\psi}$ is written for $\psi=\square\psi_1$), and so
  $\Psi^{d}_{\fr{F}}(\vp)$ has size $O(|W|^2\cdot|\vp|^2)$.  %Lastly, the levels and parameters of the
  %subformulas of $\vp$ can be computed in time polynomial in $|\vp|$, after
  %which the conjuncts of $\Psi_{\fr{F}}(\vp)$ and
  %$\Psi^{\mathit{cd}}_{\fr{F}}(\vp)$ can be written out in time polynomial in
  %$|\fr{F}|+|\vp|$.
\end{proof}

\begin{lemma}[Renaming for fluted sentences]
  \label{lem:fluted_renaming}
  Let $\fr{F} = \langle W,R \rangle$ be a finite Kripke frame, let
  $\vp \in \FLc^{k} \cap \FLc^{[0]}$, and let
  $d \in \{ \mathit{xd}, \mathit{cd} \}$.  Then, for every cardinal
  $\kappa$, the sentence $\vp$ is refuted in a $d$-model over $\fr{F}$ whose
  global domain has at most $\kappa$ elements iff $\Psi^{d}_{\fr{F}}(\vp)$ has
  a model with at most $\kappa$ elements.  In particular, $\vp$ is refuted in
  a $d$-model over $\fr{F}$ iff $\Psi^{d}_{\fr{F}}(\vp)$ is satisfiable.
\end{lemma}

\begin{proof}
  Define the formula $\mathit{Def}$ for $\vp$ as in Section~\ref{sec:two-var},
  listing the parameters $x_\ell,\ldots,x_m$ of each $\psi\in\subf\vp$ in this
  order (in Section~\ref{sec:two-var}, the order of parameters was immaterial);
  neither this definition nor the proof of Lemma~\ref{lem:renaming} uses the
  bound on the number of variables.  Then, each conjunct of $\Def^{\mathrm{fl}}$
  differs from the corresponding conjunct of $\Def$ only by the vacuous
  quantifiers $\forall x_1\cdots\forall x_{\ell-1}$, and so is equivalent to the
  said conjunct; hence, the proof of Lemma~\ref{lem:renaming} applies with
  $\mathit{Def}^{\mathrm{fl}}$ replacing $\mathit{Def}$.
\end{proof}

\begin{theorem}[Fluted fragments without counting]
\label{thm:fluted-complexity}
Let $\fr{F}$ be a finite Kripke frame and $d \in \{ \mathit{xd}, \mathit{cd} \}$.
Then,
\begin{itemize}
\item[\rmbr{$1$}] the sets $(\mathbf{ML}^d \fr{F})^1 \cap \FL$ are
  $\coNP$-complete;
\item[\rmbr{$2$}] the sets $(\mathbf{ML}^d \fr{F})^2 \cap \FL$ are
  $\coNEXP$-complete;
\item[\rmbr{$3$}] for $k \geqslant 3$, the sets $(\mathbf{ML}^d \fr{F})^k \cap \FL$
  are in $\mathrm{co}$-$(k-2)$-$\NEXP$ and are
  $\mathrm{co}$-$\lfloor k/2\rfloor$-$\NEXP$-hard; in particular,
  $(\mathbf{ML}^d \fr{F})^3 \cap \FL$ are $\coNEXP$-complete, and
  $(\mathbf{ML}^d \fr{F})^4 \cap \FL$ are $\mathrm{co}$-$2$-$\NEXP$-complete;
\item[\rmbr{$4$}] the sets $\mathbf{ML}^d \fr{F} \cap \FL$ are $\tower$-complete.
\end{itemize}
The lower bounds hold already for $\square$-free sentences, and in~\rmbr{$1$} for
propositional formulas.
\end{theorem}

\begin{proof}
  We may assume that the input formula is in $\FL^{[0]}$: its universal closure,
  with the variables renamed down as in Section~\ref{sec:prelim}, is in
  $\FL^{[0]}$, contains no more variables, and is computable in polynomial time.

  (1): The upper bound follows from Theorem~\ref{thm:one-var} since
  $(\mathbf{ML}^d \fr{F})^1 \cap \FL \subseteq (\mathbf{ML}^d_{\mathrm{C}} \fr{F})^1$;
  the lower bound is
  inherited from the classical propositional logic since propositional formulas
  are fluted.

  (2): The upper bound follows from Theorem~\ref{thm:two-var-plain} since
  $(\mathbf{ML}^d \fr{F})^2 \cap \FL \subseteq (\mathbf{ML}^d \fr{F})^2$;
  the lower bound follows from Proposition~\ref{prop:FL2} and
  Remark~\ref{rem:conserv}.

  (3) and (4): The upper bounds follow from Lemmas~\ref{lem:Psi-size}
  and~\ref{lem:fluted_renaming} together with Proposition~\ref{prop:FL}; the
  lower bounds follow from Proposition~\ref{prop:FL} and
  Remark~\ref{rem:conserv}.
\end{proof}

\begin{theorem}[Finite domain property for fluted fragments without counting]
  \label{thm:FDP-fluted}
  Let $\fr{F}=\langle W,R\rangle$ be a finite Kripke frame, $k\geqslant1$, and
  $d\in\{\mathit{xd},\mathit{cd}\}$.  If
  $\vp\in\FL^k\setminus\mathbf{ML}^d\fr{F}$, then $\vp$ is refuted in a
  $d$-model over $\fr{F}$ whose global domain has size bounded by a
  $\max(1,k-2)$-fold exponential function of $|\fr{F}|+|\vp|$.  In particular,
  $\FL$ has the finite domain property over $\fr{F}$ for $d$-models.
\end{theorem}

\begin{proof}
  As in the proof of Theorem~\ref{thm:fluted-complexity}, we may assume that
  $\vp\in\FL^{[0]}$.  By Lemmas~\ref{lem:Psi-size}
  and~\ref{lem:fluted_renaming}, $\Psi^{d}_{\fr{F}}(\vp)$ is a satisfiable
  $\mathcal{FL}^k$-sentence of size polynomial in $|\fr{F}|+|\vp|$.  By
  \cite[Theorem~6.6]{PH23} and its proof (if $k\leqslant2$) and by
  \cite[Theorem~6.10]{PH23} (if $k\geqslant3$), $\Psi^{d}_{\fr{F}}(\vp)$ has a
  model $\fr{A}$ whose size is bounded by an exponential function of
  $|\Psi^{d}_{\fr{F}}(\vp)|$ in case $k\leqslant2$ and by a $(k-2)$-fold
  exponential function of $|\Psi^{d}_{\fr{F}}(\vp)|$ in case $k\geqslant3$;
  hence, $|\fr{A}|$ is bounded by a $\max(1,k-2)$-fold exponential function of
  $|\fr{F}|+|\vp|$.  By Lemma~\ref{lem:fluted_renaming}, $\vp$ is refuted in
  a $d$-model over $\fr{F}$ with the global domain of size at most $|\fr{A}|$.
\end{proof}

\begin{theorem}[Fluted fragments with counting]
  \label{thm:fluted-counting}
  Let $\fr{F}=\langle W,R\rangle$ be a finite Kripke frame and
  $d\in\{\mathit{xd},\mathit{cd}\}$.  Then,
  \begin{itemize}
  \item[\rmbr{$1$}] the sets $(\mathbf{ML}^d_{\mathrm{C}}\fr{F})^1\cap\FLc$ are
    $\coNP$-complete;
  \item[\rmbr{$2$}] the sets $(\mathbf{ML}^d_{\mathrm{C}}\fr{F})^2\cap\FLc$ are
    $\coNEXP$-complete;
  \item[\rmbr{$3$}] for $k\geqslant3$, the sets
    $(\mathbf{ML}^d_{\mathrm{C}}\fr{F})^k\cap\FLc$ are in
    $\mathrm{co}$-$(k-1)$-$\NEXP$ and are
    $\mathrm{co}$-$\lfloor k/2\rfloor$-$\NEXP$-hard;
  \item[\rmbr{$4$}] the sets $\mathbf{ML}^d_{\mathrm{C}}\fr{F}\cap\FLc$ are
    $\tower$-complete.
  \end{itemize}
  The lower bounds hold already for $\square$-free sentences of $\FL$, that is,
  without counting quantifiers and equality, and in~\rmbr{$1$} for
  propositional formulas.
\end{theorem}

\begin{proof}
  As in the proof of Theorem~\ref{thm:fluted-complexity}, we may assume that the
  input formula is in $\FLc^{[0]}$.

  (1): The upper bound follows from Theorem~\ref{thm:one-var}, since
  $(\mathbf{ML}^d_{\mathrm{C}}\fr{F})^1\cap\FLc\subseteq(\mathbf{ML}^d_{\mathrm{C}}\fr{F})^1$;
  the lower bound is that of
  Theorem~\ref{thm:fluted-complexity}(1).

  (2): The upper bound follows from Theorem~\ref{thm:two-var}, since
  $(\mathbf{ML}^d_{\mathrm{C}}\fr{F})^2\cap\FLc\subseteq(\mathbf{ML}^d_{\mathrm{C}}\fr{F})^2$;
  the lower bound follows
  from Proposition~\ref{prop:FL2} and Remark~\ref{rem:conserv}.

  (3) and (4): The upper bounds follow from Lemmas~\ref{lem:Psi-size}
  and~\ref{lem:fluted_renaming} together with Proposition~\ref{prop:FLC};
  % ; for~(4), note  that $\tower$ is closed under polynomial-time reductions.
  the lower bounds follow from Proposition~\ref{prop:FL} and
  Remark~\ref{rem:conserv}.
\end{proof}

\begin{remark}
  The bounds in Theorems~\ref{thm:fluted-complexity}
  and~\ref{thm:fluted-counting} cannot be improved without improving the
  corresponding classical bounds: in view of Lemmas~\ref{lem:Psi-size}
  and~\ref{lem:fluted_renaming} and Remark~\ref{rem:conserv}, for every
  $k\geqslant1$ and every $d\in\{\mathit{xd},\mathit{cd}\}$, the set
  $(\mathbf{ML}^d\fr{F})^k\cap\FL$ is polynomial-time equivalent to
  $\QCL^k\cap\mathcal{FL}$, and the set
  $(\mathbf{ML}^d_{\mathrm{C}}\fr{F})^k\cap\FLc$ to
  $\mathbf{QCL}_{\mathrm{C}}^k\cap\cFLc$.  In particular, for $k\geqslant5$, closing the gap
  in Theorem~\ref{thm:fluted-complexity}(3) is equivalent to closing the
  corresponding gap for the classical fluted fragment, which is
  open~\cite[Chapter~6]{PH23}, and the gap between the upper and lower bounds
  in Theorem~\ref{thm:fluted-counting}(3) is the gap for the classical fluted
  fragment with counting quantifiers.
\end{remark}

\begin{theorem}[Finite domain property for fluted fragments with counting]
  \label{thm:fluted-FDP-counting}
  Let \mbox{$\fr{F}=\langle W,R\rangle$} be a finite Kripke frame,
  $k\geqslant1$, and $d\in\{\mathit{xd},\mathit{cd}\}$.  If
  $\vp\in\FLc^k\setminus\mathbf{ML}^d_{\mathrm{C}}\fr{F}$, then $\vp$ is
  refuted in a $d$-model over $\fr{F}$ whose global domain has size bounded by
  a $\max(1,k-1)$-fold exponential function of $|\fr{F}|+|\vp|$.  In
  particular, $\FLc$ has the finite domain property over $\fr{F}$ for
  $d$-models.
\end{theorem}

\begin{proof}
  For $k=1$, the bound is that of Theorem~\ref{thm:fdp-one-var}, as
  $\FLc^1\subseteq\CMLang^1$.  For $k\geqslant2$, we argue as in the proof of
  Theorem~\ref{thm:FDP-fluted}: we may assume that $\vp\in\FLc^{[0]}$; by
  Lemmas~\ref{lem:Psi-size} and~\ref{lem:fluted_renaming},
  $\Psi^{d}_{\fr{F}}(\vp)$ is a satisfiable $\cFLc^{k}$-sentence of size
  polynomial in $|\fr{F}|+|\vp|$; by Proposition~\ref{prop:ffc}, it has a model
  $\fr{A}$ whose size is bounded by a $(k-1)$-fold exponential function of
  $|\fr{F}|+|\vp|$; and, by Lemma~\ref{lem:fluted_renaming}, $\vp$ is refuted
  in a $d$-model over $\fr{F}$ whose global domain has at most $|\fr{A}|$
  elements.
\end{proof}

\begin{remark}
  \label{rem:fluted-FDP-counting}
  Theorem~\ref{thm:fluted-FDP-counting} contrasts with
  Remark~\ref{rem:no-FDP-counting}: the two-variable fragment with counting
  quantifiers does not have the finite domain property over $\fr{F}$, but its
  fluted subfragment does: the sentence $\chi$ of
  Remark~\ref{rem:no-FDP-counting} is not fluted, as it uses the atom $S(x,y)$
  both under quantifiers in the order $x$, $y$ and under quantifiers in the
  order $y$, $x$.
\end{remark}

\subsection{Guarded fragments}
\label{sec:guarded}

The translation $\ST^{\fr{F}}_w$ does not send guarded formulas to guarded
formulas: it relativises each quantified variable separately, translating
$\exists x\,\psi$ as $\exists x\,(D_w(x)\wedge\ST^{\fr{F}}_w(\psi))$, and the
atom $D_w(x)$ need not contain the other parameters of $\psi$, so a guarded
block $\exists \bar y\,(\gamma\wedge\psi)$ is, in general, translated as a
sequence of unguarded quantifiers.  We therefore amend the translation so that
quantifier blocks are translated as blocks, and then rename subformulas making
sure that the renaming stays within the guarded fragment.

Let $\vp\in\GFe$ be a formula whose predicate letters other than $=$ are
$P_1,\ldots,P_s$, and fix a derivation of $\vp$ by (G1)--(G4).  Its
\emph{nodes} are the occurrences in $\vp$ of the formulas produced by the
steps of the derivation, each by a single rule: an atom, by (G1);
$\psi_1\to\psi_2$ or $\square\psi_1$, by (G2), with the \emph{children}
$\psi_1,\psi_2$ or $\psi_1$; $\exists\bar y\,(\gamma\wedge\psi_1)$, by (G3),
with the children $\gamma$, the \emph{guard} of the node, and $\psi_1$; and
$\exists x\,\psi_1$, by (G4), with the child $\psi_1$.  Thus, a guarded block
is a single node: the formulas between it and its children, such as
$\gamma\wedge\psi_1$ and the suffixes $\exists y_2\cdots\exists
y_n\,(\gamma\wedge\psi_1)$ of a block with $\bar y=y_1,\ldots,y_n$, are
subformulas of $\vp$ but not nodes.  For every node $\psi$ and every $w\in W$,
define $\STg^{\fr{F}}_w(\psi)$ by recursion on the derivation:
%\begin{itemize}
%\item[\rmbr{G1}]
  $$
  \begin{array}{lcrclclclclcl}
   \mbox{(G1)} & & \STg^{\fr{F}}_w(\psi) & := &  \ST^{\fr{F}}_w(\psi) & &
                                                           \mbox{if $\psi$ is
                                                                        atomic}; \smallskip\\
  %\end{array}
  %$$
%\item[\rmbr{G2}]
  %$$
  %\begin{array}{lcl}
  \mbox{(G2)} & & \STg^{\fr{F}}_w(\psi_1\to\psi_2) & := &
    \STg^{\fr{F}}_w(\psi_1)\to\STg^{\fr{F}}_w(\psi_2);\smallskip\\
  & &   \STg^{\fr{F}}_w(\square\psi) & := & \displaystyle\bigwedge_{v\in
                                            R(w)}\STg^{\fr{F}}_v(\psi); \smallskip\\
  %\end{array}
  %$$
%\item[\rmbr{G3}]
  %$$
  %\begin{array}{lcl}
   \mbox{(G3)} & &  \STg^{\fr{F}}_w\bigl(\exists\bar y\,(\gamma\wedge\psi)\bigr) & := &
    \displaystyle \exists\bar y\,\Bigl(\gamma^{w}\wedge\bigl(\bigwedge_{y\in\bar y}D_w(y)\wedge\STg^{\fr{F}}_w(\psi)\bigr)\Bigr),%\smallskip\\
  \end{array}
  $$
  where $\gamma^{w}:=\ST^{\fr{F}}_w(\gamma)$, which is $P^w_i(\bar z)$ if
  $\gamma$ is an atom $P_i(\bar z)$, and $\gamma$ itself if $\gamma$ is an
  equality atom (we call $\gamma^{w}$ the \emph{guard atom} at $w$);
%\item[\rmbr{G4}]
  $$
  \begin{array}{lclcl}
    \mbox{(G4)} & & \STg^{\fr{F}}_w(\exists x\,\psi) & := &
    \exists x\,\bigl(D_w(x)\wedge\STg^{\fr{F}}_w(\psi)\bigr). %\smallskip\\
    %\STg^{\fr{F}}_w(\forall x\,\psi) & := &
    %\forall x\,\bigl(D_w(x)\to\STg^{\fr{F}}_w(\psi)\bigr).
  \end{array}
  $$

Note that $\STg^{\fr{F}}_w(\psi)$ is a classical guarded
$\sigma_{\fr{F}}(\vp)$-formula with
$\param{\STg^{\fr{F}}_w(\psi)}\subseteq\param\psi$: in (G3), the block
$\exists\bar y$ is guarded by the atom $\gamma^w$, which contains $\bar
y$ and all parameters of the matrix, and in (G4) the quantifier is guarded by
$D_w(x)$, which contains the only parameter of the matrix.

%\begin{remark}
%  \label{rem:why-g-all}
%  The clause for $\forall$ in (G3) is redundant up to logical equivalence:
%  $\forall\bar x\,(\gamma\to\vp)$ is equivalent to
%  $\neg\exists\bar x\,(\gamma\wedge\neg\vp)$, which belongs to $\GFe$ by (G2)
%  and the clause for $\exists$ in (G3), and has the same guard, variables and
%  modal depth.  It is not redundant outright: for instance,
%  $\forall y\,(P_1(x,y)\to P_2(y))$, which abbreviates
%  $\neg\exists y\,\neg(P_1(x,y)\to P_2(y))$, cannot be obtained without it,
%  as $\neg(P_1(x,y)\to P_2(y))$ is not of the form $\gamma\wedge\psi$ and
%  (G4) does not apply.  We include it because we use such formulas
%  throughout, in $\GFe$ and in its classical counterpart $\mathcal{GF}_=$
%  (for example, in Section~\ref{sec:guarded}).  By contrast, the clause for
%  $\forall$ in (G4) is redundant outright: $\forall x\,\vp$ is
%  $\neg\exists x\,\neg\vp$, and $\exists x\,\neg\vp\in\GFe$ by the clause for
%  $\exists$ in (G4).
%\end{remark}

\begin{remark}
  \label{rem:why-D}
  The conjunct $\bigwedge_{y\in\bar y}D_w(y)$ in the clause (G3) of the
  definition of $\STg^{\fr{F}}_w$ cannot be dropped, since nothing in
  $\Ax^{\mathit{xd}}_{\fr{F}}$ forces the tuples satisfying $\gamma^w$ into
  $D_w$.  For a guard $\gamma=P_i(\bar z)$, the interpretation
  $I^{\fr{A}}(w,P_i)$ is defined as the restriction of $(P_i^w)^{\fr{A}}$ to
  $D^{\fr{A}}_w$, so $(P_i^w)^{\fr{A}}$ itself may contain tuples with entries
  outside $D^{\fr{A}}_w$.  For instance, without these conjuncts,
  $\forall x\,(P(x)\to Q(x))$ would be translated as
  $\forall x\,(P^w(x)\to Q^w(x))$, which is false in $\fr{A}$ if some element
  outside $D^{\fr{A}}_w$ belongs to $(P^w)^{\fr{A}}$ but not to
  $(Q^w)^{\fr{A}}$, although $\forall x\,(P(x)\to Q(x))$ may well be true at $w$
  in $\fr{M}^{\fr{A}}$, where only the elements of $D^{\fr{A}}_w$ matter.
  Placing the conjuncts after the guard, rather than before it, keeps the
  translated block guarded.
\end{remark}

\begin{lemma}[Guarded translation lemma]
  \label{lem:gST}
  Let $\fr{F}=\langle W,R\rangle$ be a finite Kripke frame, let $\vp\in\GFe$
  be a formula whose predicate letters other than $=$ are $P_1,\ldots,P_s$,
  with a fixed derivation by (G1)--(G4), and let $\fr{A}$ be a classical
  $\sigma_{\fr{F}}(\vp)$-model with $\fr{A}\models\Ax^{\mathit{xd}}_{\fr{F}}$.
  Then, for every node $\psi(x_1,\ldots,x_m)$ of the derivation of $\vp$, every
  $w\in W$, and
  every $\bar a\in(D^{\fr{A}}_w)^m$,
  \[
    \begin{array}{lcl}
      \fr{M}^{\fr{A}}, w \Vdash \psi(\bar a) & \iff & \fr{A} \models \STg^{\fr{F}}_w(\psi)[\bar a].
    \end{array}
  \]
\end{lemma}

\begin{proof}
  In view of Lemma~\ref{lem:st}, it suffices to show that, for every node
  $\psi(x_1,\ldots,x_m)$, every $w\in W$ and every
  $\bar a\in(D^{\fr{A}}_w)^m$,
  \begin{equation}
    \label{eq:gST-ST}
    \fr{A}\models\STg^{\fr{F}}_w(\psi)[\bar a]
    \iff
    \fr{A}\models\ST^{\fr{F}}_w(\psi)[\bar a].
  \end{equation}
  
  This is done by induction on nodes, from the leaves of the derivation
  upwards, simultaneously for all $w$ and $\bar a$.  The cases
  (G1) and (G4) are trivial.  In (G2), the case
  $\psi = \psi_1 \to \psi_2$ is immediate from IH, and, if $\psi=\square\psi_1$
  and $v\in R(w)$, then $\bar a\in(D^{\fr{A}}_v)^m$ by
  $\Ax^{\mathit{xd}}_{\fr{F}}$, so IH applies to $\psi_1$ at $v$.

  For (G3), let $\psi=\exists y_1\ldots\exists y_k\,(\gamma\wedge\psi_1)$.
  Then $\ST^{\fr{F}}_w(\psi)$ is
  $$
  \begin{array}{c}
    \exists y_1\,(D_w(y_1)\wedge\exists y_2\,(D_w(y_2)\wedge\ldots\exists
    y_k\,(D_w(y_k)\wedge(\ST^{\fr{F}}_w(\gamma)\wedge\ST^{\fr{F}}_w(\psi_1)))\ldots)),
  \end{array}
  $$
  which is equivalent, the variables $y_1,\ldots,y_k$ being distinct, to
  $$
  \exists\bar y\,(\ST^{\fr{F}}_w(\gamma)\wedge(\bigwedge_{y\in\bar
    y}D_w(y)\wedge\ST^{\fr{F}}_w(\psi_1))).
  $$
  The latter formula is equivalent to $\STg^{\fr{F}}_w(\psi)$ since, under the
  conjuncts $D_w(y)$, the IH applies to $\ST^{\fr{F}}_w(\psi_1)$ and
  $\STg^{\fr{F}}_w(\psi_1)$.
  % , i.e., that the
  %matrices of ($\ast$) and $\STg^{\fr{F}}_w(\psi)$, which differ only in their
  %last conjuncts $\ST^{\fr{F}}_w(\psi_1)$ and $\STg^{\fr{F}}_w(\psi_1)$, are
  %satisfied by the same assignments sending each parameter $x_i$ of $\psi$ to
  %$a_i$.  Let $g$ be such an assignment.  If $g(y)\notin D^{\fr{A}}_w$ for some
  %$y\in\bar y$, then $g$~falsifies the conjunct $D_w(y)$, and so both matrices.
  %Otherwise, $g$ sends every variable in $\bar{y} \cup \param \psi$, and hence
  %every parameter of $\psi_1$, into $D^{\fr{A}}_w$.  Then, by IH, $g$~satisfies
  %$\ST^{\fr{F}}_w(\psi_1)$ iff it satisfies $\STg^{\fr{F}}_w(\psi_1)$.  Thus,
  %the two matrices agree under~$g$.
\end{proof}

Now, we bring the guarded translation down to polynomial size.  We need to make
sure that all the formulas used in renaming are guarded.  For every node
$\psi$ that is not a guard (guards are not renamed; this is not a
problem since they are atoms), let $\bar x_\psi$ list the variables of
$\param\psi$ in some fixed order.  For every $w\in W$, introduce two fresh
predicate letters $P_{w,\psi}$ and $N_{w,\psi}$ of arity $|\bar x_\psi|$
(proposition letters if $\psi$ is a sentence).  The \emph{positive name}
$\Pn{w}{\psi}$ is meant to say that $\STg^{\fr{F}}_w(\psi)$ is true, and the
\emph{negative name} $\Nn{w}{\psi}$ is meant to say that $\STg^{\fr{F}}_w(\psi)$
is false.

For every such $\psi$ and every $w\in W$, define the formulas
$\delta^+_{w,\psi}$ and $\delta^-_{w,\psi}$ by the row of the rule producing
$\psi$, whose children other than its guard are $\psi_1$ and $\psi_2$ (the
empty disjunction is $\bot$):
\[
\renewcommand{\arraystretch}{1.35}
\begin{array}{@{}lll@{}}
  \toprule
  \psi & \delta^+_{w,\psi} & \delta^-_{w,\psi}\\
  \midrule
  \text{atomic} & \STg^{\fr{F}}_w(\psi) & \neg\,\STg^{\fr{F}}_w(\psi)\\
  \psi_1\to\psi_2 & \Nn{w}{\psi_1}\vee\Pn{w}{\psi_2} & \Pn{w}{\psi_1}\wedge\Nn{w}{\psi_2}\\
  \square\psi_1 & \displaystyle \bigwedge_{v\in R(w)}\Pn{v}{\psi_1} &
                                                                      \displaystyle \bigvee_{v\in R(w)}\Nn{v}{\psi_1}\\
  \exists\bar y\,(\gamma\wedge\psi_1) &
    \displaystyle \exists\bar y\,\bigl(\gamma^w\wedge(\bigwedge\limits_{y\in\bar y}D_w(y)\wedge\Pn{w}{\psi_1})\bigr) &
   \displaystyle \forall\bar y\,\bigl(\gamma^w\to(\bigwedge\limits_{y \in \bar{y} }D_w(y)\to\Nn{w}{\psi_1})\bigr)\\
  %\forall\bar y\,(\gamma\to\psi_1) &
  %  \displaystyle \forall\bar y\,\bigl(\gamma^w\to(\bigwedge\limits_{y\in\bar y}D_w(y)\to\Pn{w}{\psi_1})\bigr) &
  %  \displaystyle \exists\bar y\,\bigl(\gamma^w\wedge(\bigwedge\limits_{y\in\bar y}D_w(y)\wedge\Nn{w}{\psi_1})\bigr)\\
  \exists x\,\psi_1 & \exists x\,\bigl(D_w(x)\wedge\Pn{w}{\psi_1}\bigr) &
    \forall x\,\bigl(D_w(x)\to\Nn{w}{\psi_1}\bigr)\\
  %\forall x\,\psi_1 & \forall x\,\bigl(D_w(x)\to\Pn{w}{\psi_1}\bigr) &
  %  \exists x\,\bigl(D_w(x)\wedge\Nn{w}{\psi_1}\bigr)\\
  \bottomrule
\end{array}
\]
%Here .
Thus, $\delta^+_{w,\psi}$ is the one-step unfolding of $\STg^{\fr{F}}_w(\psi)$,
and $\delta^-_{w,\psi}$ is the one-step unfolding of
$\neg\STg^{\fr{F}}_w(\psi)$, with the negation pushed one step inwards, where
the translations of the children are replaced by their names of
the appropriate polarity.  Note that
$\param{\delta^{\pm}_{w,\psi}}\subseteq\param\psi$, for all $\psi$.

Now, let $\Def^{\mathrm{g}}$ be the conjunction of the following sentences, for
every node $\psi$ that is not a guard and for every $w\in W$:
\begin{align}
  \forall\bar x_\psi\,\bigl(\Pn{w}{\psi}\to\delta^+_{w,\psi}\bigr)
  &\quad\text{and}\quad
    \forall\bar x_\psi\,\bigl(\Nn{w}{\psi}\to\delta^-_{w,\psi}\bigr).
  \label{eq:def-param}  
  %&&\text{if $\psi$ has parameters;}\label{eq:def-param}\\
  %P_{w,\psi}\to\delta^+_{w,\psi}
  %&\quad\text{and}\quad
  %N_{w,\psi}\to\delta^-_{w,\psi}
  %&&\text{if $\psi$ is a sentence.}\label{eq:def-sent}
\end{align}
Lastly, put
\[
\begin{array}{lclclcl}
  \Theta^{\mathit{xd}}_{\fr F}(\vp) & := & \displaystyle\Ax^{\mathit{xd}}_{\fr
                                           F}\wedge\Def^{\mathrm{g}}\wedge\bigvee_{w\in
                                           W}N_{w,\vp}, & & 
  \Theta^{\mathit{cd}}_{\fr F}(\vp) & := & \displaystyle \Ax^{\mathit{cd}}_{\fr F}\wedge\Def^{\mathrm{g}}\wedge\bigvee_{w\in W}N_{w,\vp}.
\end{array}
\]
As in Section~\ref{sec:two-var}, we take the variable of
$\Ax^{\mathit{xd}}_{\fr F}$ and $\Ax^{\mathit{cd}}_{\fr F}$ to be a variable of
$\vp$ whenever $\vp$ has any.

\begin{remark}
  \label{rem:guarded-extra-names}
  The definitions~\eqref{eq:def-param} are one-directional: they say what a name
  implies, but not what it is implied by. For every node $\psi$, we
  introduce both a positive and a negative name, without keeping track of the
  polarity of $\psi$; this is harmless, since a sentence
  of the form $\forall\bar x\,(Q(\bar x)\to\delta)$ is satisfied when $Q$ is
  interpreted as the empty relation.  Since a name contains all the free
  variables of its unfolding, it can serve as the guard of its own definition;
  this is what keeps the definitions in~\eqref{eq:def-param} guarded.  The
  converse implications $\forall\bar x\,(\delta\to Q(\bar x))$, which would make
  the definitions biconditional as in Sections~\ref{sec:two-var}
  and~\ref{sec:fluted}, are not guarded in general: for non-empty $\bar x$, the
  block $\forall\bar x$ would have to be guarded by $\delta$, which need not be
  an atom.

  Unlike in the proof of Lemma~\ref{lem:renaming}, where the equivalences of
  $\mathit{Def}$ determine the expansion satisfying $\mathit{Def}$ uniquely, a
  classical model may therefore have many expansions satisfying
  $\Def^{\mathrm{g}}$; interpreting every name as the empty relation is one of
  them.  This is harmless, since Lemma~\ref{lem:sound} below holds for all such
  expansions.  In fact, if a model $\fr B$ is the expansion of a model $\fr A$
  that interprets each $P_{w,\psi}$ as the relation defined by
  $\STg^{\fr{F}}_w(\psi)$ and each $N_{w,\psi}$ as its complement, then the
  proof of Lemma~\ref{lem:guarded-renaming} below shows that
  $\fr B\models\Def^{\mathrm{g}}$ without using any property of $\fr A$.  By
  Lemma~\ref{lem:sound}, in every expansion of $\fr A$ satisfying
  $\Def^{\mathrm{g}}$, each name denotes a subrelation of the relation it
  denotes in $\fr B$; thus, $\fr B$ is the largest expansion of $\fr A$
  satisfying $\Def^{\mathrm{g}}$, and the proof of
  Lemma~\ref{lem:guarded-renaming} uses $\fr B$ in one direction and, in the
  other, only the inclusion for the names $N_{w,\vp}$.
\end{remark}

\begin{lemma}
  \label{lem:guarded-syntax}
  Let $d\in\{\mathit{xd},\mathit{cd}\}$.
  \begin{itemize}
  \item[\rmbr{$1$}] If $\vp \in \GFe$, then
    $\Theta^{d}_{\fr F}(\vp)\in \mathcal{GF}_=$;
    moreover, $\Theta^{d}_{\fr F}(\vp)$ has a
    derivation by \rmbr{{\rm G1}}--\rmbr{{\rm G3}} in which every guard is a
    name atom $P_{w,\psi}(\bar x_\psi)$ or $N_{w,\psi}(\bar x_\psi)$, an atom
    $D_w(z)$, or the guard atom $\gamma^w$ of a guard $\gamma$ in $\vp$.
  %\item[\rmbr{$1$}] If $\vp \in \GFe$, then
  %  $\Theta^{d}_{\fr F}(\vp)\in \mathcal{GF}_=$;
  %  moreover, $\Theta^{d}_{\fr F}(\vp)$ has a
  %  derivation by \rmbr{{\rm G1}}--\rmbr{{\rm G3}} in which every guard is an atom of one
  %  of the forms $P_{w,\psi}(\bar x_\psi)$, $N_{w,\psi}(\bar x_\psi)$,
  %  $P^w(\bar z)$ and $D_w(z)$, or an equality atom that is a guard in $\vp$.
    \item[\rmbr{$2$}] If $\vp$ has at most $k$ variables, for $k\geqslant1$,
    then $\Theta^{d}_{\fr F}(\vp)\in\mathcal{L}^k_=$.
  \end{itemize}
  
\end{lemma}
\begin{proof}
  (1): Each $\delta^{\pm}_{w,\psi}$ is a classical guarded formula: in
  the row for (G3), the block is guarded by $\gamma^w$, which contains $\bar y$
  and $\bar x_{\psi_1}$ since $\param{\psi_1}\cup\bar y\subseteq\param\gamma$ by
  (G3); in the row for (G4), the quantifier is guarded by $D_w(x)$, which
  contains $\bar x_{\psi_1}\subseteq\{x\}$.

  If $\psi$ has parameters, the sentences in~\eqref{eq:def-param} are instances
  of (G3): the tuple $\bar x_\psi$ is non-empty and consists of distinct
  variables, and the guards $P_{w,\psi}(\bar x_\psi)$ and
  $N_{w,\psi}(\bar x_\psi)$ contain $\bar x_\psi$ and all the parameters of
  $\delta^{\pm}_{w,\psi}$, as these are among $\param\psi$.  If $\psi$ is a
  sentence, \eqref{eq:def-param} are implications between guarded formulas.  The
  sentence $\Ax^{d}_{\fr F}$ is guarded since we can read $\exists x\,D_w(x)$ as
  $\exists x\,(D_w(x)\wedge\top)$.  The disjunction $\bigvee_{w\in W}N_{w,\vp}$
  is propositional.  Hence $\Theta^{d}_{\fr F}(\vp)$ is guarded, with a
  derivation by (G1)--(G3) whose guards are of the listed forms.

  (2): The variables of sentences in~\eqref{eq:def-param} are those of
  $\bar x_\psi$, which are parameters of $\psi$, and those bound in
  $\delta^{\pm}_{w,\psi}$, which $\psi$ binds; all of them are variables of
  $\vp$.  The disjunction $\bigvee_{w\in W}N_{w,\vp}$ has no variables, and
  $\Ax^{d}_{\fr F}$ has one, which is a variable of $\vp$ if $\vp$ has any,
  and is allowed otherwise, as $k\geqslant1$.
\end{proof}

\begin{lemma}
  \label{lem:Theta}
  The sentences $\Theta^{d}_{\fr F}(\vp)$, for $d\in\{\mathit{xd},\mathit{cd}\}$,
  have size $O(|\fr F|\cdot|\vp|^2)$ and are computable in time polynomial in
  $|\fr F|+|\vp|$.
\end{lemma}

\begin{proof}
  Let $n\leqslant|\vp|$ be the number of nodes that are not guards.  There are
  $2n|W|$ sentences~\eqref{eq:def-param}.  The pair
  for $\psi$ and $w$ has size $O\bigl(|\vp|\cdot(1+|R(w)|)\bigr)$: each
  quantifier prefix, each guard and each name atom have size $O(|\vp|)$, and
  $\delta^{\pm}_{w,\psi}$ consists of $|R(w)|$ name atoms if
  $\psi=\square\psi_1$, and of $O(|\vp|)$ symbols otherwise.  Summing over
  $w\in W$ gives $O\bigl(|\vp|\cdot(|W|+|R|)\bigr)$ for each $\psi$, and so
  $O(|\fr F|\cdot|\vp|^2)$ in total.  The sentence $\Ax^{d}_{\fr F}$ has size
  $O(|\fr F|)$, and $\bigvee_{w\in W}N_{w,\vp}$ has size $O(|W|)$.
  %Lastly, a
  %derivation of $\vp$ by (G1)--(G4), and with it the occurrences and the tuples
  %$\bar x_\psi$, can be computed in time polynomial in $|\vp|$, after which the
  %conjuncts can be written out in time polynomial in $|\fr F|+|\vp|$.
\end{proof}

\newpage

\begin{lemma}[Soundness of names]
  \label{lem:sound}
  Let $\fr A$ be a classical model such that
  $\fr A\models\Def^{\mathrm{g}}$.  Then, for every node $\psi$ that is not a
  guard, every $w\in W$, and every assignment $g$ in $\fr A$,
  \begin{itemize}
  \item[\rmbr{$1$}] $\fr A\models^g\Pn{w}{\psi} \Rightarrow
    \fr A\models^g\STg^{\fr{F}}_w(\psi)$;
  \item[\rmbr{$2$}] $\fr A\models^g\Nn{w}{\psi} \Rightarrow \fr A\not\models^g\STg^{\fr{F}}_w(\psi)$.
  \end{itemize}
\end{lemma}

\begin{proof}
  By induction on nodes, from the leaves of the derivation upwards,
  simultaneously for all $w\in W$ and all assignments $g$.

  By~\eqref{eq:def-param}, for all $w \in W$ and all assignments $g$,
  \begin{align*}
  \fr A\models^g\Pn{w}{\psi} \Rightarrow \fr A\models^g\delta^+_{w,\psi}
  &\quad\text{and}\quad
    \fr A\models^g\Nn{w}{\psi} \Rightarrow \fr A\models^g\delta^-_{w,\psi}. 
  \end{align*}
  Thus, it suffices to show that
   \begin{align*}
  \fr A\models^g\delta^+_{w,\psi} \Rightarrow \fr A\models^g\STg^{\fr{F}}_w(\psi)
  &\quad\text{and}\quad
    \fr A\models^g\delta^-_{w,\psi} \Rightarrow \fr A\not\models^g\STg^{\fr{F}}_w(\psi).
  \end{align*}
  
  If $\psi$ is atomic, the claim is obvious.  The cases $\psi=\psi_1\to\psi_2$
  and $\psi=\square\psi_1$ are straightforward.

  Let $\psi=\exists\bar y\,(\gamma\wedge\psi_1)$.  If
  $\fr A\models^g\delta^+_{w,\psi}$, then there is an assignment
  $g'\overset{\bar y}{=}g$ such that
  $$
  \begin{array}{lclclcl}
    \fr A\models^{g'}\gamma^w,& & \displaystyle \fr A\models^{g'} \bigwedge\limits_{y \in
                                  \bar{y}}D_w(y) & ~\mbox{and}~& \fr A\models^{g'}\Pn{w}{\psi_1}.
  \end{array}
  $$
  By IH applied at $g'$, then $\fr A\models^{g'}\STg^{\fr{F}}_w(\psi_1)$, and so
  $\fr A\models^g\STg^{\fr{F}}_w(\psi)$.  Similarly, if
  $\fr A\models^g\delta^-_{w,\psi}$, then, for every $g'\overset{\bar y}{=}g$,
  $$
  \begin{array}{lcl}
    \fr A\models^{g'}\gamma^w ~~\&~~ \fr A\models^{g'} \bigwedge\limits_{y \in
    \bar{y}} D_w(y) & \Longrightarrow & \fr A\models^{g'}\Nn{w}{\psi_1}.
  \end{array}
  $$
  By IH at $g'$, then $\fr{A} \not\models^{g'} \STg^{\fr{F}}_w(\psi_1)$, and so
  $\fr A\not\models^g\STg^{\fr{F}}_w(\psi)$.

  The case (G4) is similar, with $D_w(x)$ in place of
  $\gamma^w\wedge \bigwedge\limits_{y \in \bar{y}} D_w(y)$.
\end{proof}

\begin{lemma}[Renaming for guarded sentences]
  \label{lem:guarded-renaming}
  Let $\vp\in\GFe$ be a sentence and $d\in\{\mathit{xd},\mathit{cd}\}$.  Then,
  for every cardinal $\kappa$, the sentence $\vp$ is refuted in a $d$-model
  over $\fr F$ whose global domain has at most $\kappa$ elements if, and only
  if, $\Theta^{d}_{\fr F}(\vp)$ has a model with at most $\kappa$ elements.
  In particular, $\vp$ is refuted in a $d$-model over $\fr F$ if, and only if,
  $\Theta^{d}_{\fr F}(\vp)$ is satisfiable.
\end{lemma}

\begin{proof}
  Suppose that $\fr M$ is a $d$-model over $\fr F$ with $\fr M,w_0\nsat\vp$.
  Then $\fr A:=\fr A^{\fr M}$ satisfies $\Ax^{d}_{\fr F}$, and, by
  Observation~\ref{obs:M}, $\fr M^{\fr A}=\fr M$; hence, by Lemma~\ref{lem:gST},
  $\fr A\not\models\STg^{\fr{F}}_{w_0}(\vp)$.  Let $\fr B$ be the expansion of
  $\fr A$ that interprets each name as follows:
  \[
    P_{w,\psi}^{\fr B}:=\{\bar a : \fr A\models\STg^{\fr{F}}_w(\psi)[\bar a]\},\qquad
    N_{w,\psi}^{\fr B}:=\{\bar a : \fr A\not\models\STg^{\fr{F}}_w(\psi)[\bar a]\},
  \]
  where $\bar a$ ranges over the tuples of elements of $\fr A$ of length
  $|\bar x_\psi|$, assigned to $\bar x_\psi$ in order.  Under this
  interpretation, every positive name defines the same relation as the
  translation of its node, and every negative name defines its complement.
  Hence, $\delta^+_{w,\psi}$ defines in $\fr B$ the same relation as
  $\STg^{\fr{F}}_w(\psi)$, and $\delta^-_{w,\psi}$ its complement.  Thus, every
  sentence of $\Def^{\mathrm{g}}$ is true in $\fr B$.  Moreover,
  $\fr B\models N_{w_0,\vp}$, and $\fr B\models\Ax^{d}_{\fr F}$, as
  $\Ax^{d}_{\fr F}$ contains no names.  Thus,
  $\fr B\models\Theta^{d}_{\fr F}(\vp)$; the universe of $\fr B$ is the global
  domain of $\fr M$.

  Conversely, suppose that $\fr A\models\Theta^{d}_{\fr F}(\vp)$.  Since
  $\fr A\models\Ax^{d}_{\fr F}$, it follows that $\fr M^{\fr A}$
  is a \mbox{$d$-model} over $\fr F$, whose global domain is a subset of
  the universe of $\fr A$.  Since
  \mbox{$\fr A\models\bigvee_{w\in W}N_{w,\vp}$}, there exists $w_0\in W$ such
  that $\fr A\models N_{w_0,\vp}$, and so, by Lemma~\ref{lem:sound}\rmbr{$2$},
  $\fr A\not\models\STg^{\fr{F}}_{w_0}(\vp)$.  Hence, by Lemma~\ref{lem:gST},
  $\fr M^{\fr A},w_0\nsat\vp$.

  The claim about cardinalities follows since the first part of the proof
  turns a refuting model with global domain $\Dplus$ into a model of
  $\Theta^{d}_{\fr F}(\vp)$ with universe $\Dplus$, and the second part turns
  a model of $\Theta^{d}_{\fr F}(\vp)$ into a refuting model whose global
  domain is a subset of its universe.
\end{proof}

\newpage

\begin{theorem}[Guarded fragments]
  \label{thm:guarded-complexity}
  Let $\fr F=\langle W,R\rangle$ be a finite Kripke frame and
  $d\in\{\mathit{xd},\mathit{cd}\}$.  Then,
  \begin{itemize}
  \item[\rmbr{$1$}] for every $k\geqslant2$, the sets
    $(\mathbf{ML}^d_=\fr F)^k\cap\GFe$ are $\coEXP$-complete, and so are the
    sets $(\mathbf{ML}^d\fr F)^k\cap\GFe$ of formulas without equality;
  \item[\rmbr{$2$}] the sets $\mathbf{ML}^d_=\fr F\cap\GFe$ are
    $\coTwoEXP$-complete, and so are the sets $\mathbf{ML}^d\fr F\cap\GFe$ of
    formulas without equality.
  \end{itemize}
  Both lower bounds hold already for $\square$-free sentences without equality;
  that in~\rmbr{$1$}, already for such sentences with two variables all of whose
  predicate letters, except for a single binary one, are monadic.
\end{theorem}

\begin{proof}
  We may restrict attention to sentences: if $\vp\in\GFe$ has parameters
  $\bar x$, then $\forall\bar x\,(G(\bar x)\to\vp)$, where $G$ is a fresh
  predicate letter of arity $|\bar x|$, is a guarded sentence with the same
  variables as $\vp$, and it is $d$-valid on $\fr{F}$ if, and only if, $\vp$ is,
  since $G$ may be interpreted at every world $w$ as $D_w^{|\bar x|}$.

  Upper bounds follow from Lemmas~\ref{lem:guarded-syntax}, \ref{lem:Theta}, and
  \ref{lem:guarded-renaming}, together with Proposition~\ref{prop:GF}.  The
  lower bound in (2) follows from Remark~\ref{rem:GF-variants}
  and~\cite[Theorem~3]{BKP25}.
  %The lower bound in (2) follows from Remark~\ref{rem:conserv}, since, by
  %Remark~\ref{rem:GF-variants} and~\cite[Theorem~3]{BKP25}, validity of
  %$\mathcal{GF}_=$-sentences without equality is $\coTwoEXP$-hard.
  The lower bound in (1) follows from 
  Remark~\ref{rem:GF-variants}.
  %The lower bound in (1) follows from Remark~\ref{rem:conserv}, since, by
  %Remark~\ref{rem:GF-variants}, validity of $\mathcal{GF}_=$-sentences with
  %two variables and without equality, all of whose predicate letters other
  %than a single binary one are monadic, is $\coEXP$-hard.
\end{proof}

\begin{remark}
  \label{rem:guarded-one-var}
  Theorem~\ref{thm:guarded-complexity} is stated for $k \geqslant 2$, as is
  Proposition~\ref{prop:GF}.  For $k = 1$ and $d\in\{\mathit{xd},\mathit{cd}\}$,
  the two sets of Theorem~\ref{thm:guarded-complexity}\rmbr{$1$} are
  $\coNP$-complete: a formula belongs to one of them if, and only if, it is a
  guarded formula of the corresponding language with at most one variable and
  belongs to $(\mathbf{ML}^d_{\mathrm{C}}\fr{F})^1$, which is in $\coNP$ by
  Theorem~\ref{thm:one-var}; the lower bound holds already for propositional
  formulas, which are guarded and contain no equality.
\end{remark}

\begin{theorem}[Finite domain property for guarded fragments]
  \label{thm:FDP-guarded}
  Let $\fr F=\langle W,R\rangle$ be a finite Kripke frame, let
  $d\in\{\mathit{xd},\mathit{cd}\}$, let $\vp\in\GFe$ have at most $k$
  variables, let $a$ be the maximal arity of the predicate letters of $\vp$, and
  let $\varrho:=\max(2,k,a)$.  If $\vp\notin\mathbf{ML}^d_=\fr F$, then $\vp$ is
  refuted in a $d$-model $\langle W,R,D,I\rangle$ with
  $|D^+|\leqslant2^{O(|W|\cdot|\vp|\cdot(2\varrho)^{\varrho})}$.  In particular,
  $|D^+|\leqslant2^{|W|\cdot2^{O(|\vp|\log|\vp|)}}$; if $k\geqslant2$ is fixed
  and $a\leqslant k$, then $|D^+|\leqslant2^{O(|W|\cdot|\vp|)}$, the constant
  depending on $k$.
\end{theorem}

\begin{proof}
  Follows from an estimate of the size of models of satisfiable
  $\mathcal{GF}_=$-sentences~\cite{Fiuk26} applied to $\Theta^{d}_{\fr F}(\vp)$;
  see Appendix~\ref{sec:thm:FDP-guarded}.
\end{proof}

\section{Complexity of the monadic fragment with counting and of its subfragments}
\label{sec:types}

In this section, we determine, for a finite Kripke frame $\fr{F}$, the
complexity of the monadic fragments of the logics
$\mathbf{ML}^d_{\mathrm{C}}\fr{F}$, for $d\in\{\mathit{xd},\mathit{cd}\}$, as
well as of their subfragments with a fixed number of predicate letters.  The
translation $\ST^{\fr{F}}_w$ of Section~\ref{sec:transfer} sends monadic
formulas to monadic formulas
(Observation~\ref{obs:monadicity}), but the renaming used in
Section~\ref{sec:complexity} introduces predicate letters of arity greater than
one. If we were interested only in full monadic fragments, we could obtain the
optimal complexity bounds by exploiting a connection with $\CQCL$: although the
classical formula $\ST_w^{\fr{F}}(\vp)$ may be exponentially larger than $\vp$
(Remark~\ref{rem:st}), it has at most $|W|\cdot(s+1)$ predicate letters, where
$s$ is the number of predicate letters of $\vp$, and the same variables and
counting thresholds as $\vp$, so a bound on the size of models of monadic
sentences in terms of these parameters yields refuting models of size
exponential in $|\fr{F}|+|\vp|$ (\cite[Lemma~3.20]{PH23} for sentences without
counting quantifiers, and its routine extension to counting quantifiers, cf.\
Proposition~\ref{prop:classical-counting} below); thus, a refuting model could
be guessed and checked by a non-deterministic algorithm.\footnote{The algorithm
  would be optimal since, as we shall see (Theorem~\ref{thm:nexptime}), the
  sets $(\mathbf{ML}^d_{\mathrm{C}}\fr{F})^{\mathit{mon}}$ are
  $\coNEXP$-complete.} We do, however, also consider the variants of these
problems where the number of predicate letters, and the number of variables,
allowed in $\vp$ are fixed.  While most common first-order modal logics over
classes of frames~\cite{RS19,Ryb24,RS25-trick} and over infinite
frames~\cite{RS-AiML2020,RShJLC21c,RSh26-Dedekind,RS26-ARQNL} remain undecidable
even in languages with a fixed number of monadic predicate letters and a fixed
number of variables (for many of them, one letter and two variables), over a
single finite frame a fixed number of predicate letters brings the complexity
down to $\PSPACE$-completeness, and fixing, in addition, the number of variables
brings it down to polynomial time.

To establish these results, we adapt the histogram argument for the classical
one-variable fragment with counting~\cite[Section 7.6]{PH23}, combined with
L\"owenheim's identification of elements of the same type~\cite{L1915},
\cite[Section 3.6]{PH23}: we describe models over $\fr{F}$ succinctly, by the
number of elements of each type, capped at a number $\kappa_\vp$ computed from
the input formula $\vp$.

The key observation (Lemma~\ref{lem:eval}) is that the truth of a monadic
sentence $\vp$ in a model over $\fr{F}$ depends only on the capped histogram of
the model.  This gives the finite domain property with an explicit bound
(Theorem~\ref{thm:small}), which can be lowered to one element of each type for
formulas without equality and counting quantifiers
(Proposition~\ref{prop:fdp-ML}), $\coNEXP$-completeness
(Theorem~\ref{thm:nexptime}), $\PSPACE$-completeness when the frame and the
number of predicate letters are fixed (Proposition~\ref{prop:pspace}), and
polynomial time when, in addition, the number of variables is fixed
(Proposition~\ref{prop:bounded-var}); the proofs of the last two results need
the frame to be fixed: with the frame as part of the input one predicate letter
and two variables already give $\coNEXP$-hardness
(Proposition~\ref{prop:frame-input-mon}).

Types and histograms depend on the frame and on the predicate letters of the
formula being considered.  In Sections~\ref{sec:types-hist}--\ref{sec:refute},
we therefore fix a finite Kripke frame $\fr{F}=\langle W,R\rangle$ and $s\in\N$,
and consider formulas of $\CMLmon$ whose predicate letters other than $=$ are
among $P_1,\ldots,P_s$. Since proposition letters can be eliminated without
changing any of the parameters introduced below (Remark~\ref{rem:prop-letters}),
we assume that $P_1,\ldots,P_s$ are all monadic.

For a formula $\vp$, we use the following notation.  The set of variables of
$\vp$, free or bound, is denoted by $\varset\vp$, and $v_\vp:=|{\varset\vp}|$.
We only consider formulas $\vp$ with $v_\vp\geqslant1$ (otherwise, since we
exclude proposition letters, $\vp$ is built from $\bot$ using $\to$ and
$\square$, and its validity on $\fr{F}$ can be decided in polynomial time);
Theorem~\ref{thm:small} and Proposition~\ref{prop:classical-counting}, whose
statements may be read on their own, say explicitly what happens if $v_\vp=0$.
Lastly, we put (recall that $M_\vp$ denotes the counting threshold of $\vp$)
$$
\begin{array}{lcl}
  \kappa_\vp & := & M_\vp+v_\vp-1;
\end{array}
$$
this is the cap we use for histograms (Remark~\ref{rem:cap} explains why it
cannot be lowered).  Note that $1\leqslant\kappa_\vp<2^{|\vp|+1}$.

\begin{remark}[Proposition letters]
  \label{rem:prop-letters}
  Let $d\in\{\mathit{xd},\mathit{cd}\}$, let $\vp\in\CMLmon$, let
  $Q_1,\ldots,Q_r$ be the proposition letters of
  $\vp$, let $U_1,\ldots,U_r$ be monadic predicate letters not occurring in
  $\vp$, and let $x$ be a variable of $\vp$ if $v_\vp\geqslant1$, and
  $x:=x_1$ otherwise.  Denote by $\vp^{\flat}$ the result of replacing in
  $\vp$ every occurrence of $Q_j$ with the sentence $\exists x\,U_j(x)$, for
  $j=1,\ldots,r$.  Then, 
  $$
  \begin{array}{lcl}
    \vp\in\mathbf{ML}^d_{\mathrm{C}}\fr{F} & \iff & \vp^{\flat}\in\mathbf{ML}^d_{\mathrm{C}}\fr{F},
  \end{array}
  $$
  and $\vp$ is refuted in a $d$-model over $\fr{F}$ whose global domain has
  a given cardinality if, and only if, so is $\vp^{\flat}$.
  %Indeed, given
  %$\fr{M}=\langle W,R,D,I\rangle$, let $I^{\flat}$ agree with $I$ on every
  %predicate letter other than $U_1,\ldots,U_r$, and put
  %$I^{\flat}(w,U_j):=D_w$ if $\fr{M},w\sat Q_j$ and
  %$I^{\flat}(w,U_j):=\varnothing$ otherwise; conversely, given
  %$\fr{M}'=\langle W,R,D,I'\rangle$, let $I$ agree with $I'$ on every
  %predicate letter other than $Q_1,\ldots,Q_r$ and make $Q_j$ true at $w$
  %if, and only if, $I'(w,U_j)\neq\varnothing$.  In both cases, since
  %$D_w\neq\varnothing$, the sentence $\exists x\,U_j(x)$ is true at $w$ in
  %the model interpreting $U_j$ if, and only if, $Q_j$ is true at $w$ in the
  %model interpreting $Q_j$.  As the inserted sentences have no parameters,
  %the substitution commutes with $\to$, $\square$ and the quantifiers, and a
  %routine induction on $\psi\in\subf\vp$ shows that, for every $w\in W$ and
  %every assignment $g$, the formula $\psi$ is true at $w$ under $g$ in the
  %one model if, and only if, $\psi^{\flat}$ is true at $w$ under $g$ in the
  %other.

  The formula $\vp^{\flat}$ is computable from $\vp$ in polynomial time, and
  $\param{\vp^{\flat}}=\param{\vp}$; $\varset\vp^{\flat}=\varset\vp$ if
  $v_\vp\geqslant1$ or $r=0$, and $\varset\vp^{\flat}=\{x_1\}$ if $v_\vp=0$ and
  $r\geqslant1$; $M_{\vp^{\flat}}=M_\vp$, since $M_\vp\geqslant1$ and the only
  quantifiers added by the substitution are the quantifiers $\exists (=\ex{1})$ of
  the sentences $\exists x\,U_j(x)$; the predicate letters of
  $\vp^{\flat}$ other than $=$ are monadic and as many as those of $\vp$, so
  they may be renamed to be among $P_1,\ldots,P_s$; $\vp^{\flat}\in\MLangeq$ if
  $\vp\in\MLangeq$, $\vp^{\flat}\in\MLang$ if $\vp\in\MLang$, and $\vp^{\flat}$
  is $\square$-free if $\vp$ is.  Thus, $v_{\vp^{\flat}}=\max\{v_\vp,1\}$ if
  $r\geqslant1$ and $v_{\vp^{\flat}}=v_\vp$ if $r=0$; in either case,
  $\kappa_{\vp^{\flat}}=\kappa_\vp$ whenever $v_\vp\geqslant1$.
\end{remark}

\subsection{Types and histograms}
\label{sec:types-hist}

The type of an element of a model over $\fr{F}$ records the worlds at which
the element exists and the letters among $P_1,\ldots,P_s$ it satisfies at
each of them.  We first define types abstractly.

A \emph{type} over $\fr{F}$ (for the letters $P_1,\ldots,P_s$) is a set
$t\subseteq W\times\{0,1,\ldots,s\}$ satisfying the following conditions (we
read `$\langle w,0\rangle\in t$' as `$t$ exists at $w$'):
\begin{enumerate}[label=(T\arabic*)]
\item there exists $w\in W$ such that $\langle w,0\rangle\in t$;
\item if $\langle w,0\rangle\in t$ and $wRv$, then $\langle v,0\rangle\in t$;
\item for every $i\in\{1,\ldots,s\}$, if $\langle w,i\rangle\in t$, then
  $\langle w,0\rangle\in t$.
\end{enumerate}
We denote by $\Typ(\fr{F},s)$ the set of all types over $\fr{F}$ for
$P_1,\ldots,P_s$.  If $t\in\Typ(\fr{F},s)$, then
$E_t:=\{w\in W:\langle w,0\rangle\in t\}$ is the \emph{existence set} of~$t$; by
(T1) and (T2), $E_t$ is a non-empty subset of $W$ closed under $R$ (i.e.,
$R(E_t)\subseteq E_t$).  By (T3), a type $t$ is determined by $E_t$ together
with the sets $\{i\geqslant1:\langle w,i\rangle\in t\}$, where $w\in E_t$.
Hence,
\begin{equation}
  \label{eq:numtypes}
  |\Typ(\fr{F},s)| \;=\; \sum_{E}2^{s\cdot|E|} \;\leqslant\; (2^{s}+1)^{|W|}-1
  \;<\; 2^{(s+1)|W|},
\end{equation}
where $E$ ranges over the non-empty subsets of $W$ closed under $R$; in
particular, $|\Typ(\fr{F},s)|=2^{s}$ if $|W|=1$.

Let $\fr{M}=\langle W,R,D,I\rangle$ be a Kripke model over $\fr{F}$.  The
\emph{type} of $a\in\Dplus$ in $\fr{M}$ is the set
\[
  \tp_{\fr{M}}(a):=\{\langle w,0\rangle : a\in D_w\}\cup\{\langle w,i\rangle :
  1\leqslant i\leqslant s \text{ and } a\in P_i^{I,w}\}.
\]
Then, $\tp_{\fr{M}}(a)\in\Typ(\fr{F},s)$: condition (T1) holds since
$a\in\Dplus$, condition (T2) holds by~\eqref{eq:expanding}, and condition (T3)
holds since, by definition of a Kripke model, $P_i^{I,w}\subseteq D_w$.  By
definition of $\tp_{\fr{M}}(a)$, for all $a\in\Dplus$, $w\in W$ and
$i\in\{1,\ldots,s\}$,
\begin{equation}
  \label{eq:tp}
  a\in D_w \iff w\in E_{\tp_{\fr{M}}(a)}
  \qquad\mbox{and}\qquad
  a\in P_i^{I,w} \iff \langle w,i\rangle\in\tp_{\fr{M}}(a).
\end{equation}
The \emph{histogram} of $\fr{M}$ is the function $\hist_{\fr{M}}$ sending each
$t\in\Typ(\fr{F},s)$ to the cardinal
$$
\begin{array}{lcl}
  \hist_{\fr{M}}(t) & := & |\{a\in\Dplus:\tp_{\fr{M}}(a)=t\}|.
\end{array}
$$
If
$\kappa\in\Npos$, the \emph{$\kappa$-capped histogram} of $\fr{M}$ is the
function $\hist^{\kappa}_{\fr{M}}\colon\Typ(\fr{F},s)\to\{0,1,\ldots,\kappa\}$
defined by
$$
\begin{array}{lcl}
  \hist^{\kappa}_{\fr{M}}(t) & := & \min\{\hist_{\fr{M}}(t),\kappa\}.
\end{array}
$$

Not every function on $\Typ(\fr{F},s)$ is the histogram of a model.  A
function $\hist$ from $\Typ(\fr{F},s)$ to cardinals is
\emph{$\mathit{xd}$-admissible} if
\begin{equation}
  \label{eq:nonempty-h}
  (\Forall w\in W) (\Exists t\in\Typ(\fr{F},s))\, (\hist(t)\geqslant 1 ~\&~ w\in E_t),
  %\mbox{for every $w\in W$, there exists $t\in\Typ(\fr{F},s)$ such that
  %  $\hist(t)\geqslant1$ and $w\in E_t$,}
\end{equation}
and \emph{$\mathit{cd}$-admissible} if, in addition,
\begin{equation}
  \label{eq:model-h-cd}
  (\Forall t\in\Typ(\fr{F},s)) (\Forall w,v\in W)\, (\hist(t)\geqslant1 ~\&~ wRv
  ~\&~ v\in E_t \Rightarrow w\in E_t).
  %\mbox{for all $t\in\Typ(\fr{F},s)$ and $w,v\in W$, if $\hist(t)\geqslant1$,
  %  $wRv$ and $v\in E_t$, then $w\in E_t$.}
\end{equation}
Since $E_t$ is closed under $R$, condition~\eqref{eq:model-h-cd} says that,
whenever $\hist(t)\geqslant1$, the set $E_t$ is also closed under the converse
of $R$; in other words, $E_t$ is a union of connected components of $\fr{F}$,
i.e., of equivalence classes of the least equivalence relation on $W$ containing
$R$.  We denote by $\Typ^{\mathit{cd}}(\fr{F},s)$ the set of types $t$ such that
$E_t$ is a union of connected components of $\fr{F}$; hence, if $\hist$ is
$\mathit{cd}$-admissible, then $\hist(t)=0$ for every
$t\notin\Typ^{\mathit{cd}}(\fr{F},s)$.  Thus,
\begin{equation}
  \label{eq:numtypes-cd}
  |\Typ^{\mathit{cd}}(\fr{F},s)| \;=\; \sum_{E}2^{s\cdot|E|},
\end{equation}
where $E$ ranges over the non-empty unions of connected components of
$\fr{F}$; in particular, $|\Typ^{\mathit{cd}}(\fr{F},s)|=2^{s|W|}$ if
$\fr{F}$ is connected.  We also put
$\Typ^{\mathit{xd}}(\fr{F},s):=\Typ(\fr{F},s)$; thus, for
$d\in\{\mathit{xd},\mathit{cd}\}$, if $\hist$ is $d$-admissible, then
$\hist(t)=0$ for every $t\notin\Typ^{d}(\fr{F},s)$.

It is easy to check that, if $d\in\{\mathit{xd},\mathit{cd}\}$, $\fr{M}$ is a
$d$-model over $\fr{F}$ and $\kappa\in\Npos$, then both $\hist_{\fr{M}}$ and
$\hist^{\kappa}_{\fr{M}}$ are $d$-admissible.

% FULL PROOF
%\begin{observation}
%  \label{obs:hist-admissible}
%  Let $d\in\{\mathit{xd},\mathit{cd}\}$, let $\fr{M}$ be a $d$-model over
%  $\fr{F}$ and $\kappa\in\Npos$.  Then $\hist_{\fr{M}}$ and
%  $\hist^{\kappa}_{\fr{M}}$ are $d$-admissible.
%\end{observation}

%\begin{proof}
%  Conditions~\eqref{eq:nonempty-h} and~\eqref{eq:model-h-cd} depend on a
%  function $\hist$ only through the set of types $t$ with
%  $\hist(t)\geqslant1$.  As $\kappa\geqslant1$, this set is the same for
%  $\hist_{\fr{M}}$ and $\hist^{\kappa}_{\fr{M}}$; hence, it suffices to verify
%  the conditions for $\hist_{\fr{M}}$.
%
%  For~\eqref{eq:nonempty-h}, let $w\in W$.  Since $D_w\neq\varnothing$, there
%  is some $a\in D_w$; put $t:=\tp_{\fr{M}}(a)$.  Then
%  $\hist_{\fr{M}}(t)\geqslant1$, as $a$ is an element of type $t$, and
%  $w\in E_t$ by~\eqref{eq:tp}.
%
%  For~\eqref{eq:model-h-cd}, let $\fr{M}$ be a $\mathit{cd}$-model, and let
%  $t\in\Typ(\fr{F},s)$ and $w,v\in W$ be such that
%  $\hist_{\fr{M}}(t)\geqslant1$, $wRv$ and $v\in E_t$.  Choose $a\in\Dplus$
%  with $\tp_{\fr{M}}(a)=t$.  By~\eqref{eq:tp}, $a\in D_v$;
%  by~\eqref{eq:constant}, $D_w=D_v$, so $a\in D_w$; hence, $w\in E_t$, again
%  by~\eqref{eq:tp}.
%\end{proof}

Conversely, for $d\in\{\mathit{xd},\mathit{cd}\}$, every $d$-admissible
function is the histogram of a $d$-model (we prove this only for functions with
values in $\N$, which is sufficient for our purposes).  If
$\hist\colon\Typ(\fr{F},s)\to\N$ is $\mathit{xd}$-admissible, the
\emph{canonical model} of
$\hist$ is $\fr{M}_{\hist}:=\langle W,R,D^{\hist},I^{\hist}\rangle$, where
$$
\begin{array}{lcl}
  (D^{\hist})^{+} & := & \{\langle t,j\rangle : t\in\Typ(\fr{F},s) \mbox{ and }
                         1\leqslant j\leqslant\hist(t)\};\\
  D^{\hist}_w & := & \{\langle t,j\rangle\in(D^{\hist})^{+} : w\in E_t\};\\
  I^{\hist}(w,P_i) & := & \{\langle t,j\rangle\in(D^{\hist})^{+} : 
                          \langle w,i\rangle\in t\}, \quad\mbox{with $1\leqslant i\leqslant s$,}
\end{array}
$$
and, for definiteness, every predicate letter other than $=$ and
$P_1,\ldots,P_s$ is interpreted as $\varnothing$ at every world.  By (T1),
$(D^{\hist})^{+}$ is the union of the domains $D^{\hist}_w$, i.e., the global
domain of $\fr{M}_{\hist}$.

\begin{lemma}
  \label{lem:canonical}
  Let $d\in\{\mathit{xd},\mathit{cd}\}$, and let
  $\hist\colon\Typ(\fr{F},s)\to\N$ be $d$-admissible.  Then $\fr{M}_{\hist}$
  is a $d$-model over $\fr{F}$ such that $\tp_{\fr{M}_{\hist}}(t,j)=t$ for
  every $\langle t,j\rangle\in(D^{\hist})^{+}$; hence,
  $\hist_{\fr{M}_{\hist}}=\hist$ and
  $|(D^{\hist})^{+}|=\sum_{t\in\Typ(\fr{F},s)}\hist(t)$.
\end{lemma}

\begin{proof}
  Each $D^{\hist}_w$ is non-empty by~\eqref{eq:nonempty-h}, the domains
  satisfy~\eqref{eq:expanding} due to (T2), and
  $I^{\hist}(w,P_i)\subseteq D^{\hist}_w$ by (T3); thus, $\fr{M}_{\hist}$ is an
  $\mathit{xd}$-model over $\fr{F}$.

  By definition, $\langle t,j\rangle\in D^{\hist}_w$ iff
  $\langle w,0\rangle\in t$, and $\langle t,j\rangle\in I^{\hist}(w,P_i)$ iff
  $\langle w,i\rangle\in t$; hence,
  $\tp_{\fr{M}_{\hist}}(t,j)=t$, and the elements of type $t$ in
  $\fr{M}_{\hist}$ are $\langle t,1\rangle,\ldots,\langle t,\hist(t)\rangle$.
  
  Lastly, let $d=\mathit{cd}$, so that $\hist$ is $\mathit{cd}$-admissible,
  and let $wRv$.  Every
  $\langle t,j\rangle\in(D^{\hist})^{+}$ satisfies $\hist(t)\geqslant1$, and so,
  by (T2) and~\eqref{eq:model-h-cd}, $\langle t,j\rangle\in D^{\hist}_w$ iff
  $\langle t,j\rangle\in D^{\hist}_v$; thus, $\fr{M}_{\hist}$
  satisfies~\eqref{eq:constant}, i.e., it is a $\mathit{cd}$-model.
\end{proof}

\subsection{Evaluating formulas on capped histograms}
\label{sec:eval}

Throughout this subsection, we fix a formula $\vp\in\CMLmon$ whose predicate
letters other than $=$ are among $P_1,\ldots,P_s$.  We define a satisfaction
relation for the subformulas of $\vp$ that refers only to a capped histogram and
a type assignment.  We then show that this relation agrees with the satisfaction
relation of every model with that capped histogram.

Let $V \subseteq\varset\vp$.  A \emph{type assignment} on $V$ is a pair
$\tau=\langle\bar t,\equiv\rangle$, where $\bar t$ is a function sending each
$y\in V$ to a type $t_y\in\Typ(\fr{F},s)$ and $\equiv$ is an equivalence
relation on $V$ such that $t_{y_1}=t_{y_2}$ whenever $y_1 \equiv y_2$.  We denote by
$V/{\equiv}$ the set of equivalence classes with respect to $\equiv$ and write $t_C$ for the
common type of the members of a class $C \in V/{\equiv}$; if
$t\in\Typ(\fr{F},s)$, we put
$$
\begin{array}{lcl}
  c_t(\tau) & := & |\{C\in V/{\equiv} : t_C=t\}|.
\end{array}
$$
We say that $\tau$ is \emph{at} $w\in W$ if $w\in E_{t_y}$ for every $y\in V$.
The only type assignment on $\varnothing$ is denoted by $\tau_\varnothing$; it
is at every world.

An assignment $g$ in a Kripke model $\fr{M}$ over $\fr{F}$ \emph{realises}
$\tau$ if $\tp_{\fr{M}}(g(y))=t_y$ for every $y\in V$, and $g(y_1)=g(y_2)$ iff
$y_1\equiv y_2$, for all $y_1,y_2\in V$; thus, distinct classes of $\tau$, even
of the same type, take distinct values under $g$.  Every assignment realises
$\tau_\varnothing$.  By~\eqref{eq:tp}, if $g$ realises a type assignment on $V$
that is at $w$, then $g(y)\in D_w$ for every $y\in V$; conversely, every
assignment $g$ with $g(y)\in D_w$, for every $y\in V$, realises exactly one type
assignment on $V$, and this type assignment is at $w$.

We use the restriction of $\tau$ to a set $V_0\subseteq V$, defined as
$\tau{\restriction}V_0:=\langle\bar t{\restriction} V_0,\,{\equiv}\cap(V_0\times V_0)\rangle$,
and two ways of extending $\tau$ to a variable $z\in\varset\vp\setminus V$:
\begin{itemize}
\item if $C\in V/{\equiv}$, then $\tau[z\equiv C]$ is the type assignment on
  $V\cup\{z\}$ that extends $\tau$ and adds $z$ to the class $C$, with the
  type $t_C$;
\item if $t\in\Typ(\fr{F},s)$, then $\tau[z:t]$ is the type assignment on
  $V\cup\{z\}$ that extends $\tau$ and puts $z$ into a new singleton class,
  with the type $t$.
\end{itemize}
If $\tau$ is at $w$, then the type assignments $\tau[z\equiv C]$, with
$C\in V/{\equiv}$, and $\tau[z:t]$, with $w\in E_t$, are exactly the type
assignments on $V\cup\{z\}$ at $w$ whose restriction to $V$ is $\tau$; we call
them the \emph{extensions of $\tau$ to $z$ at $w$}.

Let $\hist\colon\Typ(\fr{F},s)\to\{0,1,\ldots,\kappa_\vp\}$.  We say that a type
assignment $\tau$ on $V$ is \emph{$\hist$-compatible} if
$c_t(\tau)\leqslant\hist(t)$ for every $t\in\Typ(\fr{F},s)$, i.e., if $\hist$
provides enough elements of each type for the classes of $\tau$ to take distinct
values.

\begin{lemma}
  \label{lem:realised-compatible}
  Let $\fr{M}$ be a model over $\fr{F}$ and $g$ an assignment in $\fr{M}$.  If
  $g$ realises $\tau$ in $\fr{M}$, then $\tau$ is
  $\hist^{\kappa_\vp}_{\fr{M}}$-compatible.
\end{lemma}

\begin{proof}
  Let $t\in\Typ(\fr{F},s)$.  The classes of $\tau$ of type $t$ take distinct
  values under $g$, all of type $t$, so $c_t(\tau)\leqslant\hist_{\fr{M}}(t)$;
  moreover, $c_t(\tau)\leqslant|V|\leqslant v_\vp\leqslant\kappa_\vp$, since
  $V\subseteq\varset\vp$ and $M_\vp\geqslant1$.  Hence,
  $c_t(\tau)\leqslant\min\{\hist_{\fr{M}}(t),\kappa_\vp\}
  =\hist^{\kappa_\vp}_{\fr{M}}(t)$.
\end{proof}

Conversely, if $\hist$ is $\mathit{xd}$-admissible, then every
$\hist$-compatible $\tau$ is realised in $\fr{M}_{\hist}$: for each
$t\in\Typ(\fr{F},s)$, choose an injection $\iota_t$ from the set of classes of
$\tau$ of type $t$ into
$\{1,\ldots,\hist(t)\}$, which exists since $c_t(\tau)\leqslant\hist(t)$, and
put $g(y):=\langle t_y,\iota_{t_y}(C_y)\rangle$ for $y\in V$, where $C_y$ is
the class of $y$ (the values of $g$ outside $V$ are immaterial).  By
Lemma~\ref{lem:canonical}, $\tp_{\fr{M}_{\hist}}(g(y))=t_y$, and
$g(y_1)=g(y_2)$ iff $C_{y_1}=C_{y_2}$, i.e., iff $y_1\equiv y_2$; thus, $g$
realises $\tau$.

The type assignment $\tau_\varnothing$ is $\hist$-compatible; restrictions of
$\hist$-compatible type assignments are $\hist$-compatible, and so are the
extensions $\tau[z\equiv C]$; as $c_t(\tau[z:t])=c_t(\tau)+1$ while the other
numbers $c_{t'}$ are unchanged, the extension $\tau[z:t]$ of an
$\hist$-compatible $\tau$ is $\hist$-compatible iff $\hist(t)>c_t(\tau)$.

Let $\tau$ be an $\hist$-compatible type assignment on $V$ at $w$, and let
$z\in\varset\vp\setminus V$.  The \emph{multiplicity} in $\hist$ of an
extension $\tau'$ of $\tau$ to $z$ at $w$ is defined by
$$
\begin{array}{lclcl}
  \hist(\tau[z\equiv C]/\tau) & := & 1, & & \mbox{for every $C\in V/{\equiv}$;}\\
  \hist(\tau[z:t]/\tau) & := & \hist(t)-c_t(\tau), & & \mbox{for every
                                $t\in\Typ(\fr{F},s)$ with $w\in E_t$.}
\end{array}
$$
To see what multiplicities mean, apply the same definition to the histogram
$\hist_{\fr{M}}$ of a model $\fr{M}$, which need not take values in
$\{0,1,\ldots,\kappa_\vp\}$.  If $g$ realises $\tau$ in $\fr{M}$, then
$\hist_{\fr{M}}(\tau'/\tau)$ is the number of elements $b\in D_w$ such that
the assignment that differs from $g$ at most in the value of $z$ and sends $z$
to $b$ realises $\tau'$; this number depends on the type of $\tau'$: if
$\tau'=\tau[z\equiv C]$, then $b$ must be the common value under $g$ of the
members of $C$; if $\tau'=\tau[z:t]$, then $b$ must be one of the
$\hist_{\fr{M}}(t)$ elements of type $t$ other than the $c_t(\tau)$ values
under $g$ of the classes of $V/{\equiv}$ of type~$t$.  Note that, for
$\hist\colon\Typ(\fr{F},s)\to\{0,1,\ldots,\kappa_\vp\}$, the $\hist$-compatible
extensions of $\tau$ to $z$ at $w$ are exactly those of positive multiplicity.

For a function $\hist\colon\Typ(\fr{F},s)\to\{0,1,\ldots,\kappa_\vp\}$ (not
necessarily $\mathit{xd}$-admissible), $\psi\in\subf\vp$, $w\in W$ and an
$\hist$-compatible type assignment $\tau=\langle\bar t,\equiv\rangle$ at $w$
on a set $V\subseteq\varset\vp$ with $\param{\psi}\subseteq V$, we define the
relation $\hist,w\sat^{\tau}\psi$ by recursion on $\psi$:
\begin{itemize}
\item $\hist,w\sat^{\tau} P_i(y)$ if $\langle w,i\rangle\in t_y$;
\item $\hist,w\sat^{\tau} y_1 = y_2$ if $y_1 \equiv y_2$;
\item $\hist,w\nsat^{\tau}\bot$;
\item $\hist,w\sat^{\tau}\psi_1\to\psi_2$ if $\hist,w\nsat^{\tau}\psi_1$ or
  $\hist,w\sat^{\tau}\psi_2$;
\item $\hist,w\sat^{\tau}\square\psi_1$ if $\hist,v\sat^{\tau}\psi_1$ whenever
  $v\in R(w)$;
\item $\hist,w\sat^{\tau}\ex{M}z\,\psi_1$ if
  \[
    \sum_{\tau'\in X}\hist(\tau'/\tau_0)\;\geqslant\;M,
  \]
  where $\tau_0:=\tau{\restriction}(V\setminus\{z\})$ and $X$ is the set of
  $\hist$-compatible extensions $\tau'$ of $\tau_0$ to $z$ at $w$ such that
  $\hist,w\sat^{\tau'}\psi_1$.
\end{itemize}

The clause for $\ex{M}$ mirrors the truth clause for counting quantifiers, which
asks for at least $M$ values $b\in D_w$ of $z$ satisfying $\psi_1$.  With the
classes of $\tau_0$ standing for the values of the variables in
$V_0:=V\setminus\{z\}$, every value $b\in D_w$ for $z$, together with these
values, realises exactly one extension $\tau'$ of $\tau_0$ to $z$ at $w$;
whether $b$ satisfies $\psi_1$ depends only on $\tau'$, and
$\hist(\tau'/\tau_0)$ is the number of values $b$ realising $\tau'$.  So the sum
in the clause is the number of values of $z$ satisfying $\psi_1$, provided
$\hist$ is the exact histogram of the model.  The choice of the cap makes the
clause work for the $\kappa_\vp$-capped histogram as well: if
$\hist(t)=\kappa_\vp$, there may be more than $\hist(t)$ elements of type $t$,
and then $\hist(\tau_0[z:t]/\tau_0)$ underestimates the number of values
realising $\tau_0[z:t]$; but, as $c_t(\tau_0)\leqslant|V_0|\leqslant v_\vp-1$,
this multiplicity is at least $\kappa_\vp-(v_\vp-1)=M_\vp\geqslant M$; thus, if
$\tau_0[z:t]\in X$, the sum reaches $M$, and so does the true number of values
satisfying $\psi_1$.  Lemma~\ref{lem:eval} below makes this reading precise.

Note that the recursion for $\Vdash$ is well defined: in the clause for $\square$,
the type assignment $\tau$ is at every $v\in R(w)$, by (T2); in the clause for
$\ex{M}$, we have $\param{\psi_1}\subseteq V_0\cup\{z\}$, and $\tau_0$ and the
members of $X$ are $\hist$-compatible type assignments at $w$.  The clauses do
not refer to the cap: they evaluate $\psi$ as if $\hist$ were the histogram of
a model.  Thus, the relation $\hist,w\sat^{\tau}\psi$ depends on $\vp$ only
through $\varset\vp$ and the range $\{0,1,\ldots,\kappa_\vp\}$ of $\hist$.

\begin{example}
  \label{ex:hist}
  \rm
  Let $\fr{F}$ be the frame with $W=\{u,w\}$ and $R=\{\langle u,w\rangle\}$, and
  let $s=1$.  The non-empty $R$-closed subsets of $W$ are $\{w\}$ and $W$;
  hence, by~\eqref{eq:numtypes}, there are $2+4=6$ types.  Let
  $$
  \begin{array}{lcl}
    \vp & := & \exists x\,\bigl(P_1(x)\wedge\Diamond\,\ex{2}y\,(\neg P_1(y)\wedge\neg(x=y))\bigr),
  \end{array}
  $$
  so that $v_\vp=M_\vp=2$ and $\kappa_\vp=3$.  Let $\fr{M}$ be the model over
  $\fr{F}$ with $D_u=\{a\}$, $D_w=\{a,b\}$, $P_1^{I,u}=\{a\}$ and
  $P_1^{I,w}=\varnothing$.  Then $a$ has the type
  $t_a:=\{\langle u,0\rangle,\langle u,1\rangle,\langle w,0\rangle\}$, $b$ has
  the type $t_b:=\{\langle w,0\rangle\}$, and $\hist:=\hist^{3}_{\fr{M}}$ sends
  $t_a$ and $t_b$ to $1$ and the other four types to $0$.  The extensions of
  $\tau_\varnothing$ to $x$ at $u$ are those of the form $\tau_\varnothing[x:t]$
  with $u\in E_t$, of multiplicity $\hist(t)$; the only one of positive
  multiplicity is $\tau:=\tau_\varnothing[x:t_a]$.  As
  $\langle u,1\rangle\in t_a$ and $R(u)=\{w\}$, we get
  $\hist,u\sat^{\tau_\varnothing}\vp$ iff
  $\hist,w\sat^{\tau}\ex{2}y\,(\neg P_1(y)\wedge\neg(x=y))$.  The extensions of
  $\tau$ to $y$ at $w$ are $\tau[y\equiv\{x\}]$, of multiplicity $1$, and those
  of the form $\tau[y:t]$, with $w\in E_t$ (i.e., with $t$ being any of the six
  types), of multiplicity $\hist(t)-c_t(\tau)$; these are $\hist(t_a)-1=0$ (for
  $t=t_a$), $\hist(t_b)=1$ (for $t=t_b$), and $0$ for all types
  $t \notin \{t_a, t_b\}$, as, in this case, $\hist(t)=0$.  Hence, only
  $\tau[y\equiv\{x\}]$ and $\tau[y:t_b]$ are $\hist$-compatible.  The formula
  $\neg P_1(y)\wedge\neg(x=y)$ fails under $\tau[y\equiv\{x\}]$, since
  $x\equiv y$, and holds under $\tau[y:t_b]$, since
  $\langle w,1\rangle\notin t_b$.  Hence, the sum in the clause for $\ex{2}y$ is
  $1<2$, and so $\hist,u\nsat^{\tau_\varnothing}\vp$.  Accordingly,
  $\fr{M},u\nsat\vp$, since $b$ is the only element of $D_w$ other than~$a$.  If
  a second element of type $t_a$ is added to $\fr{M}$, then $\hist(t_a)=2$, so
  $\tau[y:t_a]$ becomes $\hist$-compatible, with multiplicity $1$, and it
  satisfies the formula, since $\langle w,1\rangle\notin t_a$; the sum becomes
  $2$, and $\vp$ becomes true at $u$.
\end{example}

\begin{lemma}[Evaluation lemma]
  \label{lem:eval}
  Let $\fr{M}=\langle W,R,D,I\rangle$ be a model over $\fr{F}$ and let
  $\hist:=\hist^{\kappa_\vp}_{\fr{M}}$.  Then, for every $\psi\in\subf\vp$,
  every $w\in W$, every type assignment $\tau$ at $w$ on a set
  $V\subseteq\varset\vp$ with $\param{\psi}\subseteq V$, and every assignment
  $g$ in $\fr{M}$ realising $\tau$,
  \[
    \fr{M},w\sat^{g}\psi \iff \hist,w\sat^{\tau}\psi.
  \]
\end{lemma}

\begin{proof}
  By induction on $\psi$, simultaneously for all $w$, $V$, $\tau$ and $g$; note
  that $\tau$ is $\hist$-compatible by Lemma~\ref{lem:realised-compatible}, so the
  right-hand side is defined.  Let $\tau=\langle\bar t,\equiv\rangle$.

  If $\psi=P_i(y)$, then, in view of~\eqref{eq:tp} and since $g$ realises
  $\tau$, so that $\tp_{\fr{M}}(g(y))=t_y$,
  $$
  \begin{array}{lclclcl}
    \fr{M},w\sat^{g} P_i(y) & \iff & g(y)\in P_i^{I,w} & \iff & \langle w,i\rangle\in t_y
    & \iff & \hist,w\sat^{\tau} P_i(y).
  \end{array}
  $$
  If $\psi= (y_1= y_2)$, then, again since $g$ realises $\tau$,
  $$
  \begin{array}{lclclcl}
    \fr{M},w\sat^{g}y_1=y_2 & \iff &
                                     g(y_1) = g(y_2) & \iff &
    y_1 \equiv y_2. 
  \end{array}
  $$
  
  The cases $\psi=\bot$ and $\psi=\psi_1\to\psi_2$ are immediate from IH.

  If $\psi=\square\psi_1$ and $v\in R(w)$, then $\tau$ is at $v$ by (T2), and so,
  by IH, $\fr{M},v\sat^{g}\psi_1$ iff $\hist,v\sat^{\tau}\psi_1$; the claim
  follows.

  Let $\psi=\ex{M}z\,\psi_1$, let $V_0:=V\setminus\{z\}$ and
  $\tau_0=\langle\bar t_0,\equiv_0\rangle:=\tau{\restriction}V_0$, and let $X$
  be as in the definition; note that $M\leqslant M_\vp$.  For $b\in D_w$, let
  $g_b$ be the assignment defined by $g_b\overset{z}{=}g$ and $g_b(z) = b$.  Then
  $g_b$ agrees with $g$ on $V_0$, and so it realises $\tau_0$; since $\tau_0$ is
  at $w$ and $b\in D_w$, the assignment $g_b$ realises exactly one type
  assignment, say $\tau_b$, on $V_0\cup\{z\}$, which is at $w$ and is an
  extension of $\tau_0$ to $z$ at $w$, and which is $\hist$-compatible by
  Lemma~\ref{lem:realised-compatible}.  By IH and definition of $X$,
  $$
  \begin{array}{lclcl}
    \fr{M},w\sat^{g_b}\psi_1 & \iff & \hist,w\sat^{\tau_b}\psi_1& \iff & \tau_b\in
  X.
  \end{array}
  \eqno{(\ast)}
  $$
  Now, let
  $$
  \begin{array}{lcl}
    S & := & \{b\in D_w : \fr{M},w\sat^{g_b}\psi_1\}
  \end{array}
  $$  
  and, for $\tau'\in X$, 
  $$
  \begin{array}{lcl}
    F_{\tau'} \;:=\; \{b\in D_w : \tau_b=\tau'\}.
  \end{array}
  $$
  Then, $\fr{M},w\sat^{g}\psi$ iff $|S|\geqslant M$.  The members of the family
  $(F_{\tau'} : \tau'\in X)$ are pairwise disjoint, as $\tau_b$ is determined by
  $b$, and their union is $S$, as, by ($\ast$), $b\in S$ iff $\tau_b\in X$; thus,
  $$
  \begin{array}{lcl}
  |S| & = & \displaystyle \sum_{\tau'\in X}|F_{\tau'}|.
  \end{array}  
  $$
  Thus, it is enough to show that, for each $\tau'\in X$, either
  $|F_{\tau'}|=\hist(\tau'/\tau_0)$, or both $|F_{\tau'}|$ and
  $\hist(\tau'/\tau_0)$ are at least~$M$.

  Let $A:=\{g(y):y\in V_0\}$.  Since $g$ realises $\tau_0$ and $\tau_0$ is at
  $w$, it follows that $A\subseteq D_w$ and the map sending each
  $C\in V_0/{\equiv_0}$ to the common value of $g$ on $C$ is a bijection from
  $V_0/{\equiv_0}$ onto $A$ that preserves types.  Hence, for every
  $t\in\Typ(\fr{F},s)$,
  \begin{equation}
    \label{eq:ct}
    c_t(\tau_0) \;=\; |\{a\in A:\tp_{\fr{M}}(a)=t\}| \;\leqslant\; |A|
    \;\leqslant\; |V_0| \;\leqslant\; v_\vp-1,
  \end{equation}
  where the last inequality holds because $z\in\varset\vp\setminus V_0$.

  If $\tau'=\tau_0[z\equiv C]$, then $\tau_b=\tau'$ iff $b$ is the value of $g$
  on $C$; thus, $|F_{\tau'}|=1=\hist(\tau'/\tau_0)$.  If $\tau'=\tau_0[z:t]$,
  then $\tau_b=\tau'$ iff $b\notin A$ and $\tp_{\fr{M}}(b)=t$.  As $w\in E_t$,
  all elements of type $t$ are in $D_w$, by~\eqref{eq:tp}; hence, by the
  equality in~\eqref{eq:ct}, $|F_{\tau'}|=\hist_{\fr{M}}(t)-c_t(\tau_0)$.  If
  $\hist(t)<\kappa_\vp$, then $\hist_{\fr{M}}(t)=\hist(t)$, and so
  $|F_{\tau'}|=\hist(\tau'/\tau_0)$.  If $\hist(t)=\kappa_\vp$, then
  $\hist_{\fr{M}}(t)\geqslant\kappa_\vp$, and so, by~\eqref{eq:ct},
  \begin{equation}
    \label{eq:capped}
    |F_{\tau'}| \;\geqslant\; \hist(\tau'/\tau_0) \;=\; \kappa_\vp-c_t(\tau_0)
    \;\geqslant\; M_\vp \;\geqslant\; M.
  \end{equation}

  Therefore,
  $|S|=\sum_{\tau'\in X}|F_{\tau'}|\geqslant\sum_{\tau'\in
    X}\hist(\tau'/\tau_0)$, with equality unless $\hist(t)=\kappa_\vp$ for some
  $\tau_0[z:t]\in X$, in which case both sums are at least $M$,
  by~\eqref{eq:capped}, since all multiplicities in question are positive.  In
  either case, $|S|\geqslant M$ iff
  $\sum_{\tau'\in X}\hist(\tau'/\tau_0)\geqslant M$.  Thus,
  $\fr{M},w\sat^{g}\psi$ iff $\hist,w\sat^{\tau}\psi$.
\end{proof}

The following consequence of Lemma~\ref{lem:eval} says that models with the
same capped histogram agree on assignments that realise the same type
assignment; we use it to prove the finite domain property.

\begin{corollary}
  \label{cor:capped-agree}
  Let $\fr{M}=\langle W,R,D,I\rangle$ and $\fr{M}'=\langle W,R,D',I'\rangle$ be
  models over $\fr{F}$ with
  $\hist^{\kappa_\vp}_{\fr{M}}=\hist^{\kappa_\vp}_{\fr{M}'}$.  Let
  $\psi\in\subf\vp$, let $w\in W$, let $V\subseteq\varset\vp$ with
  $\param{\psi}\subseteq V$, and let $g$ and $g'$ be assignments in,
  respectively, $\fr{M}$ and $\fr{M}'$ such that $g(y)\in D_w$ and
  $g'(y)\in D'_w$ whenever $y\in V$ and, for all $y,y'\in V$,
  \begin{enumerate}[label=\rmbr{\rm S\arabic*}]
  \item $\tp_{\fr{M}}(g(y))=\tp_{\fr{M}'}(g'(y))$;
  \item $g(y)=g(y')$ iff $g'(y)=g'(y')$.
  \end{enumerate}
  Then, $\fr{M},w\sat^{g}\psi$ iff $\fr{M}',w\sat^{g'}\psi$.  In particular,
  if $\psi$ is a sentence, then $\fr{M},w\sat\psi$ iff $\fr{M}',w\sat\psi$.
\end{corollary}

\begin{proof}
  By (S1) and (S2), $g$ and $g'$ realise the same type assignment $\tau$ on
  $V$, and $\tau$ is at $w$.  By Lemma~\ref{lem:eval}, applied to $\fr{M}$ and
  to $\fr{M}'$, both $\fr{M},w\sat^{g}\psi$ and $\fr{M}',w\sat^{g'}\psi$ are
  equivalent to $\hist^{\kappa_\vp}_{\fr{M}},w\sat^{\tau}\psi$.  If $\psi$ is a
  sentence, take $V=\varnothing$ and $\tau=\tau_\varnothing$.
\end{proof}

\begin{remark}[Optimality of the cap]
  \label{rem:cap}
  %The two summands of the cap $\kappa_\vp=M_\vp+v_\vp-1$ play different
  %roles in the proof of Lemma~\ref{lem:eval}: $M_\vp$ bounds the number of
  %witnesses a counting quantifier of $\vp$ asks for, and $v_\vp-1$ bounds the
  %number of elements of a type that the values of the parameters of a
  %quantified subformula can take up (see~\eqref{eq:ct}
  %and~\eqref{eq:capped}).
  The two summands of the cap $\kappa_\vp=M_\vp+v_\vp-1$ play different roles in
  the proof of Lemma~\ref{lem:eval}.  When a subformula $\ex{M}z\,\psi_1$ is
  evaluated, $M_\vp$ bounds the number $M$ of witnesses it asks for, and
  $v_\vp-1$ bounds how many elements of any one type are already values of the
  other variables; so, if a type reaches the cap, at least $M_\vp$ of its
  elements differ from these values, which is enough (see~\eqref{eq:ct}
  and~\eqref{eq:capped}).  The cap cannot be lowered.  Let $M,v\geqslant1$ with
  $M+v\geqslant3$, and let
  $$
  \begin{array}{lcl}
    \theta & := & \displaystyle \exists x_1\cdots\exists x_{v-1}
                  \Bigl(\bigwedge_{1\leqslant i<j<v}\neg(x_i=x_j)\wedge
                  \ex{M}x_v\bigwedge_{1\leqslant i<v}\neg(x_v=x_i)\Bigr).
  \end{array}
  $$
  Then $M_{\theta}=M$ and $v_{\theta}=v$, and $\theta$ is true at a world $w$
  of a model iff $|D_w|\geqslant M+v-1$.  Hence, $\theta$ is false in the
  $\mathit{cd}$-model over $\fr{F}$ in which every world has the domain
  $\{1,\ldots,M+v-2\}$ and every predicate letter is interpreted as
  $\varnothing$, and true in the analogous model with the domain
  $\{1,\ldots,M+v-1\}$, although the $(M+v-2)$-capped histograms of the two
  models coincide.
\end{remark}

\subsection{Refutations on capped histograms}
\label{sec:refute}

\begin{theorem}[Finite domain property for the monadic fragment with counting]
  \label{thm:small}
  Let $d\in\{\mathit{xd},\mathit{cd}\}$, and let $\vp\in\CMLmon$ be a formula
  whose predicate letters other than $=$ are among $P_1,\ldots,P_s$, with
  $\kappa_\vp:=M_\vp+v_\vp-1$ even if $v_\vp=0$ \rmbr{thus, the maximum below
  matters only if $\vp$ has no variables}.  If $\vp$ is refuted in a $d$-model
  over $\fr{F}$, then it is refuted in a $d$-model
  $\fr{M}'=\langle W,R,D',I'\rangle$ with
  \[
    |D'^{+}| \;\leqslant\; \max\{1,\kappa_\vp\}\cdot|\Typ^{d}(\fr{F},s)| \;<\;
    \max\{1,\kappa_\vp\}\cdot2^{(s+1)|W|}
  \]
  \rmbr{in particular, if $d=\mathit{cd}$ and $\fr{F}$ is connected, then
  $|D'^{+}|\leqslant\max\{1,\kappa_\vp\}\cdot2^{s|W|}$}.  Thus, $\CMLmon$ has
  the finite domain property over $\fr{F}$ for $d$-models.
\end{theorem}

\begin{proof}
  %A formula is refuted in a model iff its universal closure is refuted at some
  %world of that model, and the universal closure of $\vp$ has the same
  %variables and the same counting threshold as $\vp$, hence the same cap
  %$\kappa_\vp$; thus,
  We may assume that $\vp$ is a sentence.  If $v_\vp=0$, then $\vp$ contains
  neither quantifiers nor atoms with variables, so its truth at a world of a
  model over $\fr{F}$ does not depend on the domains; hence, if $\vp$ is
  refuted at a world of a $d$-model over $\fr{F}$, it is refuted at that world
  of the $\mathit{cd}$-model over $\fr{F}$ in which every world has the
  one-element domain $\{a\}$, every monadic letter is interpreted as
  $\varnothing$ and every proposition letter, if any, as in the given model;
  this model is a $d$-model, and its global domain has
  $1=\max\{1,\kappa_\vp\}$ elements, as required.

  So assume that $v_\vp\geqslant1$; then, $\max\{1,\kappa_\vp\}=\kappa_\vp$.  Let
  $\fr{M}$ be a $d$-model over $\fr{F}$ and $w_0$ be a world from $W$ with
  $\fr{M},w_0\nsat\vp$, and let $\hist:=\hist^{\kappa_\vp}_{\fr{M}}$.  Then,
  $\hist$ is $d$-admissible.  Hence, by Lemma~\ref{lem:canonical},
  $\fr{M}':=\fr{M}_{\hist}$ is a $d$-model over $\fr{F}$ with
  $\hist_{\fr{M}'}=\hist$.  As
  $\hist(t)\leqslant\kappa_\vp$ for every $t$, it follows that
  $\hist^{\kappa_\vp}_{\fr{M}'}=\hist=\hist^{\kappa_\vp}_{\fr{M}}$, and so, by
  Corollary~\ref{cor:capped-agree}, $\fr{M}',w_0\nsat\vp$.  Lastly,
  $$
  \begin{array}{lcl}
    |D'^{+}| & = & \displaystyle \sum_t\hist(t)\leqslant\kappa_\vp\cdot|\{t:\hist(t)\geqslant1\}|,
  \end{array}  
  $$
  and every $t$ with $\hist(t)\geqslant1$ belongs to $\Typ^{d}(\fr{F},s)$.  It
  remains to apply~\eqref{eq:numtypes} and~\eqref{eq:numtypes-cd}.
\end{proof}

Without equality and counting quantifiers, one element of each type
suffices, since identifying the elements of the same type then preserves the
truth of formulas.

\begin{proposition}[One element per type]
  \label{prop:fdp-ML}
  Let $d\in\{\mathit{xd},\mathit{cd}\}$, and let $\vp\in\MLang^{\mathit{mon}}$
  be a formula whose predicate letters are among $P_1,\ldots,P_s$.  If $\vp$ is
  refuted in a $d$-model $\fr{M}$ over $\fr{F}$, then it is refuted in a
  $d$-model $\fr{M}'$ over $\fr{F}$ whose global domain has exactly one
  element of type $t$ for each $t$ with $\hist_{\fr{M}}(t)\geqslant1$, and no
  other elements; hence, $\fr{M}'$ has at most $|\Typ^{d}(\fr{F},s)|$ elements
  \rmbr{in particular, at most $2^{s|W|}$ elements if $d=\mathit{cd}$ and
  $\fr{F}$ is connected}.
\end{proposition}

\begin{proof}
  We may assume that $\vp$ is a sentence.  Let $\fr{M}=\langle W,R,D,I\rangle$
  and $w_0\in W$ be such that $\fr{M},w_0\nsat\vp$, let
  $\hist:=\hist^{1}_{\fr{M}}$, which is $d$-admissible, and let
  $\fr{M}':=\fr{M}_{\hist}$.  By Lemma~\ref{lem:canonical}, $\fr{M}'$ is a
  $d$-model over $\fr{F}$ whose global domain consists of the
  elements $\langle t,1\rangle$ with $\hist_{\fr{M}}(t)\geqslant1$, the element
  $\langle t,1\rangle$ being of type $t$.  Let $\pi$ be a map sending each
  $a\in\Dplus$ to $\langle\tp_{\fr{M}}(a),1\rangle$.  Then $\pi$ maps $\Dplus$
  onto $(D^{\hist})^{+}$, and, by~\eqref{eq:tp} and the definition of
  $\fr{M}_{\hist}$, for all $a\in\Dplus$, $w\in W$ and $i\in\{1,\ldots,s\}$,
  $$
  a\in D_w \iff \pi(a)\in D^{\hist}_w
  \qquad\mbox{and}\qquad
  a\in P_i^{I,w} \iff \pi(a)\in I^{\hist}(w,P_i);
  $$
  in particular, $\pi$ maps $D_w$ onto $D^{\hist}_w$.  A straightforward
  induction on $\psi$ then shows that, for every $\psi\in\subf\vp$, every
  $w\in W$ and every assignment $g$ in $\fr{M}$,
  $$
  \fr{M},w\sat^{g}\psi \iff \fr{M}',w\sat^{\pi\circ g}\psi;
  $$
  the case $\psi=\exists z\,\psi_1$ uses that $\pi$ maps $D_w$ onto
  $D^{\hist}_w$.  Hence, $\fr{M}',w_0\nsat\vp$.  The bounds on the number of
  elements of $\fr{M}'$ follow as in the proof of Theorem~\ref{thm:small}.
\end{proof}

Lemma~\ref{lem:eval} also means that the refutability of a sentence can be
checked on capped histograms alone.

\begin{proposition}[Refutations on capped histograms]
  \label{prop:capped-hist}
  Let $d\in\{\mathit{xd},\mathit{cd}\}$, let $\vp\in\CMLmon$ be a sentence
  whose predicate letters other than $=$ are among $P_1,\ldots,P_s$, and let
  $w_0\in W$.  Then the following are equivalent:
  \begin{itemize}
  \item[\rmbr{$1$}] $\vp$ is refuted at $w_0$ in a $d$-model over $\fr{F}$;
  \item[\rmbr{$2$}] $\hist,w_0\nsat^{\tau_\varnothing}\vp$ for some
    $d$-admissible $\hist\colon\Typ(\fr{F},s)\to\{0,1,\ldots,\kappa_\vp\}$;
  \item[\rmbr{$3$}] $\vp$ is refuted at $w_0$ in the canonical model
    $\fr{M}_{\hist}$ of some $d$-admissible
    $\hist\colon\Typ(\fr{F},s)\to\{0,1,\ldots,\kappa_\vp\}$.
  \end{itemize}
\end{proposition}

\begin{proof}
  (1)$\Rightarrow$(2): If $\fr{M},w_0\nsat\vp$ for a $d$-model $\fr{M}$ over
  $\fr{F}$, then $\hist:=\hist^{\kappa_\vp}_{\fr{M}}$ is $d$-admissible;
  moreover,
  $\hist,w_0\nsat^{\tau_\varnothing}\vp$ by Lemma~\ref{lem:eval}, since every
  assignment realises $\tau_\varnothing$.

  (2)$\Rightarrow$(3): By Lemma~\ref{lem:canonical}, $\fr{M}_{\hist}$ is a
  $d$-model over $\fr{F}$ with $\hist_{\fr{M}_{\hist}}=\hist$; since
  $\hist(t)\leqslant\kappa_\vp$ for every $t$, also
  $\hist^{\kappa_\vp}_{\fr{M}_{\hist}}=\hist$, and so
  $\fr{M}_{\hist},w_0\nsat\vp$ by Lemma~\ref{lem:eval}.

  (3)$\Rightarrow$(1): This is trivial, as $\fr{M}_{\hist}$ is a $d$-model
  over $\fr{F}$ by Lemma~\ref{lem:canonical}.
\end{proof}

The relation $\hist,w_0\sat^{\tau_\varnothing}\vp$ can be decided by
implementing its definition.  The following lemma estimates the cost; we use
it in Section~\ref{sec:mon-complexity}.

\begin{lemma}
  \label{lem:cost}
  Let $\vp\in\CMLmon$ be a sentence whose predicate letters other than $=$ are
  among $P_1,\ldots,P_s$, let
  $\hist\colon\Typ(\fr{F},s)\to\{0,1,\ldots,\kappa_\vp\}$ be
  $\mathit{xd}$-admissible, and let $w_0\in W$.  Given $\fr{F}$, $\vp$,
  $\hist$ and $w_0$, whether
  $\hist,w_0\sat^{\tau_\varnothing}\vp$ can be decided
  \begin{itemize}
  \item[\rmbr{$1$}] in time
    $2^{O(v_\vp\cdot((s+1)|W|+\log v_\vp))}\cdot|\vp|^{O(1)}$, and
  \item[\rmbr{$2$}] in space
    $O\bigl(|\vp|\cdot(v_\vp\cdot((s+1)|W|+\log v_\vp)+|\vp|)\bigr)$, in
    addition to the space taken by the input.
  \end{itemize}
\end{lemma}

\begin{proof}
  Let $N:=|\Typ(\fr{F},s)|$; then, $N<2^{(s+1)|W|}$, by~\eqref{eq:numtypes}.  We
  first show that, for every $\psi\in\subf\vp$ and every $\hist$-compatible type
  assignment $\tau$ at $w$ on a set $V\subseteq\varset\vp$ with
  $\param{\psi}\subseteq V$,
  \begin{equation}
    \label{eq:restrict}
    \hist,w\sat^{\tau}\psi \iff \hist,w\sat^{\tau{\restriction}\param{\psi}}\psi.
  \end{equation}
  By Lemma~\ref{lem:canonical}, $\fr{M}_{\hist}$ is a model over $\fr{F}$
  with $\hist_{\fr{M}_{\hist}}=\hist$, and so
  $\hist^{\kappa_\vp}_{\fr{M}_{\hist}}=\hist$.  As $\tau$ is
  $\hist$-compatible, it is realised by an assignment $g$ in
  $\fr{M}_{\hist}$, which then also realises $\tau{\restriction}\param{\psi}$;
  hence, by Lemma~\ref{lem:eval}, applied twice, both sides
  of~\eqref{eq:restrict} are equivalent to $\fr{M}_{\hist},w\sat^{g}\psi$.

  (1): By~\eqref{eq:restrict}, it suffices to compute, for all $\psi\in\subf\vp$,
  all $w\in W$, and all $\hist$-compatible type assignments $\tau$ at $w$ on
  $\param{\psi}$, whether $\hist,w\sat^{\tau}\psi$.  We do so in order of
  increasing size of $\psi$, using the clause for the main symbol of $\psi$ with
  $V:=\param{\psi}$, and finding the values for the immediate subformulas
  $\psi_1$ of $\psi$, restricted to $\param{\psi_1}$ by~\eqref{eq:restrict},
  among those already computed.  There are at most $|\vp|$ subformulas, $|W|$
  worlds, and $(N\cdot v_\vp)^{v_\vp}$ type assignments on a set of at most
  $v_\vp$ variables.  Each value is computed from at most
  $\max\{2,|W|,v_\vp-1+N\}$ values in time $(N\cdot|W|\cdot|\vp|)^{O(1)}$, which
  suffices to compute the multiplicities and to add up at most $v_\vp-1+N$
  numbers of at most $|\vp|+1$ bits.  As $N<2^{(s+1)|W|}$ and $v_\vp\geqslant1$,
  this gives the time bound.

  (2): A depth-first implementation of the recursion has depth at most
  $|\vp|$.  For each level, it stores a subformula of $\vp$; a world; a type
  assignment on a subset of $\varset\vp$, given, for each of at most $v_\vp$
  variables, by a type and a representative of its class, which takes
  $O(v_\vp\cdot((s+1)|W|+\log v_\vp))$ bits; the position reached in the
  enumeration of the at most $|W|$ successors of the world or of the at most
  $v_\vp-1+N$ extensions of $\tau_0$; and a running sum of $O(|\vp|)$
  bits, which can be stopped as soon as it reaches the threshold $M$.  This
  gives the space bound.
\end{proof}

\subsection{Complexity}
\label{sec:mon-complexity}

\begin{theorem}[Monadic fragment with counting]
  \label{thm:nexptime}
  Let $\fr{F}$ be a finite Kripke frame.  Then the sets
  $(\mathbf{ML}^d_{\mathrm{C}}\fr{F})^{\mathit{mon}}$, for
  $d\in\{\mathit{xd},\mathit{cd}\}$, are $\coNEXP$-complete.  The lower bound
  holds already for $\square$-free sentences with two variables and without
  equality and counting quantifiers.
\end{theorem}

\begin{proof}
  For the upper bound, we describe a nondeterministic algorithm that, given
  \mbox{$\vp\in\CMLmon$}, has an accepting run iff $\vp$ is refuted in a
  $d$-model over $\fr{F}$.  We may assume that $\vp$ is a sentence, and that the
  predicate letters of $\vp$ other than $=$ are $P_1,\ldots,P_s$, for some
  $s\leqslant|\vp|$; we apply Sections~\ref{sec:types-hist}--\ref{sec:refute}
  with this $s$.

  The algorithm lists the set $\Typ(\fr{F},s)$, checking (T1)--(T3) for each
  subset of $W\times\{0,1,\ldots,s\}$; guesses a function
  $\hist\colon\Typ(\fr{F},s)\to\{0,1,\ldots,\kappa_\vp\}$, which consists of
  fewer than $2^{(s+1)|W|}$ numbers of at most $|\vp|+1$ bits each; checks that
  $\hist$ is $d$-admissible; and accepts if
  $\hist,w_0\nsat^{\tau_\varnothing}\vp$ for some $w_0\in W$.  By
  Proposition~\ref{prop:capped-hist}, the algorithm has an accepting run iff
  $\vp$ is refuted in a $d$-model over $\fr{F}$.

  The first three steps take time $2^{O((s+1)|W|)}\cdot|\vp|^{O(1)}$, and, by
  Lemma~\ref{lem:cost}\rmbr{$1$}, the last one takes time
  \[
    2^{O(v_\vp\cdot((s+1)|W|+\log v_\vp))}\cdot|\vp|^{O(1)}.
  \]
  As
  $s \leqslant |\vp|$ and $v_\vp\leqslant|\vp|$, the algorithm runs in
  nondeterministic time $2^{O(|W|\cdot|\vp|^2)}$, i.e., $2^{O(|\vp|^2)}$ for the
  fixed frame $\fr{F}$; hence, $(\mathbf{ML}^d_{\mathrm{C}}\fr{F})^{\mathit{mon}}$
  is in $\coNEXP$.

  The lower bound, already for the sentences described in the statement,
  follows from Proposition~\ref{prop:lewis} and Remark~\ref{rem:conserv}.
\end{proof}

\begin{corollary}[Monadic fragments without counting]
  \label{cor:complexity-monadic}
  Let $\fr{F}$ be a finite Kripke frame.  The sets
  $(\mathbf{ML}^d\fr{F})^{\mathit{mon}}$ and
  $(\mathbf{ML}^d_=\fr{F})^{\mathit{mon}}$, for
  $d\in\{\mathit{xd},\mathit{cd}\}$, are $\coNEXP$-complete.  The lower
  bounds hold already for $\square$-free sentences with two variables and
  without equality.
\end{corollary}

\begin{proof}
  The upper bounds follow from Theorem~\ref{thm:nexptime}; the lower bounds,
  already for the sentences described in the statement, from
  Proposition~\ref{prop:lewis} and Remark~\ref{rem:conserv}.
\end{proof}

Since we are not aware of the following result stated anywhere in the
literature, we include it here; \rmbr{$1$} and~\rmbr{$2$} are consequences of
Theorems~\ref{thm:small} and~\ref{thm:nexptime}, and \rmbr{$3$} shows that,
for sentences with at least one variable, the bound in~\rmbr{$2$} is
parameterwise optimal, that is, it cannot be lowered for any value of $s$,
$M_\vp$ and $v_\vp$; for such sentences without counting quantifiers, where
$\kappa_\vp=v_\vp$, it is the bound of~\cite[Lemma~3.20]{PH23}.

\begin{proposition}[Classical monadic fragment with counting]
  \label{prop:classical-counting}
  \mbox{}
  \begin{itemize}
  \item[\rmbr{$1$}] The set $\CQCL^{\mathit{mon}}$ is $\coNEXP$-complete;
    the lower bound holds already for sentences with two variables and
    without equality and counting quantifiers.
  \item[\rmbr{$2$}] Every satisfiable sentence
    $\vp\in\mathcal{L}_{\mathrm{C}}^{\mathit{mon}}$ whose predicate letters other than $=$
    are among $P_1,\ldots,P_s$ has a model with at most
    $\max\{1,\kappa_\vp\}\cdot2^{s}$ elements, with $\kappa_\vp:=M_\vp+v_\vp-1$
    even if $v_\vp=0$.
  \item[\rmbr{$3$}] The bound in~\rmbr{$2$} is parameterwise optimal, that
    is, it is attained for every value of the parameters $s$, $M$ and
    $v\geqslant1$:
    for all $s\in\N$ and $M,v\geqslant1$, there is a satisfiable sentence
    $\vp$ as in~\rmbr{$2$} with $M_\vp=M$ and $v_\vp=v$, every model of which
    has at least $\kappa_\vp\cdot2^{s}$ elements.
  \end{itemize}
\end{proposition}

\begin{proof}
  See Appendix~\ref{sec:prop:classical-counting}.
\end{proof}

We next consider the case where the number of predicate letters is fixed in
advance.  In the classical setting, \cite[Lemma~3.20 and Theorem~3.9]{PH23} give
a $\PSPACE$ upper bound for monadic sentences without counting quantifiers over
a fixed finite set of predicate letters: such a sentence, if satisfiable, has a
model of polynomial size, and model checking is in $\PSPACE$.  With counting
quantifiers, models may have to be exponentially large, as thresholds are
written in binary.  The following proposition covers counting quantifiers and
logics of finite frames, and gives matching lower bounds.

\begin{proposition}[Fixed number of predicate letters]
  \label{prop:pspace}
  Let $\fr{F}$ be a finite Kripke frame, $s\in\N$, and
  $d\in\{\mathit{xd},\mathit{cd}\}$.  The sets
  $(\mathbf{ML}^d_{\mathrm{C}}\fr{F})^{\mathit{mon}}$, restricted to formulas
  whose predicate letters other than $=$ are among $P_1,\ldots,P_s$, are
  $\PSPACE$-complete.  If $s\geqslant1$, the lower bound holds already for
  $\square$-free sentences whose only predicate letter is $P_1$, without equality
  and counting quantifiers; if $s=0$, it holds for $\square$-free sentences
  without counting quantifiers.  Consequently, the corresponding restrictions
  of $(\mathbf{ML}^d\fr{F})^{\mathit{mon}}$, for every $s\geqslant1$, and of
  $(\mathbf{ML}^d_=\fr{F})^{\mathit{mon}}$, for every $s \in \N$, are
  $\PSPACE$-complete.
\end{proposition}

\begin{proof}
  For the upper bound, let $\vp$ be such a formula; we may assume that $\vp$ is
  a sentence.  The following deterministic algorithm decides whether $\vp$ is
  refuted in a $d$-model over $\fr{F}$.  It enumerates, one at a time, the pairs
  $\langle w_0,\hist\rangle$ with $w_0\in W$ and
  $\hist\colon\Typ(\fr{F},s)\to\{0,1,\ldots,\kappa_\vp\}$, and reports that
  $\vp$ is refuted iff, for some pair, $\hist$ is $d$-admissible and
  $\hist,w_0\nsat^{\tau_\varnothing}\vp$; the algorithm is correct by
  Proposition~\ref{prop:capped-hist}.  Since $\fr{F}$ and $s$ are fixed,
  $|\Typ(\fr{F},s)|$ is a constant; hence, each pair can be written with
  $O(|\vp|)$ bits, and, by Lemma~\ref{lem:cost}\rmbr{$2$}, it can be checked in
  space $O(|\vp|\cdot(v_\vp\log v_\vp+|\vp|))\subseteq O(|\vp|^2\log|\vp|)$.
  
  For the lower bound, in view of Remark~\ref{rem:conserv}, it suffices to
  reduce the truth problem for closed quantified Boolean formulas, which is
  $\PSPACE$-complete~\cite{SM73}, to the validity of classical sentences of the
  two kinds described.  Let
  $\beta=\mathscr{Q}_1p_1\cdots \mathscr{Q}_np_n\,\chi$ be a closed quantified
  Boolean formula, with
  $\mathscr{Q}_1,\ldots,\mathscr{Q}_n\in\{\forall,\exists\}$ and $\chi$
  quantifier-free.  For $s\geqslant1$, let $\beta^{\ast}$ be obtained from
  $\mathscr{Q}_1x_1\cdots \mathscr{Q}_nx_n\,\chi$ by replacing each $p_i$ with
  $P_1(x_i)$, and put
  $$
  \begin{array}{lcl}
    \beta_1 & := & \bigl(\exists x_1\,P_1(x_1)\wedge\exists x_1\,\neg
                   P_1(x_1)\bigr)\to\beta^{\ast}.
  \end{array}
  $$
  For $s=0$, let $\beta^{=}$ be obtained from $\mathscr{Q}_1x_1\cdots \mathscr{Q}_nx_n\,\chi$ by
  replacing each $p_i$ with $x_i=x_0$, and put
  $$
  \begin{array}{lcl}
    \beta_2 & := & \exists x_1\exists x_2\,\neg(x_1=x_2)\to\forall
                   x_0\,\beta^{=}.
  \end{array}
  $$
  In a classical model in which $P_1$ is neither empty nor the whole universe,
  each atom $P_1(x_i)$ takes both truth values as $x_i$ ranges over the
  universe; so does each atom $x_i=x_0$, for every value of $x_0$, in a model
  with at least two elements.  Hence, a backward induction on $i$, as in the
  proof of~\cite[Theorem~3.9]{PH23}, shows that $\beta^{\ast}$ and
  $\forall x_0\,\beta^{=}$ are true in such models iff $\beta$ is true.  As
  the antecedents of $\beta_1$ and $\beta_2$ say exactly that the model is of
  the respective kind, $\beta_1$ and $\beta_2$ are valid iff $\beta$ is true;
  clearly, they are computable from $\beta$ in polynomial time.

  The last claim follows from the bounds established above.
  %: since
  %$\mathbf{ML}^d\fr{F}=\mathbf{ML}^d_{\mathrm{C}}\fr{F}\cap\MLang$ and
  %$\mathbf{ML}^d_=\fr{F}=\mathbf{ML}^d_{\mathrm{C}}\fr{F}\cap\MLangeq$, the
  %upper bound applies to the restrictions of
  %$(\mathbf{ML}^d\fr{F})^{\mathit{mon}}$ and
  %$(\mathbf{ML}^d_=\fr{F})^{\mathit{mon}}$; since
  %$\beta_1\in\MLang\subseteq\MLangeq$ and $\beta_2\in\MLangeq$, the lower bound
  %applies to the restriction of $(\mathbf{ML}^d\fr{F})^{\mathit{mon}}$ for
  %$s\geqslant1$ and to that of $(\mathbf{ML}^d_=\fr{F})^{\mathit{mon}}$ for
  %every $s$.
\end{proof}

The reductions in the proof of Proposition~\ref{prop:pspace} use unboundedly
many variables.  If the number of variables is bounded as well, the complexity
drops to polynomial time.  The reason is that, by the clauses of
Section~\ref{sec:eval}, whether a sentence is true on a capped histogram
$\hist$ depends only on the outcomes of comparisons between sums
$\sum_{t\in S}\hist(t)$, for sets $S$ of types, and thresholds $M+j$, where
$\ex{M}$ occurs in the sentence and $|j|$ is less than its number of variables.
When the frame and the numbers of predicate letters and variables are fixed,
there are polynomially many such comparisons, and we show, using a sensitivity
theorem for integer linear programming, that polynomially many histograms,
computable in polynomial time, realise every combination of outcomes that a
histogram can realise; it then suffices to run the algorithm of
Proposition~\ref{prop:pspace} on these histograms only.

\begin{proposition}[Fixed numbers of predicate letters and variables]
  \label{prop:bounded-var}
  Let $\fr{F}$ be a finite Kripke frame, $s,v\in\N$, and
  $d\in\{\mathit{xd},\mathit{cd}\}$.  The set
  $(\mathbf{ML}^d_{\mathrm{C}}\fr{F})^{\mathit{mon}}$, restricted to formulas
  with at most $v$ variables whose predicate letters other than $=$ are among
  $P_1,\ldots,P_s$, is decidable in polynomial time.
\end{proposition}

\begin{proof}
  Let $\vp$ be such a formula; since the universal closure of $\vp$ has the same
  variables as $\vp$, we may assume that $\vp$ is a sentence.  Since $\fr{F}$
  and $s$ are fixed, so is $N:=|\Typ(\fr{F},s)|$.  Let $\cl{M}$ be the set of
  numbers $M$ such that $\ex{M}$ occurs in $\vp$.  We call a pair
  $\langle S,c\rangle$, where $S$ is a non-empty subset of $\Typ(\fr{F},s)$ and
  $c$ is an integer, a \emph{constraint} if either $S$ is a singleton and
  $c\in\{0,1,\ldots,v\}\cup\{\kappa_\vp\}$, or $c=M+j$ for some $M\in\cl{M}$ and
  some integer $j$ with $|j|<v$.  As $N$ and $v$ are constants, the set $L$ of
  constraints has $O(|\vp|)$ elements.  If
  $\hist\colon\Typ(\fr{F},s)\to\{0,1,\ldots,\kappa_\vp\}$ and
  $\langle S,c\rangle\in L$, let $\sigma_{\hist}(S,c)\in\{-1,0,1\}$ be the sign
  of $\sum_{t\in S}\hist(t)-c$.

  \smallskip\noindent
  \textit{Claim 1.}  Let $\hist,\hist'\colon\Typ(\fr{F},s)\to\{0,1,\ldots,\kappa_\vp\}$
  be such that $\sigma_{\hist}=\sigma_{\hist'}$.  Then \rmbr{a} $\hist$ is
  $d$-admissible iff $\hist'$ is;
  \rmbr{b} a type assignment is $\hist$-compatible iff it is
  $\hist'$-compatible; and \rmbr{c} $\hist,w\sat^{\tau}\psi$ iff
  $\hist',w\sat^{\tau}\psi$, for all $\psi\in\subf\vp$, all $w\in W$, and all
  $\hist$-compatible type assignments $\tau$ at $w$ on sets
  $V\subseteq\varset\vp$ with $\param{\psi}\subseteq V$.

  \smallskip\noindent
  \textit{Proof of Claim 1.}  Whether $\hist$ is $d$-admissible depends only
  on which types $t$ satisfy $\hist(t)\geqslant1$, that is, on the values
  $\sigma_{\hist}(\{t\},0)$; this proves \rmbr{a}.  As
  $c_t(\tau)\leqslant|V|\leqslant v$, a type assignment $\tau$ is
  $\hist$-compatible iff $\sigma_{\hist}(\{t\},c_t(\tau))\geqslant0$ for every
  $t$; this proves \rmbr{b}.  Statement \rmbr{c} is proved by induction on
  $\psi$; only the case $\psi=\ex{M}z\,\psi_1$ needs comment.  By \rmbr{b} and
  IH, the set $X$ in the clause for $\ex{M}$ is the same for $\hist$ and
  $\hist'$.  Let $n$ be the number of members of $X$ of the form
  $\tau_0[z\equiv C]$, and let $U$ be the set of types $t$ with
  $\tau_0[z:t]\in X$, so that the sum in the clause is
  $n+\sum_{t\in U}(\hist(t)-c_t(\tau_0))$.  If $U=\varnothing$, the clause
  says that $n\geqslant M$; otherwise, it says that
  $\sigma_{\hist}(U,M')\geqslant0$, where
  $M':=M-n+\sum_{t\in U}c_t(\tau_0)$.  Both $n$ and $\sum_{t\in U}c_t(\tau_0)$
  lie between $0$ and the number of classes of $\tau_0$, which is at most
  $|V\setminus\{z\}|\leqslant v-1$; hence, $|M'-M|<v$, and so
  $\langle U,M'\rangle\in L$.  This proves Claim 1.

  \smallskip

  \smallskip Since $\kappa_\vp$ may be exponential in $|\vp|$, we cannot try all
  functions $\hist$; by Claim~1, it suffices to try one for each value
  of~$\sigma_{\hist}$.  To find such representatives, we regard functions
  $\Typ(\fr{F},s)\to\mathbb{R}$ as vectors
  $\mathbf{x}=(x_t)_{t\in\Typ(\fr{F},s)}$ and use two facts about a finite
  system $\Sigma$ of linear inequalities $\mathbf{a}\cdot\mathbf{x}\leqslant b$
  in these $N$ unknowns, with $\mathbf{a}\in\{-1,0,1\}^{\Typ(\fr{F},s)}$ and
  $b\in\mathbb{Z}$ (we write $\mathbf{a}\cdot\mathbf{x}\geqslant b$ as
  $(-\mathbf{a})\cdot\mathbf{x}\leqslant-b$, and an equation as two
  inequalities).
  \begin{itemize}
  \item[\rmbr{F1}] If the real solutions of $\Sigma$ form a non-empty bounded
    set, then one of them is the unique solution of some $N$ of the
    inequalities of $\Sigma$ taken as equations (it is a vertex of that
    set).  Indeed, choose a real solution $\mathbf{y}$ for which the vectors
    $\mathbf{a}$ of the inequalities holding with equality at $\mathbf{y}$
    span a space of maximal dimension.  If this dimension were less than
    $N$, then moving from $\mathbf{y}$ in the direction of a non-zero vector
    $\mathbf{d}$ orthogonal to all these $\mathbf{a}$ would preserve these
    equalities and, as the set is bounded, would reach a solution at which a
    further inequality, whose $\mathbf{a}$ is not orthogonal to $\mathbf{d}$,
    holds with equality, contradicting the choice of $\mathbf{y}$; thus, $N$
    of these $\mathbf{a}$ are linearly independent.
  \item[\rmbr{F2}] If $\Sigma$ has an integer solution, then, for every real
    solution $\mathbf{y}$, it has an integer solution $\mathbf{z}$ with
    $|z_t-y_t|\leqslant N\cdot N!$ for all $t$.  This is the case of the
    objective function $0$ of~\cite[Theorem~1]{CGST86}, which gives the distance
    $N\Delta$ for any bound $\Delta$ on the absolute values of the determinants
    of the square submatrices of the matrix whose rows are the vectors
    $\mathbf{a}$ of $\Sigma$; we can take $\Delta:=N!$: indeed, this matrix has
    $N$ columns, one for each unknown, so its square submatrices have order
    $k\leqslant N$; and the determinant of a $k\times k$ matrix is a sum of $k!$
    terms, one for each permutation of $\{1,\ldots,k\}$, each of which is $\pm1$
    times a product of $k$ entries of the matrix (cf.~\cite[Section~7.1]{PH23}).
    As the entries are in $\{-1,0,1\}$, so is each term, and hence the absolute
    value of such a determinant is at most $k!\leqslant N!$.
  \end{itemize}
  Let $Z$ be the set of vectors
  $\mathbf{z}\in\{0,1,\ldots,\kappa_\vp\}^{\Typ(\fr{F},s)}$ such that
  $|z_t-y_t|\leqslant N\cdot N!$ for all $t$, for some $\mathbf{y}$ that is
  the unique real solution of a system of $N$ equations of the form
  $\sum_{t\in S}x_t=c+j$, with $\langle S,c\rangle\in L$ and
  $j\in\{-1,0,1\}$.

  \smallskip\noindent
  \textit{Claim 2.}  For every
  $\hist\colon\Typ(\fr{F},s)\to\{0,1,\ldots,\kappa_\vp\}$, there exists
  $\mathbf{z}\in Z$ such that $\sigma_{\mathbf{z}}=\sigma_{\hist}$.

  \smallskip\noindent
  \textit{Proof of Claim 2.}  Let $\Sigma_{\hist}$ consist, for every
  $\langle S,c\rangle\in L$, of $\sum_{t\in S}x_t\leqslant c-1$, of
  $\sum_{t\in S}x_t=c$, or of $\sum_{t\in S}x_t\geqslant c+1$, according as
  $\sigma_{\hist}(S,c)$ is $-1$, $0$ or $1$.  As $L$ contains
  $\langle\{t\},0\rangle$ and $\langle\{t\},\kappa_\vp\rangle$ for every $t$
  (and $0\leqslant\hist(t)\leqslant\kappa_\vp$), the system $\Sigma_{\hist}$
  forces $0\leqslant x_t\leqslant\kappa_\vp$; so its real solutions form a
  bounded set, and its integer solutions are the vectors
  $\mathbf{z}\in\{0,1,\ldots,\kappa_\vp\}^{\Typ(\fr{F},s)}$ with
  $\sigma_{\mathbf{z}}=\sigma_{\hist}$, among them $\hist$.  By~\rmbr{F1},
  some real solution $\mathbf{y}$ of $\Sigma_{\hist}$ is the unique solution
  of $N$ equations of the form $\sum_{t\in S}x_t=c+j$ with
  $\langle S,c\rangle\in L$ and $j\in\{-1,0,1\}$; by~\rmbr{F2},
  $\Sigma_{\hist}$ has an integer solution $\mathbf{z}$ with
  $|z_t-y_t|\leqslant N\cdot N!$ for all $t$.  Then $\mathbf{z}\in Z$ and
  $\sigma_{\mathbf{z}}=\sigma_{\hist}$.  This proves Claim 2.

  \smallskip Lastly, $Z$ can be computed in time polynomial in $|\vp|$: there
  are at most $(3|L|)^N$ systems of $N$ equations to consider, Gaussian
  elimination decides in polynomial time whether such a system has a unique
  solution and computes it, and each solution has at most $(2N\cdot N!+1)^N$
  integer vectors within the required distance.  The following algorithm
  therefore runs in polynomial time: for each $\mathbf{z}\in Z$ and each
  $w_0\in W$, check whether $\mathbf{z}$ is $d$-admissible and whether
  $\mathbf{z},w_0\nsat^{\tau_\varnothing}\vp$, which, as $\fr{F}$, $s$ and $v$
  are fixed, takes polynomial time by Lemma~\ref{lem:cost}\rmbr{$1$}; report
  that $\vp$ is refuted iff some pair passes the check.  By
  Proposition~\ref{prop:capped-hist} and Claims~1 and~2, some pair passes the
  check iff $\vp$ is refuted in a $d$-model over $\fr{F}$.
\end{proof}

\begin{remark}
  \label{rem:mon-frame-input}
  As for the results of Section~\ref{sec:complexity}
  (Remark~\ref{rem:frame-input}), Theorem~\ref{thm:nexptime} and
  Corollary~\ref{cor:complexity-monadic} hold in the same form for the problems
  in which $\fr{F}$ is part of the input: the algorithm in the proof of
  Theorem~\ref{thm:nexptime} runs in nondeterministic time
  $2^{O(|W|\cdot|\vp|^2)}$, which is exponential in $|\fr{F}|+|\vp|$; and the
  lower bounds carry over from any fixed frame $\fr{F}_0$, for instance the
  one-world frame, as $\vp\mapsto\langle\fr{F}_0,\vp\rangle$ reduces the problem
  for $\fr{F}_0$ to the problem with the frame as part of the input.  In
  contrast, the proofs of Propositions~\ref{prop:pspace}
  and~\ref{prop:bounded-var} need $\fr{F}$ to be fixed, since $|\Typ(\fr{F},s)|$
  grows exponentially with $|W|$; moreover, by the following proposition, a
  single predicate letter and two variables suffice for $\coNEXP$-hardness when
  $\fr{F}$ is part of the input.
\end{remark}

\begin{proposition}[One predicate letter, frame as part of the input]
  \label{prop:frame-input-mon}
  The problem of deciding, given a finite Kripke frame $\fr{F}$ and a formula
  $\vp\in\CMLmon$ whose only predicate letter other than $=$ is $P_1$, whether
  $\vp\in\mathbf{ML}^d_{\mathrm{C}}\fr{F}$, for
  $d \in \{ \mathit{xd}, \mathit{cd} \}$, is
  $\coNEXP$-complete.  The lower bound holds already for sentences with two
  variables and without equality and counting quantifiers, and for frames of
  the form $\langle\{w_0,\ldots,w_r\},\{\langle w_{i-1},w_i\rangle :
  1\leqslant i\leqslant r\}\rangle$, with $r\in\N$.
\end{proposition}

\begin{proof}
  The upper bound is given by Remark~\ref{rem:mon-frame-input}.  For the lower
  bound, let $\vp\in\mathcal{L}^{\mathit{mon}}$ be a sentence whose predicate
  letters are monadic, say $P_1,\ldots,P_r$, let $\fr{F}_r$ be the frame with the worlds
  $w_0,\ldots,w_r$ and the accessibility relation
  $\{\langle w_{i-1},w_i\rangle:1\leqslant i\leqslant r\}$, and let
  $\vp^{\circ}$ be obtained from $\vp$ by replacing each atom $P_i(x)$ with
  $\square^{i}P_1(x)$.  We show that
  $$
  \begin{array}{lclcl}
    \vp\in\QCL & \iff & \vp^{\circ}\in\ML\fr{F}_r & \iff & \vp^{\circ}\in\MLcd\fr{F}_r.
  \end{array}
  $$

  At a world $w_j$ of a model over $\fr{F}_r$, the formula $\square^{i}P_1(x)$
  defines a subset of $D_{w_j}$ (the set of elements of $D_{w_j}$ satisfying
  $P_1$ at $w_{i+j}$ if $i+j\leqslant r$, and $D_{w_j}$ otherwise), and the
  quantifiers of $\vp^{\circ}$ range over $D_{w_j}$; hence, $\vp^{\circ}$ is
  true at $w_j$ if $\vp$ is classically valid.

  Conversely, if $\vp$ is false in a classical model $\fr{A}$ with universe $A$,
  then $\vp^{\circ}$ is false at $w_0$ in the $\mathit{cd}$-model over
  $\fr{F}_r$ with $D_w:=A$, for every world $w$, with
  $I(w_i,P_1):=P_i^{\fr{A}}$, for $1\leqslant i\leqslant r$, and with
  $I(w_0,P_1):=\varnothing$.  Since $\fr{F}_r$ and $\vp^{\circ}$ are computable
  from $\vp$ in polynomial time, and $\vp^{\circ}$ has the same variables as
  $\vp$ and, like $\vp$, contains neither equality nor counting quantifiers, the
  map $\vp\mapsto\langle\fr{F}_r,\vp^{\circ}\rangle$ is a polynomial-time
  reduction of the set of $\mathcal{L}^{\mathit{mon}}$-sentences in $\QCL$ to
  the problem in the statement; by Proposition~\ref{prop:lewis}, that set is
  $\coNEXP$-hard already for sentences with two variables and with monadic
  predicate letters only, which, as $\fr{F}_r$ is a frame of the form described
  in the statement, gives the last claim.
\end{proof}

\section{Individual constants}
\label{sec:constants}

The languages considered so far have no individual constants.  Constants are,
however, used in applications.
%: e.g.,
%first-order standpoint logic uses rigid constants and constant
%domains~\cite{GRS22}.
In this section, we show that rigid constants can be added to our languages
without affecting the complexity results of Sections~\ref{sec:complexity}
and~\ref{sec:types}, with two qualifications: the polynomial-time bound for a
fixed number of variables (Proposition~\ref{prop:bounded-var}) requires the
number of constants to be fixed as well (Theorem~\ref{thm:rigid}\rmbr{$4$} and
Proposition~\ref{prop:const-fixed}), and the fluted fragments with counting
quantifiers are left aside (see Section~\ref{sec:limits}).  The
classical counterparts are now the fragments with constants, and the extension
proceeds in two steps.  First, the translation of Section~\ref{sec:transfer} and
the renamings of Section~\ref{sec:complexity} carry constants along unchanged,
so that each of the two-variable, one-variable, fluted and guarded fragments
with rigid constants is polynomial-time equivalent to its classical counterpart
with constants; for the two-variable, one-variable and guarded fragments, the
complexity classes then follow from the known results on the classical fragments
with constants (Proposition~\ref{prop:classical-const}).  Second, for the fluted
fragments, whose complexity is known only for formulas without constants, we
eliminate constants from classical fluted sentences without increasing the
number of variables (Lemma~\ref{lem:fluted-const}); this step is polynomial for
each fixed number of variables and elementary in general, which is all that the
$\tower$ bound requires.  For the monadic fragments, neither step is needed:
rigid constants are universally quantified parameters
(Lemma~\ref{lem:const-param}), and the histogram method applies to the result.
%We do not treat non-rigid constants~\cite{AHKMW25,KShW26} in general; in
%one-variable logics with equality, they amount to counting up to
%one~\cite{KShW26}.

\paragraph{Syntax.}  We extend the language $\CMLang$ with countably many
\emph{individual constants}, denoted by $c_1,c_2,\ldots$; a \emph{term} is a
variable or a constant, and atoms are formulas of the form $\bot$,
$P(t_1,\ldots,t_n)$ and $t_1=t_2$, where $P$ is an $n$-ary predicate letter and
$t_1,\ldots,t_n$ are terms.  The resulting language is denoted by $\CMLangrc$;
its sublanguages $\MLangeqrc$ and $\MLangrc$, and the fragments $\Gamma^k$ and
$\Gamma^{\mathit{mon}}$, are defined analogously to Section~\ref{sec:prelim}
(constants do not count as variables).  As before, the \emph{parameters} of a
formula are its free variables.  The set of constants occurring in a formula
$\vp$ is denoted by $\Const{\vp}$, and $c_\vp:=|\Const{\vp}|$.  When measuring
the size of formulas, we count every constant as a single symbol, as we do 
predicate letters and variables.

The \emph{fluted fragment} $\FLrc$ of $\MLangrc$ is defined as $\FL$ in
Section~\ref{sec:prelim}, with the following notion of atom: a \emph{fluted
  $m$-atom} is $\bot$ or an atom $P(\bar t)$ whose variables, read from left to
right, are $x_\ell,\ldots,x_m$ for some $\ell\leqslant m+1$; constants may occur
in any argument places.  Thus, $(\FLrc)^{[m]}$ is the least set containing the
fluted $m$-atoms in this sense, closed under $\to$ and $\square$, and containing
$\exists x_{m+1}\,\vp$ whenever $\vp\in(\FLrc)^{[m+1]}$, and
$\FLrc:=\bigcup_{m\geqslant0}(\FLrc)^{[m]}$; as in Section~\ref{sec:prelim}, we
assume that fluted sentences are members of $(\FLrc)^{[0]}$.  The \emph{guarded
  fragment} $\GFerc$ of $\MLangeqrc$ is defined by (G1)--(G4) of
Section~\ref{sec:prelim}, with atoms that may contain constants; the conditions
$\param\vp\cup\bar x\subseteq\param\gamma$ in (G3) and $\param\vp\subseteq\{x\}$
in (G4) refer to variables only: thus, the matrix of a quantifier may contain
constants that do not occur in the guard, and guards themselves may contain
constants.  The classical fragments $\mathcal{FL}^{\rc}$ and
$\mathcal{GF}_{=}^{\rc}$ are defined in the same way; the latter is the guarded
fragment of~\cite{PH23}.

\paragraph{Semantics.}  A \emph{Kripke model with \rmbr{rigid} constants} over a
frame $\fr{F}=\langle W,R\rangle$ is a tuple $\fr{M}=\langle W,R,D,I\rangle$ as
in Section~\ref{sec:prelim} whose interpretation function $I$ additionally sends
every constant $c$ of the formula under consideration to an element $I(c)$ of
$\Dplus$ such that $I(c)\in D_w$, for every $w\in W$.  Thus, constants are
rigid, i.e., they have the same value in every world, and this value exists at
every world.  The value $t^{g}$ of a term $t$ under an assignment $g$
is $g(x)$ if $t$ is a variable $x$, and $I(c)$ if $t$ is a constant $c$.  The
truth clauses for atoms are as follows:
\begin{itemize}
\item $\fr{M},w\sat^{g}P(t_1,\ldots,t_n)$ if
  $\langle t_1^{g},\ldots,t_n^{g}\rangle\in P^{I,w}$;
\item $\fr{M},w\sat^{g}t_1=t_2$ if $t_1^{g}=t_2^{g}$;
\end{itemize}
the remaining clauses are as in Section~\ref{sec:prelim}.  The following
definitions are analogous to those for languages without constants: truth in a
model, validity, refutation, the logics $(\mathbf{ML}_{\mathrm{C}}^{\rc})^d\fr{F}$,
$(\mathbf{ML}_{=}^{\rc})^{d}\fr{F}$ and $(\mathbf{ML}^{\rc})^{d}\fr{F}$, for
$d\in\{\mathit{xd},\mathit{cd}\}$, the finite domain property, and the problems
in which the frame is part of the input.  Remark~\ref{rem:conserv} holds for the
languages with constants.

The following lemma says that rigid constants are universally quantified
parameters; it holds for arbitrary classes of frames and for both domain
conditions.

\begin{lemma}[Constants as parameters]
  \label{lem:const-param}
  Let $\cl{C}$ be a class of Kripke frames and
  \mbox{$d\in\{\mathit{xd},\mathit{cd}\}$}.  Let $\vp\in\CMLangrc$ with
  $\Const{\vp}=\{c_1,\ldots,c_m\}$, let $y_1,\ldots,y_m$ be distinct variables
  not occurring in $\vp$, and let $\vp(y_1,\ldots,y_m)$ be the result of
  replacing each $c_i$ in $\vp$ with~$y_i$.  Then
  \[
  \begin{array}{lcl}  
    \vp\in(\mathbf{ML}_{\mathrm{C}}^{\rc})^d\cl{C} & \iff &
                                                   \forall y_1\cdots\forall y_m\,\vp(y_1,\ldots,y_m)\in\mathbf{ML}^d_{\mathrm{C}}\cl{C}.
  \end{array}                                              
  \]
  The formula $\forall y_1\cdots\forall y_m\,\vp(y_1,\ldots,y_m)$ has the same
  predicate letters, modal depth and counting threshold as $\vp$, and $m$ more
  variables.
\end{lemma}

\begin{proof}
  Suppose that $\forall\bar y\,\vp(\bar y)$ is refuted at a world $w_0$ of a
  $d$-model $\fr{M}=\langle W,R,D,I\rangle$ over $\fr{F}\in\cl{C}$.  Then
  \mbox{$\fr{M},w_0\nsat\vp(a_1,\ldots,a_m)$} for some
  $a_1,\ldots,a_m\in D_{w_0}$.  Let $\fr{M}'$ be a Kripke model obtained from
  $\fr{M}$ by adding $a_1,\ldots,a_m$ to the domain of every world that does not
  already contain them and leaving the interpretation of the predicate letters
  unchanged; then, $\fr{M}'$ is a $d$-model over~$\fr{F}$.
  By~\eqref{eq:expanding}, $a_1,\ldots,a_m\in D_v$ for every $v\in R^{*}(w_0)$,
  so the domains of the worlds in $R^{*}(w_0)$ are unchanged, and a
  straightforward induction shows that $\fr{M}$ and $\fr{M}'$ agree on all
  formulas at all worlds in $R^{*}(w_0)$.  Thus,
  $\fr{M}',w_0\nsat\vp(a_1,\ldots,a_m)$. Interpreting each $c_i$ as $a_i$, we
  obtain a $d$-model with constants refuting $\vp$ at~$w_0$.

  Conversely, if $\vp$ is refuted at $w_0$ in a $d$-model with constants, then
  $\vp(\bar y)$ is refuted at $w_0$ under the assignment sending each $y_i$ to
  $I(c_i)\in D_{w_0}$; hence, so is $\forall\bar y\,\vp(\bar y)$.
\end{proof}

Lemma~\ref{lem:const-param} reduces the languages with constants to the
languages without constants, which is all that is needed for the monadic
fragments of Section~\ref{sec:types}.  The reduction, however, increases the
number of variables, and it preserves neither flutedness nor guardedness, so for
the fragments of Section~\ref{sec:complexity} we extend the translation of
Section~\ref{sec:transfer}.

\paragraph{The translation with constants.}  Let $\fr{F}=\langle W,R\rangle$ be
a finite Kripke frame, $\vp\in\CMLangrc$ and $C:=\Const{\vp}$.  The signature
$\sigma^{C}_{\fr{F}}(\vp)$ is $\sigma_{\fr{F}}(\vp)$ enriched with the constants
from $C$. We define
\[
  (\Ax^{\mathit{xd}}_{\fr{F}})^{C}:=\Ax^{\mathit{xd}}_{\fr{F}}\wedge\bigwedge_{w\in
    W}\bigwedge_{c\in C}D_w(c), \qquad (\Ax^{\mathit{cd}}_{\fr{F}})^C
  :=\Ax^{\mathit{cd}}_{\fr{F}}\wedge\bigwedge_{w\in W}\bigwedge_{c\in C}D_w(c).
\]
The classical model $\fr{A}^{\fr{M}}$ defined from a Kripke model $\fr{M}$ with
constants interprets each $c\in C$ as $I(c)$, so that
\mbox{$\fr{A}^{\fr{M}}\models(\Ax^{\mathit{xd}}_{\fr{F}})^{C}$}; the Kripke model
$\fr{M}^{\fr{A}}$ defined from a classical model
$\fr{A}\models(\Ax^{\mathit{xd}}_{\fr{F}})^{C}$ puts
$I^{\fr{A}}(c):=c^{\fr{A}}$; by the added conjuncts, $c^{\fr{A}}$ belongs to
every $D^{\fr{A}}_w$.  The translation $\ST^{\fr{F}}_w$ is extended as follows:
\[
  \ST^{\fr{F}}_w(P_i(t_1,\ldots,t_n)):=P^{w}_i(t_1,\ldots,t_n);\qquad
  \ST^{\fr{F}}_w(t_1=t_2):=(t_1=t_2).
\]

An analogue of Lemma~\ref{lem:st} holds for formulas with constants and
classical models satisfying $(\Ax^{\mathit{xd}}_{\fr{F}})^{C}$: the atomic case
relies on the fact that the values of all terms of an atom are in
$D^{\fr{A}}_w$, which is ensured for constants by
$(\Ax^{\mathit{xd}}_{\fr{F}})^{C}$.  Hence, Observation~\ref{obs:M} and
Corollary~\ref{cor:st} hold for models with constants and formulas with
constants.  Observations~\ref{obs:tr-kvar} and~\ref{obs:monadicity} also hold,
as the translation changes neither the variables nor the arities of
atoms. Proposition~\ref{prop:flutedness} holds for $\FLrc$ and
$\mathcal{FL}^{\rc}$, as the translation sends fluted atoms to fluted atoms and
the conjuncts $D_w(c)$ are ground atoms.

The renamings of Section~\ref{sec:complexity} are unaffected, up to one
convention: a formula without variables, such as $P(c)$, is a fluted $m$-atom
for every $m$ and so, like a propositional formula, may occur at several levels
of a fluted sentence; accordingly, we take the level of every variable-free
formula (not just a propositional one) to be~$0$; we also adjust the convention
about the variables of sentences $\Phi^{d}_{\fr{F}}(\vp)$ of
Section~\ref{sec:two-var} to apply to all variable-free formulas (see
Remark~\ref{rem:var-conv}). Apart from this convention, nothing changes:
names take as arguments the parameters of the formula they name, not its
constants; the name of an atom $\psi$ at $w$ is still defined by
$\ST^{\fr{F}}_w(\psi)$, which keeps its constants; and, in
Section~\ref{sec:guarded}, a guard $\gamma$ still has the guard atom
$\gamma^{w}:=\ST^{\fr{F}}_w(\gamma)$, and the translation of an instance
$\exists x\,\theta$ of (G4) is still guarded by $D_w(x)$, even if $\gamma$ or
$\theta$ contains constants.  Thus,
Lemmas~\ref{lem:Phi}, \ref{lem:renaming}, \ref{lem:Psi-size},
\ref{lem:fluted_renaming}, \ref{lem:gST}, \ref{lem:guarded-syntax},
\ref{lem:Theta}, \ref{lem:sound} and~\ref{lem:guarded-renaming}, and
Observation~\ref{obs:ML2} hold for sentences with constants, with
$(\Ax^{d}_{\fr{F}})^{C}$ in place of $\Ax^{d}_{\fr{F}}$; since the axioms have
$|W|\cdot c_\vp$ additional ground conjuncts, the size bounds of
Lemmas~\ref{lem:Phi}, \ref{lem:Psi-size} and~\ref{lem:Theta} hold as stated.  
In particular, by Lemma~\ref{lem:guarded-syntax}\rmbr{$1$}, every guard of
$\Theta^{\mathit{xd}}_{\fr{F}}(\vp)$ and $\Theta^{\mathit{cd}}_{\fr{F}}(\vp)$ is
a name atom $P_{u,\psi}(\bar x_\psi)$ or $N_{u,\psi}(\bar x_\psi)$, an atom
$D_u(z)$, or the guard atom $\gamma^u$ of a guard $\gamma$ in $\vp$; only the
guard atoms may contain constants.

\paragraph{Classical fragments with constants.}  We write
$\mathcal{L}^{\rc}$, $\mathcal{L}_{=}^{\rc}$ and $\mathcal{L}_{\mathrm{C}}^{\rc}$ for the
classical languages with constants, identified, as in Section~\ref{sec:QCL},
with the sets of $\square$-free formulas of $\MLangrc$, $\MLangeqrc$ and
$\CMLangrc$.  The two-variable fragments with and without counting quantifiers
and the one-variable fragment with counting quantifiers, as defined
in~\cite{PH23}, admit individual constants \cite[pp.~61, 197 and~222]{PH23},
and so does the guarded fragment of~\cite{PH23}, which is $\mathcal{GF}_{=}^{\rc}$
(Remark~\ref{rem:GF-variants}); the results on these fragments that we need
are stated below for the languages with constants:

\begin{proposition}
  \label{prop:classical-const}
  \mbox{}
  \begin{itemize}
  \item[\rmbr{$1$}] Satisfiability of sentences of $(\mathcal{L}_{=}^{\rc})^2$ and
    of $(\mathcal{L}_{\mathrm{C}}^{\rc})^2$ is in $\NEXP$ {\rm \cite[Theorems~3.18
      and~8.10]{PH23}}, and satisfiability of sentences of
    $(\mathcal{L}_{\mathrm{C}}^{\rc})^1$ is in $\NP$ {\rm \cite[Theorem~7.18]{PH23}}.  Every
    satisfiable sentence of $(\mathcal{L}_{=}^{\rc})^2$ has a finite model
    {\rm \cite[Lemmas~3.3, 3.4, 3.6 and~3.8]{PH23}}.
  \item[\rmbr{$2$}] Satisfiability of sentences of $\mathcal{GF}_{=}^{\rc}$ is in
    $\TwoEXP$ {\rm \cite{Gradel99}, \cite[Theorem~1]{tCF05}; cf.\
      \cite[Theorem~4.9 and Chapter~4, Exercise~3]{PH23}}.
  \end{itemize}
\end{proposition}

In~\cite{PH23}, constants are eliminated from one- and two-variable sentences by
\cite[Lemmas~3.4 and~3.6]{PH23}; in the presence of counting quantifiers, they
are eliminated using the same argument once it has been guessed which constants
denote the same element \cite[p.~61]{PH23}.  For the guarded fragment, the proof
of the upper bound in~\cite{PH23} assumes that there are no constants and leaves
the general case as \cite[Chapter~4, Exercise~3]{PH23}, noting that existentially
quantifying the constants preserves the $\TwoEXP$ bound but not the $\EXP$ bound
for a fixed number of variables \cite[p.~99]{PH23}; Gr\"adel's elimination of
constants~\cite{Gradel99} likewise increases the number of variables by the
number of constants \cite[Section~3]{tCF05}.  The $\TwoEXP$ bound with constants
is thus Gr\"adel's~\cite{Gradel99}, as restated in \cite[Theorem~1]{tCF05}; the
$\EXP$ bound for a fixed number of variables and an unbounded number of
constants is due to ten Cate and Franceschet~\cite[Theorem~2]{tCF05}---we use it
in the proof of Theorem~\ref{thm:rigid}\rmbr{$3$}.

By contrast, the fluted fragment is only defined in the literature over purely
relational signatures \cite[Section~6.1]{PH23}, \cite{PH21,PST19}, and we are
not aware of any published treatment of fluted formulas with constants.  The
following lemma eliminates constants from fluted formulas without increasing the
number of variables.  Its construction is close to the elimination of constants
from guarded formulas with a bounded number of variables
in~\cite[Section~3]{tCF05}.  A witness for $\exists x\,\psi$ is either
\emph{named}, that is, denoted by some constant $c$, or \emph{unnamed}.
Accordingly, $\exists x\,\psi$ is replaced by the disjunction of the instances
$\psi[c/x]$, one for each constant $c$, and of $\exists x\,(U(x)\wedge\psi)$,
where $U$ is a fresh monadic letter meant to hold of the unnamed elements.  As
all quantifiers now range over unnamed elements, constants can be eliminated
from atoms: for instance, $P(x,c,y)$ becomes $P'(x,y)$, for a fresh letter $P'$
standing for $P$ with its second argument fixed to $c$.  To the resulting
constant-free sentence we add the conjunct $\forall x\,U(x)$, so that its models
correspond to the unnamed parts of models of the original sentence.  What is
specific to fluted formulas is that an instantiated variable must be removed
from the sequence of variables, which is done by the shift.  The proof of the
following lemma can be found in Appendix~\ref{sec:fluting-lemma}:

\begin{lemma}[Eliminating constants from fluted sentences]
  \label{lem:fluted-const}
  Let $k\geqslant 1$.  For every sentence $\chi\in(\mathcal{FL}^{\rc})^{k}$ with
  $m$ constants, one can compute, in time polynomial in $(m+1)^{k}\cdot|\chi|$,
  a sentence $\chi^{\circ}\in\mathcal{FL}^{k}$ without constants such that
  \begin{itemize}
  \item[\rmbr{$1$}] $\chi^{\circ}$ is satisfiable if, and only if, $\chi$ is;
  \item[\rmbr{$2$}] every model of $\chi^{\circ}$ with $n$ elements yields a
    model of $\chi$ with $n+m$ elements, and every model of $\chi$ with $n$
    elements yields a model of $\chi^{\circ}$ with at most $2n$ elements.
  \end{itemize}
\end{lemma}

\begin{proof}
  See Appendix~\ref{sec:fluting-lemma}.
\end{proof}

\paragraph{Results.}  We are now in a position to extend the results of
Sections~\ref{sec:complexity} and~\ref{sec:types} to languages with constants.
By the analogues of Lemmas~\ref{lem:renaming}, \ref{lem:fluted_renaming}
and~\ref{lem:guarded-renaming} for sentences with constants and by
Remark~\ref{rem:conserv}, each of the two-variable, one-variable, fluted and
guarded fragments with constants is polynomial-time equivalent to its classical
counterpart with constants, as in Theorem~\ref{thm:main}\rmbr{$1$}; the
following theorem gives the resulting complexity classes, whose upper bounds
come from Proposition~\ref{prop:classical-const} and, for the fluted fragments,
from Lemma~\ref{lem:fluted-const}.

\begin{theorem}[Rigid constants]
  \label{thm:rigid}
  Let $\fr{F}=\langle W,R\rangle$ be a finite Kripke frame and
  \mbox{$d\in\{\mathit{xd},\mathit{cd}\}$}.
  \begin{itemize}
  \item[\rmbr{$1$}] The fragments $((\mathbf{ML}_{\mathrm{C}}^{\rc})^d\fr{F})^2$,
    $((\mathbf{ML}_{=}^{\rc})^{d}\fr{F})^2$ and $((\mathbf{ML}^{\rc})^{d}\fr{F})^2$
    are $\coNEXP$-complete, and $((\mathbf{ML}_{\mathrm{C}}^{\rc})^d\fr{F})^1$ is
    $\coNP$-complete.
  \item[\rmbr{$2$}] Theorem~\ref{thm:fluted-complexity} holds for the fluted
    fragments $((\mathbf{ML}^{\rc})^{d}\fr{F})^{k}\cap\FLrc$, with $k\geqslant 1$, and
    $(\mathbf{ML}^{\rc})^{d}\fr{F}\cap\FLrc$.
  \item[\rmbr{$3$}] The fragment $(\mathbf{ML}_{=}^{\rc})^{d}\fr{F}\cap\GFerc$ is
    $\coTwoEXP$-complete, and so is its restriction to formulas without
    equality; for every $k\geqslant 2$, the restriction of
    $(\mathbf{ML}_{=}^{\rc})^{d}\fr{F}\cap\GFerc$ to formulas with at most $k$
    variables, and any number of constants, is $\coEXP$-complete, and so is
    its restriction to formulas without equality.
  \item[\rmbr{$4$}] Theorem~\ref{thm:nexptime},
    Corollary~\ref{cor:complexity-monadic} and Proposition~\ref{prop:pspace}
    hold for the monadic fragments of the languages with constants, and
    Proposition~\ref{prop:bounded-var} holds for formulas with at most $m$
    constants, for every $m\in\N$.
  \item[\rmbr{$5$}] Items \rmbr{$1$}--\rmbr{$3$} hold for the problems in which
    $\fr{F}$ is part of the input, and so do Theorem~\ref{thm:nexptime} and
    Corollary~\ref{cor:complexity-monadic} for the monadic fragments of the
    languages with constants; Proposition~\ref{prop:frame-input-mon} holds true
    for languages with constants.
  \item[\rmbr{$6$}] The fragments in \rmbr{$1$}--\rmbr{$4$}, except the
    two-variable fragments with counting quantifiers
    $((\mathbf{ML}_{\mathrm{C}}^{\rc})^d\fr{F})^2$, have the finite domain property over
    $\fr{F}$.  Moreover, Theorem~\ref{thm:fdp-two-var} holds with the bound
    $2^{O(|W|\cdot|\vp|)}$ replaced by $2^{p(|\fr{F}|+|\vp|)}$, for a
    polynomial $p$ that depends neither on $\fr{F}$ nor on~$\vp$;
    Theorem~\ref{thm:fdp-one-var} holds with its bound increased by $c_\vp$;
    Theorem~\ref{thm:FDP-fluted} holds with its bound increased by $c_\vp$, the
    exponential function now depending on~$k$;
    Theorem~\ref{thm:FDP-guarded} holds with $(2\varrho)^{\varrho}$ replaced
    by $(2(\varrho+c_\vp))^{\varrho}$, so that its last bound needs the number
    of constants to be fixed as well; Theorem~\ref{thm:small} holds with
    $|D'^{+}|\leqslant\max\{1,\kappa_\vp+c_\vp\}\cdot|\Typ^{d}(\fr{F},s)|$, with
    $\kappa_\vp:=M_\vp+v_\vp-1$ even if $v_\vp=0$ \rmbr{so that the maximum
    matters only if $\vp$ has neither variables nor constants}; and
    Proposition~\ref{prop:fdp-ML} holds with the bound $|\Typ^{d}(\fr{F},s)|$ on
    the number of elements.
  \end{itemize}
  In \rmbr{$1$}--\rmbr{$5$}, the lower bounds hold already for formulas
  without constants and, except for the lower bound of
  Proposition~\ref{prop:frame-input-mon} in~\rmbr{$5$}, for $\square$-free ones.
\end{theorem}

\begin{proof}
  All lower bounds are those of the corresponding results of
  Sections~\ref{sec:complexity} and~\ref{sec:types}; for the upper bounds, see
  Appendix~\ref{sec:thm:rigid}.
  %and the bounds in~(6).
%
%  
\end{proof}

The reductions in the proof of Proposition~\ref{prop:pspace} use unboundedly
many variables; the proof of Proposition~\ref{prop:bounded-var} shows that,
without this assumption, the complexity drops to polynomial time.  Constants
play the same role as variables here, and the bound on their number in
Theorem~\ref{thm:rigid}(4) is necessary, unless $\mathrm{P}=\NP$: unboundedly
many constants bring the complexity up to $\coNP$ already for formulas without
variables.

\begin{proposition}[Fixed numbers of predicate letters and variables, with
  constants]
  \label{prop:const-fixed}
  Let $\fr{F}$ be a finite Kripke frame, $s,v\in\N$, and
  $d\in\{\mathit{xd},\mathit{cd}\}$.
  \begin{itemize}
  \item[\rmbr{$1$}] For every $m\in\N$, the set
    $((\mathbf{ML}_{\mathrm{C}}^{\rc})^d\fr{F})^{\mathit{mon}}$, restricted to formulas
    with at most $v$ variables and at most $m$ constants whose predicate
    letters other than $=$ are among $P_1,\ldots,P_s$, is decidable in
    polynomial time.
  \item[\rmbr{$2$}] If $s\geqslant 1$, the set
    $((\mathbf{ML}_{\mathrm{C}}^{\rc})^d\fr{F})^{\mathit{mon}}$, restricted to formulas
    with at most $v$ variables whose predicate letters other than $=$ are
    among $P_1,\ldots,P_s$, is $\coNP$-complete; the lower bound holds already
    for $\square$-free formulas without variables, equality and counting
    quantifiers, with the single predicate letter $P_1$.  If $s=0$, the same
    holds, the lower bound holding for $\square$-free formulas with equality,
    without variables and counting quantifiers.
  \end{itemize}
\end{proposition}

\begin{proof}
  See Appendix~\ref{sec:prop:const-fixed}.
\end{proof}

\begin{remark}
  \label{rem:const-frame-input}
  Proposition~\ref{prop:const-fixed}, like Propositions~\ref{prop:pspace}
  and~\ref{prop:bounded-var}, needs the frame to be fixed, since its algorithms
  enumerate or guess histograms with $|\Typ(\fr{F},s)|$ entries, and with
  constants the frame-as-input problem is hard already for formulas without
  variables: given a propositional formula $\beta(Q_1,\ldots,Q_n)$, let
  $\fr{F}_n$ be the frame with the worlds $w_0,\ldots,w_n$ and the accessibility
  relation $\{\langle w_{i-1},w_i\rangle:1\leqslant i\leqslant n\}$ of the proof
  of Proposition~\ref{prop:frame-input-mon}, and let $\beta^{\circ}$ be obtained
  from $\beta$ by replacing each $Q_i$ with $\Diamond^{i}P_1(c)$, for a single
  constant $c$.  At $w_0$, the formula $\Diamond^{i}P_1(c)$ is true if, and only
  if, $I(c)\in I(w_i,P_1)$; so $\beta^{\circ}$ is valid on $\fr{F}_n$ if, and
  only if, $\beta$ is a tautology, a falsifying assignment being realised by the
  $\mathit{cd}$-model with the domain $\{a\}$ and with the interpretation defined
  as follows: $I(c)=a$ and $I(w_i,P_1)=\{a\}$ iff $Q_i$ is true.  Hence,
  deciding, given a finite frame $\fr{F}$ and a variable-free formula $\vp$ with
  one predicate letter and one constant, whether
  $\vp\in(\mathbf{ML}^{\rc})^{d}\fr{F}$, for $d\in\{\mathit{xd},\mathit{cd}\}$,
  is $\coNP$-hard.  By Theorem~\ref{thm:rigid}(5), this problem is in $\coNEXP$;
    in fact, it is in $\coNP$, even for variable-free formulas of $\CMLangrc$
  with any number of letters and constants.  Such a formula $\vp$ has no
  quantifiers, so its truth at the worlds of a $d$-model $\fr{M}$ over
  $\fr{F}$ depends only on the equivalence relation ${\equiv}$ on $\Const\vp$
  given by $c_1 \equiv c_2$ iff $I(c_1)=I(c_2)$, and on the set $S$ of pairs
  $\langle w,\alpha\rangle$ such that $w\in W$, $\alpha$ is an atom of $\vp$
  other than an equality, and $\fr{M},w\sat\alpha$; whether
  $\langle w,\alpha\rangle\in S$ does not change if constants in $\alpha$ are
  replaced with $\equiv$-equivalent ones.  Conversely, every such pair
  $\langle{\equiv},S\rangle$ arises from a $\mathit{cd}$-model over $\fr{F}$
  whose domain at every world is the set of $\equiv$-classes (or a single
  element, if $\vp$ has no constants), each constant denoting its class.  A
  nondeterministic algorithm can therefore guess ${\equiv}$ and $S$, which
  have size polynomial in $|\fr{F}|+|\vp|$, check the condition on $S$,
  compute the truth values of the subformulas of $\vp$ at all worlds, and
  accept if $\vp$ is false at some world; thus, refutability is in $\NP$.
\end{remark}

\section{Frame-definable operators}
\label{sec:operators}

The translation of Section~\ref{sec:transfer} and the renamings of
Section~\ref{sec:complexity} are not restricted to languages with a single unary
modal operator~$\square$: they handle any operator whose meaning at a world of a
finite Kripke frame $\fr{F} = \langle W, R \rangle$ is a Boolean combination of
truth values at the worlds of $W$ (this idea generalises a single $\square$
whose meaning at a world $w$ of a finite Kripke frame is a conjunction over
truth values at worlds accessible from $w$; see the clause for $\square \psi_1$
in the definition of the translation $\ST_w^{\fr{F}}$ of Section~\ref{sec:transfer}).

Let $\fr{F}=\langle W,R\rangle$ be a finite Kripke frame.  An
\emph{$\fr{F}$-definable operator} is a symbol $O$ of arity
$\mathrm{ar}(O)\in\N$ together with, for every $w\in W$, a formula
$\beta^{O}_w$, called the \emph{definition} of $O$ at $w$, of classical
propositional logic built from the set of propositional variables
$\{ p^{j}_{v} : v\in W, 1\leqslant j\leqslant\mathrm{ar}(O) \}$; if a variable
$p^{j}_{v}$, for some $1\leqslant j\leqslant\mathrm{ar}(O)$, occurs in
$\beta^{O}_w$, we say that $\beta^{O}_w$ \emph{consults} the world $v$.

If $\Ops$ is a finite set of $\fr{F}$-definable operators, we define the
language $\CMLang(\Ops)$ to contain operators from $\Ops$ as well as all the
symbols of $\CMLang$ except $\square$.  Formulas of $\CMLang(\Ops)$ are defined
by replacing, in the definition of \mbox{$\CMLang$-formulas}, the clause for
$\square$ with the following one: if $\psi_1,\ldots,\psi_{\mathrm{ar}(O)}$ are
formulas and $O\in\Ops$, then $O(\psi_1,\ldots,\psi_{\mathrm{ar}(O)})$ is a
formula.  The truth clause for $O \in \Ops$ is as follows:
\begin{itemize}
\item $\fr{M},w\sat^{g}O(\psi_1,\ldots,\psi_{\mathrm{ar}(O)})$ if $\beta^{O}_w$
  is true under the assignment sending $p^{j}_{v}$ to the truth value of
  $\fr{M},v\sat^{g}\psi_j$.
\end{itemize}
The \emph{size} of $\Ops$ is
$\|\Ops\|:=\sum_{O\in\Ops}\sum_{w\in W}|\beta^{O}_w|$.

Here are some examples of $\fr{F}$-definable operators (we write $\beta_w$ for
the definition of a unary operator at $w$, and $p_v$ for $p^{1}_{v}$):
\begin{itemize}
\item $\square$, with $\beta_w=\bigwedge_{v\in R(w)}p_v$, and its dual,
  $\Diamond$; the transitive closure operator $\square^{+}$, with
  $\beta_w=\bigwedge_{v\in R^{+}(w)}p_v$; the boxes $\square_i$ of a polymodal
  frame $\langle W,R_1,\ldots,R_m\rangle$, with domains of $d$-models expanding
  along every $R_i$, with $\beta_w=\bigwedge_{v\in R_i(w)}p_v$;
\item the universal modality, with $\beta_w=\bigwedge_{v\in W}p_v$; the
  difference modality, with $\beta_w=\bigwedge_{v\neq w}p_v$; a nominal
  naming $u\in W$, a $0$-ary operator with $\beta_u=\top$ and $\beta_w=\bot$
  whenever $w\neq u$; the graded modalities $\Diamond_{\geqslant k}$ for fixed $k$,
  with $\beta_w$ the disjunction of the conjunctions of $k$ distinct
  $p_v$, with $v\in R(w)$;
\item the temporal operators on a chain $0R1R\cdots R(n-1)$: $\Next$ (next),
  with $\beta_i=p_{i+1}$ for $i<n-1$ and $\beta_{n-1}=\bot$; $\Until$ (until),
  with
  $\beta_i=\bigvee_{i\leqslant j<n}\bigl(p^{2}_j\wedge\bigwedge_{i\leqslant
    k<j}p^{1}_k\bigr)$; the past counterparts $\Prev$ (previously) and
  $\Since$ (since).
\end{itemize}
The definitions of all these operators have size polynomial in $|W|$.

The logics $\CMLO{d}\fr{F}$, $\MLeqO{d}\fr{F}$ and $\MLO{d}\fr{F}$, for
$d\in\{\mathit{xd},\mathit{cd}\}$, in, respectively, the language
$\CMLang(\Ops)$, its sublanguage without counting quantifiers, and its
sublanguage without counting quantifiers and equality are defined similarly to
Section~\ref{sec:prelim}: the same goes for $k$-variable and monadic fragments.
The fluted fragment is defined with the clause
\mbox{$O(\psi_1,\ldots,\psi_{\mathrm{ar}(O)})\in\FLc^{[m]}$} whenever
$\psi_1,\ldots,\psi_{\mathrm{ar}(O)}\in\FLc^{[m]}$, and the guarded fragment
with clause (G2) for all $O\in\Ops$.

We say that $\Ops$ is \emph{forward} if, for every $O\in\Ops$ and $w\in W$, the
definition $\beta^{O}_w$ consults only worlds $v\in R^{*}(w)$; thus, boxes,
transitive closure operators, graded modalities and the future temporal
operators are all forward, and so, vacuously, are nominals; on the other hand,
the universal and difference modalities and the past temporal operators are, in
general, not forward.

The results of this section hold if $\Ops$ is forward, or if $d=\mathit{cd}$ and
$\fr{F}$ is connected (then, every $\mathit{cd}$-model over $\fr{F}$ has a
single constant domain).  These hypotheses are used in exactly one place: to
ensure that the values of the parameters of a formula exist at every world that
an operator consults.  We state this as a lemma:

\begin{lemma}[Consulted worlds]
  \label{lem:consulted}
  Let $\fr{F}=\langle W,R\rangle$ be a finite Kripke frame, $\Ops$ a finite
  set of $\fr{F}$-definable operators and $d\in\{\mathit{xd},\mathit{cd}\}$,
  and suppose that $\Ops$ is forward, or that $d=\mathit{cd}$ and $\fr{F}$ is
  connected.
  \begin{itemize}
  \item[\rmbr{$1$}] If $\fr{M}=\langle W,R,D,I\rangle$ is a $d$-model over
    $\fr{F}$, and if $O\in\Ops$, $w\in W$ and $v$ is a world consulted by
    $\beta^{O}_w$, then $D_w\subseteq D_v$.
  \item[\rmbr{$2$}] If $\fr{A}$ is a classical model of $\Ax^{d}_{\fr{F}}$ or of
    $(\Ax^{d}_{\fr{F}})^{C}$, and if $O\in\Ops$, $w\in W$ and $v$ is a world
    consulted by $\beta^{O}_w$, then $D^{\fr{A}}_w\subseteq D^{\fr{A}}_v$.
  \item[\rmbr{$3$}] Let $w_0\in W$, and let $S$ be the least subset of $W$
    containing $w_0$ such that $v\in S$ whenever $w\in S$ and $\beta^{O}_w$
    consults $v$, for some $O\in\Ops$.  If two $d$-models over $\fr{F}$ have,
    at every world in $S$, the same domain and the same interpretation of
    every predicate letter, then every formula of $\CMLang(\Ops)$ has the
    same truth value at $w_0$ in the two models under every assignment in
    both.  Moreover, $D_{w_0}\subseteq D_v$ for every $v\in S$ and every
    $d$-model $\langle W,R,D,I\rangle$ over $\fr{F}$.
  \end{itemize}
\end{lemma}

\begin{proof}
  (1): If $\Ops$ is forward, then $v\in R^{*}(w)$, and $D_w\subseteq D_v$
  follows from~\eqref{eq:expanding} by induction on the length of a path from
  $w$ to $v$ along $R$.  If $d=\mathit{cd}$ and $\fr{F}$ is connected, then
  $D_w=D_v$ follows from~\eqref{eq:constant} by induction on the length of a
  path from $w$ to $v$ along $R\cup R^{-1}$.

  (2): The same argument as in (1) applies, with the conjuncts
  $\forall x\,(D_u(x)\to D_{v}(x))$ of $\Ax^{\mathit{xd}}_{\fr{F}}$, for $uRv$,
  in place of~\eqref{eq:expanding} and the conjuncts
  $\forall x\,(D_{v}(x)\to D_u(x))$ of $\Ax^{\mathit{cd}}_{\fr{F}}$ in place
  of~\eqref{eq:constant}; the remaining conjuncts of $(\Ax^{d}_{\fr{F}})^{C}$
  are not needed.

  (3): The first claim, which does not use the hypothesis of the lemma, is
  proved by induction on the formula, simultaneously for all worlds in $S$;
  the case of $O(\psi_1,\ldots,\psi_{\mathrm{ar}(O)})$ at $w\in S$ uses IH at
  the worlds consulted by $\beta^{O}_w$, which are in $S$.  The second claim
  follows from (1) by induction along a sequence of consultations leading
  from $w_0$ to $v$.
\end{proof}

\paragraph{Translation and renaming with operators.}  The translation
$\ST^{\fr{F}}_w$ of Section~\ref{sec:transfer} and its extension with constants
of Section~\ref{sec:constants} are modified for $\CMLang(\Ops)$ by replacing the
clause for $\square$ with the following:
\begin{equation}
  \label{eq:ST-O}
  \ST^{\fr{F}}_w\bigl(O(\psi_1,\ldots,\psi_{\mathrm{ar}(O)})\bigr):=
  \beta^{O}_w\bigl[p^{j}_v/\ST^{\fr{F}}_v(\psi_j)\bigr],
\end{equation}
where $\beta^{O}_w\bigl[p^{j}_v/\ST^{\fr{F}}_v(\psi_j)\bigr]$ is the result of
substituting $\ST^{\fr{F}}_v(\psi_j)$ for $p^{j}_v$ in $\beta^{O}_w$; for
$\square$, this is exactly the clause of Section~\ref{sec:transfer}. The guarded
translation $\STg^{\fr{F}}_w$ of Section~\ref{sec:guarded} is modified with an
analogous clause, with $\STg^{\fr{F}}$ in place of $\ST^{\fr{F}}$.  We change,
in the renamings of Sections~\ref{sec:two-var} and~\ref{sec:fluted}, the
definition of the name of a subformula
$\psi=O(\psi_1,\ldots,\psi_{\mathrm{ar}(O)})$ at $w$ to
\begin{equation}
  \label{eq:delta-O}
  \delta_{w,\psi}:=\beta^{O}_w\bigl[p^{j}_v/Q_{v,\psi_j}(\bar x_j)\bigr],
\end{equation}
where $\bar x_j$ lists the parameters of $\psi_j$ (for fluted fragments, in the
prescribed order). In the renaming of Section~\ref{sec:guarded}, we change the
definitions of names to
\begin{equation}
  \label{eq:delta-O-pm}
  \delta^{\pm}_{w,\psi}:=\beta^{\pm}\bigl[p^{j}_v/\Pn{v}{\psi_j},\
  \neg p^{j}_v/\Nn{v}{\psi_j}\bigr],
\end{equation}
where $\beta^{+}$ and $\beta^{-}$ are negation normal forms of, respectively,
$\beta^{O}_w$ and of $\neg\beta^{O}_w$, of size $O(|\beta^{O}_w|)$, and the
substitution replaces the positive literals $p^{j}_v$ and the negative literals
$\neg p^{j}_v$ as indicated.  For $\square$, \eqref{eq:delta-O} is the
conjunction given in Sections~\ref{sec:two-var} and~\ref{sec:fluted}, and
\eqref{eq:delta-O-pm} are the conjunction and the disjunction of the table of
Section~\ref{sec:guarded}.  Lastly, the relation $\hist,w\sat^{\tau}\psi$ of
Section~\ref{sec:eval} gets the clause
\begin{equation}
  \label{eq:hist-O}
  \begin{array}{l}
    \hist,w\sat^{\tau} O(\psi_1,\ldots,\psi_{\mathrm{ar}(O)})
    \ \mbox{ if $\beta^{O}_w$ is true under the interpretation}\\
    \qquad\mbox{sending $p^{j}_{v}$ to the truth value of
    $\hist,v\sat^{\tau}\psi_j$;}
  \end{array}
\end{equation}
this clause is well defined for every $\hist$ if $\Ops$ is forward, and for
$\mathit{cd}$-admissible $\hist$ if $d=\mathit{cd}$ and $\fr{F}$ is connected,
as shown in the proof of Theorem~\ref{thm:operators} below.

\begin{theorem}[Frame-definable operators]
  \label{thm:operators}
  Let $\fr{F}=\langle W,R\rangle$ be a finite Kripke frame, $\Ops$ a finite set
  of $\fr{F}$-definable operators and $d\in\{\mathit{xd},\mathit{cd}\}$.
  Suppose that $\Ops$ is forward, or that $d=\mathit{cd}$ and $\fr{F}$ is
  connected.  Then the following results hold for $\CMLO{d}\fr{F}$,
  $\MLeqO{d}\fr{F}$ and $\MLO{d}\fr{F}$ in place of
  $\mathbf{ML}^d_{\mathrm{C}}\fr{F}$, $\mathbf{ML}^d_{=}\fr{F}$ and
  $\mathbf{ML}^d\fr{F}$ \rmbr{where constants are involved, for the language
    $\CMLang(\Ops)$ with constants in place of $\CMLangrc$}, with
  $|\fr{F}|+\|\Ops\|$ and $|W|+\|\Ops\|$ in place of $|\fr{F}|$ and $|W|$ in all
  running times and bounds:
  \begin{itemize}
  \item[\rmbr{$1$}] the decidability and complexity results of
    Theorems~\ref{thm:two-var}, \ref{thm:two-var-plain}, \ref{thm:one-var},
    \ref{thm:fluted-complexity}, \ref{thm:fluted-counting},
    \ref{thm:guarded-complexity} and~\ref{thm:nexptime},
    Corollary~\ref{cor:complexity-monadic},
    Propositions~\ref{prop:pspace}, \ref{prop:bounded-var}
    and~\ref{prop:const-fixed}, and Theorem~\ref{thm:rigid}\rmbr{$1$}--\rmbr{$5$},
    the lower bounds holding, as there, already for $\Ops$-free formulas,
    except for the lower bound of Proposition~\ref{prop:frame-input-mon} in
    Theorem~\ref{thm:rigid}\rmbr{$5$}, which is for formulas with $\square$ and
    holds whenever $\square\in\Ops$;
  \item[\rmbr{$2$}] the finite domain properties of
    Theorems~\ref{thm:fdp-two-var}, \ref{thm:fdp-one-var},
    \ref{thm:FDP-fluted}, \ref{thm:fluted-FDP-counting}, \ref{thm:FDP-guarded}
    and~\ref{thm:small}, and of Theorem~\ref{thm:rigid}\rmbr{$6$}.
  \end{itemize}
  In fact, the bounds of Theorems~\ref{thm:fdp-two-var},
  \ref{thm:fdp-one-var}, \ref{thm:FDP-guarded} and~\ref{thm:small}, and the
  bounds in Theorem~\ref{thm:rigid}\rmbr{$6$} other than those for
  Theorems~\ref{thm:fdp-two-var} and~\ref{thm:FDP-fluted}, hold as stated.
  The results in \rmbr{$1$}, other than Propositions~\ref{prop:pspace},
  \ref{prop:bounded-var} and~\ref{prop:const-fixed}, which are stated for a
  fixed frame, hold in the same form for the problems in which $\fr{F}$ and
  $\Ops$ are part of the input, as in Remarks~\ref{rem:frame-input},
  \ref{rem:mon-frame-input} and~\ref{rem:const-frame-input}.
\end{theorem}

\begin{proof}[Sketch of proof]
  \textit{Translation.}  With the clause~\eqref{eq:ST-O}, the proof of
  Lemma~\ref{lem:st} acquires the case
  $\psi=O(\psi_1,\ldots,\psi_{\mathrm{ar}(O)})$: for
  $\bar a\in(D^{\fr{A}}_w)^n$, IH is applied to $\psi_j$ at the worlds $v$
  consulted by $\beta^{O}_w$, for which $\bar a\in(D^{\fr{A}}_v)^n$ by
  Lemma~\ref{lem:consulted}\rmbr{$2$}; this is where the hypotheses of the
  theorem are used.  Under the second hypothesis, Lemmas~\ref{lem:st}
  and~\ref{lem:gST} are thus stated for classical models of
  $\Ax^{\mathit{cd}}_{\fr F}$, which is all that their applications to
  $\mathit{cd}$-models in Sections~\ref{sec:complexity} and~\ref{sec:constants}
  use.  The other cases are unchanged, and so are Observation~\ref{obs:M},
  Corollary~\ref{cor:st} \rmbr{for $\mathit{cd}$-models, under the second
  hypothesis} and Observations~\ref{obs:tr-kvar}
  and~\ref{obs:monadicity}; for Proposition~\ref{prop:flutedness}, note that
  $\beta^{O}_w[p^{j}_v/\ST^{\fr{F}}_v(\psi_j)]$ is a Boolean combination of
  formulas of $\cFLc^{[m]}$ if
  $\psi_1,\ldots,\psi_{\mathrm{ar}(O)}\in\FLc^{[m]}$.

  \smallskip\noindent
  \textit{Renamings.}  In the renamings of Section~\ref{sec:complexity}, the
  names of subformulas (in Section~\ref{sec:guarded}, of nodes) of the form
  $O(\psi_1,\ldots,\psi_{\mathrm{ar}(O)})$ are defined by~\eqref{eq:delta-O}
  and~\eqref{eq:delta-O-pm}.  In
  Section~\ref{sec:fluted}, $\delta_{w,\psi}\in\cFLc^{[m]}$, as
  in~\eqref{eq:delta-fluted}, since~\eqref{eq:delta-O} is a Boolean combination
  of fluted $m$-atoms: the $\psi_j$ have the level $m$ of $\psi$, so their
  parameters are final segments of $x_1,\ldots,x_m$.  The proofs of $(\ast)$ in
  Lemma~\ref{lem:renaming} and of its analogue for
  Lemma~\ref{lem:fluted_renaming} only use that $\delta_{w,\psi}$ is obtained
  from $\ST^{\fr{F}}_w(\psi)$ by replacing the translations of the immediate
  subformulas at the relevant worlds by their names; this remains true.  With
  $\STg^{\fr{F}}_w$ extended by the clause analogous to~\eqref{eq:ST-O}, the
  proof of Lemma~\ref{lem:gST} acquires the case
  $\psi=O(\psi_1,\ldots,\psi_{\mathrm{ar}(O)})$; it is treated as the similar
  case of Lemma~\ref{lem:st}: for $\bar a\in(D^{\fr{A}}_w)^m$, IH is applied to
  $\psi_j$ at the worlds $v$ consulted by $\beta^{O}_w$, for which
  $\bar a\in(D^{\fr{A}}_v)^m$ by Lemma~\ref{lem:consulted}\rmbr{$2$}; thus,
  $\beta^{O}_w$ is evaluated under the same valuation of $p^{j}_v$ on both sides
  of the equivalence.  The proof of Lemma~\ref{lem:sound} acquires the same
  case, which uses no hypothesis on domains.  Let
  $\fr{A}\models^{g}\delta^{+}_{w,\psi}$, and let $V$ be the valuation sending
  $p^{j}_v$ to the truth value of $\STg^{\fr{F}}_v(\psi_j)$ at $g$.  By IH for
  the children $\psi_j$ of $\psi$ at the worlds $v$ consulted by
  $\beta^{O}_w$, at the
  same assignment $g$, every literal of $\beta^{+}$ whose replacing name atom
  in~\eqref{eq:delta-O-pm} is true at $g$ is true under $V$: if $\Pn{v}{\psi_j}$
  is true, then so is $\STg^{\fr{F}}_v(\psi_j)$, and if $\Nn{v}{\psi_j}$ is
  true, then $\STg^{\fr{F}}_v(\psi_j)$ is false.  As $\beta^{+}$ is in negation
  normal form, and so monotone in its literals, it is true under $V$, and hence
  so is the equivalent formula $\beta^{O}_w$; thus,
  $\fr{A}\models^{g}\STg^{\fr{F}}_w(\psi)$.  The case of $\delta^{-}_{w,\psi}$
  is symmetric, with $\beta^{-}$ and $\neg\beta^{O}_w$.
  Lemma~\ref{lem:guarded-syntax} is unaffected, as formulas
  $\delta^{\pm}_{w,\psi}$ are quantifier-free, and the expansion $\fr{B}$ in the
  proof of Lemma~\ref{lem:guarded-renaming} still satisfies $\Def^{\mathrm{g}}$,
  as $\delta^{+}_{w,\psi}$ and $\delta^{-}_{w,\psi}$ define in $\fr{B}$ the
  relation defined by $\STg^{\fr{F}}_w(\psi)$ and its complement.  The size
  bounds of Lemma~\ref{lem:Phi}, of the proof of Lemma~\ref{lem:Psi-size}, and
  of Lemma~\ref{lem:Theta} become, respectively,
  $O(|W|\cdot|\vp|\cdot(|W|+|\vp|)+|\vp|\cdot\|\Ops\|)$,
  $O(|\vp|^2\cdot(|W|^2+\|\Ops\|))$ and $O((|\fr{F}|+\|\Ops\|)\cdot|\vp|^2)$,
  since the definition of a name at $w$ consists of $O(|\beta^{O}_w|)$ name
  atoms, each of size $O(1)$ in $\Phi^{d}_{\fr{F}}(\vp)$ and $O(|\vp|)$ in
  $\Psi^{d}_{\fr{F}}(\vp)$ and $\Theta^{d}_{\fr{F}}(\vp)$, and
  $\sum_{w\in W}|\beta^{O}_w|\leqslant\|\Ops\|$.  The complexity and
  finite domain results of Section~\ref{sec:complexity} follow as before, with
  $|W|+\|\Ops\|$ and $|\fr{F}|+\|\Ops\|$ in place of $|W|$ and $|\fr{F}|$ in the
  bounds; the bounds of Theorems~\ref{thm:fdp-two-var}, \ref{thm:fdp-one-var}
  and~\ref{thm:FDP-guarded} in fact survive as stated, since they depend only on
  the numbers of predicate letters and distinct atoms of the renamed sentence,
  on the numbers $\nu$ and $\mu$ in the proof of the first, on the numbers
  $r_\chi$ and $\varrho_\chi$ in the proof of the third, and on the numbers in
  counting quantifiers, none of which is affected by $\Ops$: the
  definitions~\eqref{eq:delta-O} and~\eqref{eq:delta-O-pm} are quantifier-free
  Boolean combinations of name atoms.  For $\fr{F}$ and $\Ops$ as part of the
  input, the translated sentences are computed in time polynomial in
  $|\fr{F}|+\|\Ops\|+|\vp|$, and the upper bounds follow as in
  Remark~\ref{rem:frame-input}.

  \smallskip\noindent
  \textit{Histograms.}  In the proofs for monadic fragments
  (Section~\ref{sec:types}), the relation $\hist,w\sat^{\tau}\psi$ is extended
  by the clause~\eqref{eq:hist-O}.  The
  clause is well defined, as $\tau$ is at every world $v$ consulted by
  $\beta^{O}_w$: if $\Ops$ is forward, then $v\in R^{*}(w)$ and the existence
  sets of types are closed under $R$ by (T2).  If $d=\mathit{cd}$ and $\fr{F}$
  is connected, we define the relation for $\mathit{cd}$-admissible $\hist$
  only; then, by~\eqref{eq:model-h-cd}, the existence set of every type $t$ with
  $\hist(t)\geqslant1$, and hence of every type of an $\hist$-compatible $\tau$,
  is a union of connected components, that is, $W$, so $\tau$ is at every world.
  This restriction is harmless: for $\mathit{cd}$-models,
  Proposition~\ref{prop:capped-hist} refers to the relation only for
  $\mathit{cd}$-admissible histograms, the algorithms of
  Theorem~\ref{thm:nexptime} and Propositions~\ref{prop:pspace},
  \ref{prop:bounded-var} and~\ref{prop:const-fixed} evaluate it only after
  checking \mbox{$\mathit{cd}$-admissibility}, and Lemma~\ref{lem:cost} is
  applied to $\mathit{cd}$-admissible histograms only; Lemma~\ref{lem:eval}
  concerns the histogram $\hist^{\kappa_\vp}_{\fr{M}}$ of a model $\fr{M}$,
  which is $\mathit{cd}$-admissible if $\fr{M}$ is a $\mathit{cd}$-model; and,
  in the proof of Proposition~\ref{prop:bounded-var}, Claim~1\rmbr{c} may be
  restricted to $\mathit{cd}$-admissible $\hist$ and $\hist'$, by
  Claim~1\rmbr{a}.  With this clause, the case of $O$ in the proof of
  Lemma~\ref{lem:eval} is immediate from IH, as the case of $\square$ is.  The
  bound of Theorem~\ref{thm:small} is unaffected, as it depends only on
  $\Typ^{d}(\fr{F},s)$ and $\kappa_\vp$.  The cost of the clause in
  Lemma~\ref{lem:cost} is polynomial in $\|\Ops\|$, which does not affect the
  complexity classes of Theorem~\ref{thm:nexptime} and
  Propositions~\ref{prop:pspace} and~\ref{prop:bounded-var}, the latter two
  for fixed $\fr{F}$ and $\Ops$; in the proof of
  Proposition~\ref{prop:bounded-var}, the truth value of
  $O(\psi_1,\ldots,\psi_{\mathrm{ar}(O)})$ depends only on those of $\psi_j$,
  which, by IH, depend only on the sign vector $\sigma_{\hist}$.  The analogues
  of the results of Section~\ref{sec:types} follow; with $\fr{F}$ and $\Ops$ as
  part of the input, the algorithm of Theorem~\ref{thm:nexptime} still runs in
  nondeterministic exponential time, whereas the proofs of
  Propositions~\ref{prop:pspace} and~\ref{prop:bounded-var} need $\fr{F}$ to be
  fixed, as in Remark~\ref{rem:mon-frame-input}.

  \smallskip\noindent
  \textit{Constants.}  The translation and the renamings with constants
  (Section~\ref{sec:constants}) extend as above.  Lemma~\ref{lem:const-param}
  holds for $\CMLang(\Ops)$ with constants: in its proof, let $S$ be the set of
  Lemma~\ref{lem:consulted}\rmbr{$3$} for $w_0$, and add $a_1,\ldots,a_m$ to the
  domains of the worlds outside $S$ only.  By that lemma, the domains of the
  worlds in $S$ already contain $a_1,\ldots,a_m$; hence, the result is a
  $d$-model \rmbr{check the pairs $wRv$ with exactly one of $w$, $v$ in $S$},
  and, by the same lemma, it agrees with the original model on
  $\vp(a_1,\ldots,a_m)$ at $w_0$.  Theorem~\ref{thm:rigid}
  and Proposition~\ref{prop:const-fixed} follow, the latter for fixed $\fr{F}$
  and $\Ops$, as in Remark~\ref{rem:const-frame-input}.

  \smallskip\noindent
  \textit{Lower bounds.}  All the lower bounds in the results listed in the
  theorem are for $\square$-free formulas, except the lower bound of
  Proposition~\ref{prop:frame-input-mon}, whose reduction uses the formulas
  $\square^{i}P_1(x)$ and so applies whenever $\square\in\Ops$;
  Remark~\ref{rem:conserv} applies verbatim to $\Ops$-free formulas.
  %See Appendix~\ref{sec:thm:operators}.
\end{proof}

\begin{remark}
  \label{rem:operators-xd}
  The proof of Theorem~\ref{thm:operators} uses its hypotheses only to ensure
  that, in a $d$-model, every world $v$ consulted by $\beta^{O}_w$ has a domain
  including $D_w$, so that the values of the parameters exist at $v$.  Under the
  $\mathit{cd}$-condition, this holds for every $v$ in the connected component
  of $w$; hence, the hypotheses may be weakened to: for every $O\in\Ops$ and
  $w\in W$, the definition $\beta^{O}_w$ consults only worlds $v$ with
  $v\in R^{*}(w)$ or, if $d=\mathit{cd}$, with $v$ in the connected component of
  $w$.  This covers, for instance, the universal and difference modalities
  relativised to connected components, with $\mathit{cd}$-models over an
  arbitrary finite frame.  Without such a hypothesis, an operator may consult a
  world whose domain does not contain the values of the parameters.  Translating
  an atom $P_i(\bar x)$ at $v$ as
  $P^{v}_i(\bar x)\wedge\bigwedge_{x\in\bar x}D_v(x)$ restores
  Lemma~\ref{lem:st} for all tuples from the global domain, and keeps the
  two-variable, one-variable, guarded and monadic shapes, but not the fluted
  one, since $D_v(x_i)$ is not a fluted atom unless $x_i$ is the last variable;
  the histogram method would also need the counting clause of
  Lemma~\ref{lem:eval} adjusted to the parameters absent at $v$.
  %We do not
  %pursue this, as the hypotheses hold in the cases considered below.
\end{remark}

\section{Conclusion}
\label{sec:limits}

Over a single finite Kripke frame, modal predicate logic reduces to classical
first-order logic with one copy of each predicate letter for each world, and
the reduction can be made to respect the syntactic restrictions behind the
classical decidability results: the renaming of Section~\ref{sec:complexity}
keeps the unfolding of the modal operators polynomial while preserving the
number of variables, the order of variables of fluted formulas and, through
positive and negative names constrained by guarded implications, the
guardedness of guarded formulas.  Thus, the one-variable, two-variable, fluted
and guarded fragments of the logics of a finite frame are polynomial-time
equivalent to their classical counterparts, with no restriction on the free
variables of modalised formulas, and inherit their complexity
(Table~\ref{tab:summary}); all but the two-variable fragment with counting
quantifiers inherit the finite model property as a finite domain property with
bounds on the size of domains.  For the monadic fragments, where renaming is
not an option, the histogram method of Section~\ref{sec:types}, which
describes a model by the number of elements of each world-dependent type,
capped at $M_\vp+v_\vp-1$, gives $\coNEXP$-completeness, a finite domain
property with an explicit bound, $\PSPACE$-completeness for a fixed frame and
a fixed number of predicate letters, and polynomial time when the number of
variables is fixed as well; over the one-world frame, it yields the
complexity and a parameterwise optimal small-model bound for the classical
monadic fragment with counting quantifiers
(Proposition~\ref{prop:classical-counting}).  Most of these results hold when
the frame, listed explicitly, is part of the input; those for a fixed number
of predicate letters need a fixed frame (Proposition~\ref{prop:frame-input-mon}).
The results survive the addition of rigid constants
(Section~\ref{sec:constants}), the classical counterparts then being the
fragments with constants, and the replacement of the box by any operators
definable on the frame that consult only worlds accessible from the current
one, or are evaluated over constant domains on a connected frame
(Section~\ref{sec:operators}).

\paragraph{Classes of frames.}  The results of
Sections~\ref{sec:transfer}--\ref{sec:types} carry over to logics of classes of
frames along the lines of~\cite{RS24}.  The logic of a finite class of finite
frames is the logic of their disjoint union, so every result above holds for
such a class.  For an infinite class $\cl{C}$ of $n$-alternative frames closed
under subframes, the subframe argument of~\cite{RS24} applies: a formula $\vp$
refuted at a world of a frame is refuted at that world of the subframe of worlds
reachable from it in at most $\md\vp$ steps, a frame of $\cl{C}$ with at most
$1+n+\cdots+n^{\md\vp}$ worlds.  Hence, if the set of finite frames in $\cl{C}$
is decidable, all the fragments considered above remain decidable over $\cl{C}$;
some such hypothesis is needed, as a bound on the size of refuting frames does
not by itself give decidability: there are classes of finite $2$-alternative
frames refuting every non-valid formula on a frame of linear size whose logics
have variable-free fragments of any degree of
unsolvability~\cite[Section~3]{RS23}.  If membership of a finite frame in
$\cl{C}$ is decidable in polynomial time, then, as the frames to be examined
have size exponential in $|\vp|$, the results for the frame as part of the input
yield upper bounds over $\cl{C}$ at most one exponential higher: for instance,
$\cotwoNEXP$ for the two-variable and the monadic fragments, and still $\tower$
for the fluted fragments.  Both hypotheses hold for the classes of all
$n$-alternative frames and of all reflexive $n$-alternative frames, over which
the logics $\lo{QAlt}_n$ and $\lo{QTAlt}_n$ are Kripke complete~\cite{SS26}.
Whether the exponential can be avoided we leave for a separate study.

\paragraph{Open problems.}  For the fluted fragments with at least five
variables, and with at least three in the presence of counting quantifiers, the
gaps between the bounds of Theorems~\ref{thm:fluted-complexity}
and~\ref{thm:fluted-counting} are those of the classical
fragments~\cite{PST19,PH21}, and will close with them.  Whether the fluted
fragments with counting quantifiers retain their complexity with rigid constants
is open, as the duplication of elements in the proof of
Lemma~\ref{lem:fluted-const} does not preserve counting quantifiers.  Constants
have been read as rigid designators that exist at every world; the free reading,
under which a constant need not denote an element of every domain, and non-rigid
designators, which make several monodic fragments undecidable over the usual
classes of frames~\cite{AHKMW25,KShW26}, remain to be examined over a single
finite frame.  Operators outside the scope of Theorem~\ref{thm:operators} and
Remark~\ref{rem:operators-xd} remain to be treated; the fluted and monadic
fragments are the obstacles.  Lastly, a frame given succinctly, say a chain of
length $n$ with $n$ in binary, is an exponentially smaller input, for which our
results, as for classes of frames, only yield upper bounds one exponential
higher; whether the bounds for the listed frame survive is open.

\paragraph{Equality as congruence.}  We have interpreted equality as identity.
Under the alternative interpretation, as a congruence relation that may identify
distinct elements, the translation of Section~\ref{sec:transfer} would have to
render equality at each world $w$ by a binary letter $\mathit{Eq}_w$,
interpreted as an equivalence relation that the atoms of the letters $P^{w}_i$
respect.  This interpretation is investigated in the companion
paper~\cite{Shkatov-eq}.

\appendix

\section{Proof of Proposition~\ref{prop:ffc}}
\label{sec:prop:ffc}

Let $k\geqslant2$, and let $\chi$ be a satisfiable $\cFLc^{k}$-sentence with
counting threshold $M_\chi$.  If $\chi$ has a model with fewer than
$M_\chi\leqslant2^{|\chi|}$ elements, there is nothing to prove; so let
$\fr{A}$ be a model of $\chi$ with at least $M_\chi$ elements.  The proof
of~\cite[Theorem~1]{PH21} brings $\chi$ into the normal form $\chi'$
of~\cite[Lemma~3]{PH21}, of size polynomial in $|\chi|$, and then bounds the
size of models of $\chi'$.  We show that $\chi'$ entails $\chi$ and that
$\fr{A}$ expands to a model of $\chi'$; it then suffices to show that $\chi'$
has a model whose size is bounded by a $(k-1)$-fold exponential function of
$|\chi|$.

Lemma~3 of~\cite{PH21} states that $\chi'$ is satisfiable over the same domains
as $\chi$.  Indeed, each step of its construction turns a sentence $\vartheta$
into a sentence $\vartheta'$ that entails $\vartheta$ and such that every model
of $\vartheta$ expands to a model of $\vartheta'$, with one exception: in the
last step, each conjunct $\vartheta=\forall^{m}(\mu\to\exle{M}\zeta)$ is
replaced, for a fresh letter $q$ of arity $m+1$, by
$$
  \vartheta'=\forall^{m}(\mu\to\exeq{M}q)\wedge\forall^{m+1}(\zeta\to q).
$$
This $\vartheta'$ still entails $\vartheta$, as $\zeta$ implies $q$, but a
model of $\vartheta$ is only guaranteed to expand to a model of $\vartheta'$
over a domain with at least $M$ elements, where $q$ can be interpreted, for
each $m$-tuple satisfying $\mu$, as its at most $M$ witnesses for $\zeta$
together with further elements to a total of exactly $M$, and as the whole
domain for the other $m$-tuples.  For instance,
$\forall x_1\,\exle{2}x_2\,\top$ holds in every one-element structure, but its
replacement, which requires two witnesses for $q$, has no one-element model;
the same qualification appears in Stage~3 of the proof
of~\cite[Lemma~8.3]{PH23}.  As $\exle{M}$ contributes $M+1$ to $M_\chi$, and
moving negations inward turns $\neg\ex{M}$ into $\exle{M-1}$, every such $M$
is less than $M_\chi$.  Hence, $\chi'$ entails $\chi$, and $\fr{A}$ expands to
a model of $\chi'$.

For $k=2$, the proof of~\cite[Lemma~4]{PH21}, applied to $\chi'$, which is
already in normal form, gives a model of $\chi'$ of size exponential in
$|\chi'|$, and so in $|\chi|$.  For $k\geqslant3$, the proof
of~\cite[Theorem~1]{PH21} turns $\chi'$ into an equivalent formula $\chi_k$ of
the class $\mathrm{FLC}^{k}_{\mathrm{ep}}$ of~\cite{PH21}, whose effective
size~\cite{PH21} is polynomial in $|\chi'|$.  For $2\leqslant j<k$,
the construction of~\cite[Lemmas~9--11]{PH21} turns a formula $\chi_{j+1}$ of
$\mathrm{FLC}^{j+1}_{\mathrm{ep}}$ into a formula $\chi_j$ of
$\mathrm{FLC}^{j}_{\mathrm{ep}}$ that is satisfiable over the same domains and
whose effective size is exponential in that of $\chi_{j+1}$.  Lastly,
by~\cite[Lemma~7]{PH21}, $\chi_2$, which is satisfiable as $\chi'$ is and whose
effective size is bounded by a $(k-2)$-fold exponential function of $|\chi|$,
has a model whose size is bounded by a $(k-1)$-fold exponential function of
$|\chi|$; hence, so do $\chi_k$ and $\chi'$, as required.

\section{Proof of Theorem~\ref{thm:fdp-two-var}}
\label{sec:thm:fdp-two-var}

We assume that $\vp$ is a sentence and put $\Phi:=\Phi^{d}_{\fr{F}}(\vp)$.  By
Observation~\ref{obs:ML2}, $\Phi$ is a sentence of $\mathcal{L}^2_=$.  As $\vp$
is refuted in a $d$-model over $\fr{F}$, by Lemma~\ref{lem:renaming} (`only
if'), $\Phi$ is satisfiable.  Also by Lemma~\ref{lem:renaming} (`if'), if $\Phi$
has a model with $n$ elements, then $\vp$ is refuted in a $d$-model over
$\fr{F}$ whose global domain has at most $n$ elements.  Thus, it suffices to
show that $\Phi$ has a model with $2^{O(|W|\cdot|\vp|)}$ elements.

  By~\cite[Lemmas 3.8, 3.4, and 3.3]{PH23}, every satisfiable
  $\mathcal{L}^2_=$-sentence $\chi$ has a model with at most $3\cdot2^{9N}$
  elements, where $N:=\lambda+20\nu+\mu$ and $\lambda$ is the number of
  predicate letters of $\chi$, $\nu$ is the number of distinct atoms of $\chi$
  whose predicate letter has arity greater than $2$, and $\mu$ is the number of
  occurrences of quantifiers in $\chi$ that begin a subformula which has a
  parameter or contains an atom of arity greater than $2$.
  Thus, it remains to estimate $\lambda$, $\nu$ and $\mu$ for
  $\chi:=\Phi$.

  The letters of $\Phi$ are the $|W|$ letters $D_w$, the
  $|W|\cdot s$ letters $P^w_i$, where $s\leqslant|\vp|$ is the number of
  predicate letters of $\vp$, and the
  $|W|\cdot|{\subf\vp}|\leqslant|W|\cdot|\vp|$ names $Q_{w,\psi}$; hence,
  $\lambda\leqslant|W|\cdot(2|\vp|+1)$.

  As the letters $D_w$ are monadic and the names have arity at most $2$, the atoms
  counted by $\nu$ are among the atoms $P^w_i(\bar x)$ with $w\in W$ and
  $P_i(\bar x)\in\subf\vp$ (each of them occurs twice in $\Phi$, in the conjunct
  of $\mathit{Def}$ for $P_i(\bar x)$, as $\leftrightarrow$ is an abbreviation,
  but is only counted once since $\nu$ counts distinct atoms rather than
  occurrences); hence, $\nu\leqslant|W|\cdot|\vp|$.

  The quantifiers of $\Phi$ occur in $\Ax^{d}_{\fr{F}}$, where each of them
  begins a sentence without atoms of arity greater than $2$ (recall that
  $\Ax^{\mathit{cd}}_{\fr{F}}$ differs from $\Ax^{\mathit{xd}}_{\fr{F}}$ only by
  sentences without parameters over the letters $D_w$), and in the conjuncts
  $\forall\bar x\,(Q_{w,\psi}(\bar x)\leftrightarrow\delta_{w,\psi})$ of
  $\mathit{Def}$.  In each conjunct of $\mathit{Def}$, the quantifiers are
  those of the prefix $\forall\bar x$, at most two, and, possibly, the
  quantifier of $\delta_{w,\psi}$, which then occurs twice since
  $\leftrightarrow$ is an abbreviation.

  If $\psi$ is atomic, then $\delta_{w,\psi}$ is quantifier-free, so at most two
  occurrences of quantifiers are counted.  Otherwise, the conjunct contains no
  atom of arity greater than $2$, so an occurrence of a quantifier is counted
  only if it begins a subformula with a parameter.  The outer quantifier of the
  prefix always begins a sentence, so it is never counted.  If $\psi$ does not
  begin with a quantifier, then $\delta_{w,\psi}$ contains no quantifiers, so
  at most one occurrence of a quantifier is counted (that of the inner
  quantifier of the prefix, in case $\psi$ has two parameters). Otherwise, we
  count, twice, the quantifier of $\delta_{w,\psi}$.  Thus, at most two
  occurrences are counted in each conjunct, and so $\mu\leqslant2|W|\cdot|\vp|$.

  Therefore,
  $$
  \begin{array}{lcl}
    N & \leqslant & |W|\cdot(2|\vp|+1)+20|W|\cdot|\vp|+2|W|\cdot|\vp|
        \;=\; |W|\cdot(24|\vp|+1),
  \end{array}
  $$
  
  Hence, $\Phi$ has a model with at most
  $3\cdot2^{9|W|\cdot(24|\vp|+1)}=2^{O(|W|\cdot|\vp|)}$ elements, as required.

\section{Proof of Theorem~\ref{thm:fdp-one-var}}
\label{sec:thm:fdp-one-var}
We assume that $\vp$ is a sentence and put $\Phi:=\Phi^{d}_{\fr{F}}(\vp)$.  As
in the proof of Theorem~\ref{thm:fdp-two-var}, it suffices to show that $\Phi$
has a model with at most $|W|\cdot(1+|\vp|\cdot M_\vp)$ elements; the second
inequality in the theorem holds as $M_\vp\leqslant2^{|\vp|}$, numbers being
written in binary.  As $\Phi$ is already essentially in the normal form
of~\cite[Lemma~7.15]{PH23}, we can apply directly the selection of witnesses in
the proof of~\cite[Lemma~7.16]{PH23}.

  Since $\Phi$ has a single variable, its subformulas of the form
  $\ex{M}x\,\theta$ are sentences.  Apart from the sentences
  $\ex{1}x\,\neg\theta$ in the conjuncts $\forall x\,\theta$ of $\Phi$, which
  are false in every model of $\Phi$, they are the $|W|$ conjuncts
  $\exists x\,D_w(x)$ of $\Ax^{d}_{\fr{F}}$ and the sentences
  $\delta_{w,\psi}=\ex{M}x\,\bigl(D_w(x)\wedge Q_{w,\psi_1}(\bar x_1)\bigr)$,
  for $w\in W$ and $\psi=\ex{M}x\,\psi_1\in\subf\vp$, of which there are at
  most $|W|\cdot|\vp|$, each with $M\leqslant M_\vp$.

  Let $\fr{A}$ be a model with universe $A$ such that $\fr{A}\models\Phi$.  For
  each of these sentences $\ex{M}x\,\theta$ that is true in $\fr{A}$, choose $M$
  elements of $A$ satisfying $\theta$ in $\fr{A}$, and let $\fr{B}$ be the
  substructure of $\fr{A}$ whose universe $B$ consists of the chosen elements.
  Then $B\neq\varnothing$, as $\fr{A}\models\exists x\,D_w(x)$, and
  $|B|\leqslant|W|\cdot(1+|\vp|\cdot M_\vp)$.  We show, by induction on
  $\theta\in\subf\Phi$, that $\fr{B}\models\theta[a]$ iff
  $\fr{A}\models\theta[a]$, for every $a\in B$; for $\theta=\Phi$, this gives
  $\fr{B}\models\Phi$, as required.  The cases of atoms and of $\to$ are
  immediate.  Let $\theta=\ex{M}x\,\theta_1$, and let
  $S:=\{b\in A:\fr{A}\models\theta_1[b]\}$.  By IH,
  $\{b\in B:\fr{B}\models\theta_1[b]\}=S\cap B$.  If $\fr{A}\models\theta$, then
  $S\cap B$ contains the $M$ elements chosen for $\theta$; otherwise,
  $|S\cap B|\leqslant|S|<M$.  In either case, $\fr{B}\models\theta$ iff
  $\fr{A}\models\theta$.

\section{Proof of Theorem~\ref{thm:FDP-guarded}}
\label{sec:thm:FDP-guarded}

  We may assume that $\vp$ is a sentence: as in the proof of
  Theorem~\ref{thm:guarded-complexity}, a model refuting
  $\forall\bar x\,(G(\bar x)\to\vp)$ also refutes $\vp$, and this sentence
  has size $O(|\vp|)$, the same variables as $\vp$, and one new letter, of
  arity at most $k$.  As $\vp$ is refuted in a $d$-model over $\fr F$, by
  Lemma~\ref{lem:guarded-renaming} (`only if'),
  $\Theta:=\Theta^{d}_{\fr F}(\vp)$ is satisfiable; by the same lemma (`if'),
  if $\Theta$ has a model with $n$ elements, then $\vp$ is refuted in a
  $d$-model over $\fr F$ whose global domain has at most $n$ elements.  By
  Lemma~\ref{lem:guarded-syntax}\rmbr{$1$}, $\Theta$ is a
  $\mathcal{GF}_=$-sentence.  As~\cite{Fiuk26} does not admit proposition
  letters, let $\Theta'$ be the result of replacing in $\Theta$ every
  proposition letter~$P$ by the guarded sentence $\exists x\,U_P(x)$, where
  $U_P$ is a fresh monadic letter; since $P$ and $\exists x\,U_P(x)$ can be made
  to agree in every non-empty universe, $\Theta'$ is satisfiable over the same
  domains as $\Theta$.  Thus, it suffices to show that $\Theta'$ has a model
  with $2^{O(|W|\cdot|\vp|\cdot(2\varrho)^{\varrho})}$ elements.

  By~\cite[Lemmas~3.2 and~3.4, Proposition~4.1, and Claim~2.1]{Fiuk26}, there is an
  absolute constant $c$ such that every satisfiable $\mathcal{GF}_=$-sentence
  $\chi$ without constants and proposition letters has a model with at most
  $$
  c\cdot\bigl((\varrho_\chi+1)\cdot2^{r_\chi\cdot \varrho_\chi^{\varrho_\chi}}\bigr)^{2^{\varrho_\chi}}
  \;\leqslant\;c\cdot2^{2r_\chi\cdot(2\varrho_\chi)^{\varrho_\chi}}
  $$
  elements, where $\varrho_\chi$ is the maximal arity of the predicate letters of
  $\chi$, and $r_\chi$ is the number of predicate letters of $\chi$ other than
  $=$ plus the number of occurrences in $\chi$ of maximal quantifier blocks
  beginning a subformula that has a parameter.  (The guarded fragment
  of~\cite{Fiuk26} admits equality, also as a guard, so it contains
  $\mathcal{GF}_=$.  The estimate comes from the small model property
  of~\cite[Proposition~4.1]{Fiuk26}, which bounds the size of a model of a
  satisfiable normal-form sentence by $c\cdot|T|^{2^{\varrho_\chi}}$, where $T$ is a
  set of $j$-types, $j\leqslant \varrho_\chi$, over its signature; by~\cite[Claim~2.1]{Fiuk26},
  $|T|\leqslant(\varrho_\chi+1)\cdot2^{r_\chi\cdot \varrho_\chi^{\varrho_\chi}}$.  The normal
  form is obtained by~\cite[Lemma~3.2]{Fiuk26}, which adds one letter for each
  block counted in $r_\chi$, does not increase the maximal arity, and yields
  an equisatisfiable sentence every model of which is, by reduct, a model of
  the original.)

  Thus, it remains to estimate $\varrho_\chi$ and $r_\chi$ for $\chi:=\Theta'$.
  The letters of $\Theta'$ are the $|W|$ letters $D_w$, at most $|W|\cdot|\vp|$
  letters $P^w_i$, and at most $2|W|\cdot|\vp|$ names, some of them replaced by
  letters $U_P$.  The maximal quantifier blocks of $\Theta'$ that begin a
  subformula with a parameter are among those of the formulas
  $\delta^{\pm}_{w,\psi}$, which contain at most one block each, since the
  blocks of $\Ax^{d}_{\fr F}$, of the prefixes $\forall\bar x_\psi$
  of~\eqref{eq:def-param}, and of the sentences $\exists x\,U_P(x)$ begin
  sentences; thus, there are at most $2|W|\cdot|\vp|$ of such blocks.  Hence,
  $r_\chi\leqslant|W|\cdot(5|\vp|+1)\leqslant6|W|\cdot|\vp|$.  The letters $D_w$
  and $U_P$ are monadic, each $P^w_i$ has the arity of $P_i$, and each name has
  arity $|\bar x_\psi|\leqslant k$; hence, $\varrho_\chi\leqslant\varrho$.
  Therefore, $\Theta'$ has a model with at most
  $c\cdot2^{12|W|\cdot|\vp|\cdot(2\varrho)^{\varrho}}$ elements.

  Lastly, as $k,a\leqslant|\vp|$, we know that
  $(2\varrho)^{\varrho}\leqslant2^{O(|\vp|\log|\vp|)}$, which gives the first of the
  particular bounds; if $k\geqslant2$ is fixed and $a\leqslant k$, then $\varrho=k$
  and $(2\varrho)^{\varrho}$ is a constant, which gives the second.

\section{Proof of Proposition~\ref{prop:classical-counting}}
\label{sec:prop:classical-counting}

  Let $\fr{F}_1$ be a Kripke frame with a single world $w$.

  (1): By Remark~\ref{rem:conserv}, a $\square$-free sentence belongs to $\CQCL$
  iff it belongs to $\CML\fr{F}_1$; hence, the upper bound follows from
  Theorem~\ref{thm:nexptime}, and the lower bound from
  Proposition~\ref{prop:lewis}.

  (2): Let $\vp\in\CLang^{\mathit{mon}}$ be a satisfiable sentence whose
  predicate letters other than $=$ are among $P_1,\ldots,P_s$.  Then
  $\neg\vp\notin\CQCL$, and so, by Remark~\ref{rem:conserv}, $\neg\vp$ is
  refuted in a model over $\fr{F}_1$.  As $\neg\vp$ has the same variables and
  the same counting threshold as $\vp$, Theorem~\ref{thm:small}, applied to
  $\fr{F}_1$, gives a model $\fr{M}'$ over $\fr{F}_1$ with $\fr{M}',w\sat\vp$
  and
  $|D'^{+}|\leqslant\max\{1,\kappa_\vp\}\cdot|\Typ(\fr{F}_1,s)|=\max\{1,\kappa_\vp\}\cdot2^{s}$,
  by~\eqref{eq:numtypes}.  By the argument of Remark~\ref{rem:conserv}, $\vp$ is
  true in the classical model with universe $D'_w$ interpreting each $P_i$ as
  $P_i^{I',w}$, which has at most $\max\{1,\kappa_\vp\}\cdot2^{s}$ elements.

  (3): Let $s\in\N$ and $M,v\geqslant1$.  For $S\subseteq\{1,\ldots,s\}$, let
  $$
  \begin{array}{lcl}
    \chi_S(x) & := & \displaystyle \bigwedge_{i\in S}P_i(x)\wedge
                     \bigwedge_{i\in\{1,\ldots,s\}\setminus S}\neg P_i(x),
  \end{array}
  $$
  and let $\theta^{S}$ be the sentence $\theta$ of Remark~\ref{rem:cap} with
  every quantifier relativised to $\chi_S$:
  $$
  \begin{array}{lcl}
    \theta^{S} & := & \displaystyle \exists x_1\cdots\exists x_{v-1}
                            \Bigl(\bigwedge_{1\leqslant i<v}\chi_S(x_i)\wedge
                            \bigwedge_{1\leqslant i<j<v}\neg(x_i=x_j)\\
                     &    & \displaystyle \qquad\qquad\wedge\;
                            \ex{M}x_v\bigl(\chi_S(x_v)\wedge
                            \bigwedge_{1\leqslant i<v}\neg(x_v=x_i)\bigr)\Bigr).
  \end{array}
  $$
  Then, $\theta^{S}$ is true in a classical model iff at least $M+v-1$ of its
  elements satisfy $\chi_S$.  Put
  $\vp:=\bigwedge_{S\subseteq\{1,\ldots,s\}}\theta^{S}$.  Then $\vp$ is a
  sentence of $\CLang^{\mathit{mon}}$ whose predicate letters other than
  $=$ are among $P_1,\ldots,P_s$, with $M_\vp=M$ and $v_\vp=v$, so that
  $\kappa_\vp=M+v-1$.  It is satisfiable: it is true in the classical model with
  universe
  $\{\langle S,j\rangle : S\subseteq\{1,\ldots,s\},\ 1\leqslant j\leqslant
  M+v-1\}$ in which $P_i$ is interpreted as the set of pairs
  $\langle S,j\rangle$ with $i\in S$, since there $\chi_S$ is satisfied by
  exactly the $M+v-1$ elements
  $\langle S,1\rangle,\ldots,\langle S,M+v-1\rangle$.  Lastly, since every
  element of a classical model satisfies exactly one of the $2^{s}$ formulas
  $\chi_S$, every model of $\vp$ has at least
  $(M+v-1)\cdot2^{s}=\kappa_\vp\cdot2^{s}$ elements.

\section{Proof of Lemma~\ref{lem:fluted-const}}
\label{sec:fluting-lemma}

If $m=0$, we put $\chi^{\circ}:=\chi$; so let $m\geqslant1$, and let $C$ be the
set of constants of $\chi$.  As $\chi\in(\mathcal{FL}^{\rc})^{[0]}$ has its
variables among $x_1,\ldots,x_k$, every subformula of $\chi$ in
$(\mathcal{FL}^{\rc})^{[j]}$ has its parameters among $x_1,\ldots,x_j$ and its
bound variables among $x_{j+1},\ldots,x_k$ (see Section~\ref{sec:prelim}), and
the quantifiers of $\chi$ are nested at most $k$ deep.  Following the outline
before Lemma~\ref{lem:fluted-const}, we construct $\chi^{\circ}$ in two steps,
using a fresh monadic letter $U$ for the elements not denoted by constants: we
first relativise the quantifiers of $\chi$ to $U$, adding the instances for
the constants, and then replace every atom with constants by an atom of a
fresh letter.  Instantiating a variable of a fluted formula requires
renumbering the variables that follow it; this is done by the shift.  We write
$g[x_i\mapsto e]$ for the assignment that differs from $g$ at most in sending
$x_i$ to $e$.

\smallskip\noindent\textit{The shift.}  Let $c\in C$ and $p\geqslant1$.  For
$\theta\in(\mathcal{FL}^{\rc})^{[j]}$ with $j\geqslant p$, let
$\theta[c/x_p]^{\downarrow}$ be the result of replacing in $\theta$ every
occurrence of $x_p$ with $c$ and every occurrence of $x_i$, for $i>p$, with
$x_{i-1}$.  For an assignment $g$ in a structure $\fr{A}$ interpreting $c$, let
$g^{\uparrow}$ be the assignment that sends $x_i$ to $g(x_i)$ if $i<p$, $x_p$
to $c^{\fr{A}}$, and $x_i$ to $g(x_{i-1})$ if $i>p$.  We claim that, for all
$j\geqslant p$ and $\theta\in(\mathcal{FL}^{\rc})^{[j]}$ with variables among
$x_1,\ldots,x_k$,
\begin{itemize}
\item[\rmbr{A}] $\theta[c/x_p]^{\downarrow}\in(\mathcal{FL}^{\rc})^{[j-1]}$, its
  variables are among $x_1,\ldots,x_{k-1}$, it has the size of $\theta$, and all
  its quantifiers have the form $\exists x_i\,(U(x_i)\wedge\ldots)$ if those of
  $\theta$ do;
\item[\rmbr{A$'$}] $\fr{A}\models^{g}\theta[c/x_p]^{\downarrow}$ iff
  $\fr{A}\models^{g^{\uparrow}}\theta$, for all $\fr{A}$ and $g$ as above.
\end{itemize}
Both are proved by induction on $\theta$, simultaneously for all
$j\geqslant p$.  If $\theta$ is a fluted $j$-atom, its variables, read from
left to right, are $x_\ell,\ldots,x_j$, for some $\ell\leqslant j+1$; after
the shift, they are $x_\ell,\ldots,x_{j-1}$ if $\ell\leqslant p$, and
$x_{\ell-1},\ldots,x_{j-1}$ if $\ell>p$, with constants in the remaining
places, so $\theta[c/x_p]^{\downarrow}$ is a fluted $(j-1)$-atom of the same
size; and, by the definition of $g^{\uparrow}$, each of its terms has under $g$
the value that the corresponding term of $\theta$ has under $g^{\uparrow}$.
The case of $\to$ is immediate.  If $\theta=\exists x_{j+1}\,\theta_1$, then
$\theta[c/x_p]^{\downarrow}=\exists x_j\,\theta_1[c/x_p]^{\downarrow}$, as
$j+1>p$, and the claims follow from IH, since
$(g[x_j\mapsto e])^{\uparrow}=g^{\uparrow}[x_{j+1}\mapsto e]$ for every
element $e$.

\smallskip\noindent\textit{Relativisation.}  For a subformula $\psi$ of $\chi$
with $\psi\in(\mathcal{FL}^{\rc})^{[j]}$, define $\psi^{U}$ by recursion on the
derivation of $\chi$: $\psi^{U}:=\psi$ if $\psi$ is an atom,
$(\psi_1\to\psi_2)^{U}:=\psi_1^{U}\to\psi_2^{U}$, and
\[
  (\exists x_{j+1}\,\psi_1)^{U}:=
  \exists x_{j+1}\,\bigl(U(x_{j+1})\wedge\psi_1^{U}\bigr)\ \vee\
  \bigvee_{c\in C}\psi_1^{U}[c/x_{j+1}]^{\downarrow};
\]
thus, $(\forall x_{j+1}\,\psi_1)^{U}$ is equivalent to
$\forall x_{j+1}\,(U(x_{j+1})\to\psi_1^{U})\wedge
\bigwedge_{c\in C}\psi_1^{U}[c/x_{j+1}]^{\downarrow}$.  Let $d(\psi)$ be the
depth of nesting of quantifiers in $\psi$.  We claim that there is a constant
$a\geqslant1$, independent of $\chi$, $m$ and $k$, such that, for every
subformula $\psi$ of $\chi$ with $\psi\in(\mathcal{FL}^{\rc})^{[j]}$,
\begin{itemize}
\item[\rmbr{B}] $\psi^{U}\in(\mathcal{FL}^{\rc})^{[j]}$, its variables are
  among $x_1,\ldots,x_k$, all its quantifiers have the form
  $\exists x_i\,(U(x_i)\wedge\ldots)$, and
  $|\psi^{U}|\leqslant a\cdot(m+1)^{d(\psi)}\cdot|\psi|$;
\item[\rmbr{B$'$}] $\fr{A}\models^{g}\psi^{U}$ iff $\fr{A}\models^{g}\psi$,
  for every structure $\fr{A}$ for the signature of $\chi$ extended by $U$
  with $U^{\fr{A}}=A\setminus\{c^{\fr{A}}:c\in C\}$ and every assignment $g$
  in $\fr{A}$.
\end{itemize}
We again use induction on $\psi$; only the case $\psi=\exists x_{j+1}\,\psi_1$
needs comment.  For~(B), the first disjunct of $\psi^{U}$ is in
$(\mathcal{FL}^{\rc})^{[j]}$, as $\psi_1^{U}\in(\mathcal{FL}^{\rc})^{[j+1]}$ by
IH and $U(x_{j+1})$ is a fluted $(j+1)$-atom, and so are the others, by~(A)
with $p:=j+1$; the other properties follow from IH and~(A).  As each disjunct
has size at most $|\psi_1^{U}|+b$, for a constant $b$, IH and
$d(\psi)=d(\psi_1)+1$ give
$|\psi^{U}|\leqslant(m+1)(|\psi_1^{U}|+b)\leqslant
a(m+1)^{d(\psi)}(|\psi_1|+1)\leqslant a(m+1)^{d(\psi)}|\psi|$ if
$a\geqslant b$.  For~(B$'$), note that, by~(A$'$),
$\fr{A}\models^{g}\psi_1^{U}[c/x_{j+1}]^{\downarrow}$ iff
$\fr{A}\models^{g[x_{j+1}\mapsto c^{\fr{A}}]}\psi_1^{U}$, as $g^{\uparrow}$,
formed with $p:=j+1$, agrees with $g[x_{j+1}\mapsto c^{\fr{A}}]$ on the
parameters of $\psi_1^{U}$.  Hence, by IH, $\fr{A}\models^{g}\psi^{U}$ iff
$\psi_1$ has a witness for $x_{j+1}$ in $U^{\fr{A}}$ or among the elements
$c^{\fr{A}}$, $c\in C$; as these are all the elements of $\fr{A}$, this is the
case iff $\fr{A}\models^{g}\psi$.

Let $\chi_1:=\chi^{U}$.  By~(B), $\chi_1$ is a sentence of
$(\mathcal{FL}^{\rc})^{k}$ of size $O((m+1)^{k}\cdot|\chi|)$, as
$d(\chi)\leqslant k$, all of whose quantifiers are relativised to $U$; as
every shift is computed in linear time, $\chi_1$ is computed in time
polynomial in its size.

\smallskip\noindent\textit{Pattern letters.}  The \emph{pattern} of an atom
$P(\bar t)$ of $\chi_1$, with $P$ of arity $n$, is the partial function $\pi$
from $\{1,\ldots,n\}$ to $C$ that sends every place of $\bar t$ occupied by a
constant to that constant.  For every letter $P$ of $\chi_1$ and every pattern
$\pi$ of an atom of $P$ in $\chi_1$, let $P_\pi$ be a fresh letter of arity
$n-|\dom\pi|$, where $P_\emptyset:=P$; in particular, $U$ is unchanged.  For a
subformula $\psi$ of $\chi_1$, let $\psi^{\circ}$ be the result of replacing
in $\psi$ every atom $P(\bar t)$, with pattern $\pi$ and variables $\bar x$
(from left to right), by $P_\pi(\bar x)$, and let
\[
  \chi^{\circ}:=\chi_1^{\circ}\wedge\forall x_1\,U(x_1).
\]
As $P_\pi(\bar x)$ has the same variables as $P(\bar t)$, in the same order,
$\chi^{\circ}$ is a sentence of $\mathcal{FL}^{k}$ without constants, of size
at most $|\chi_1|+O(1)$, and it is computed from $\chi_1$ in linear time.  This
completes the construction and proves the time bound.  The conjunct
$\forall x_1\,U(x_1)$ makes the quantifiers of $\chi_1^{\circ}$, all
relativised to $U$, range over the whole universe of a model of
$\chi^{\circ}$.

For a structure $\fr{A}$ interpreting the constants in $C$, a pattern $\pi$ of
an $n$-ary letter and a tuple $\bar b$ of $n-|\dom\pi|$ elements, let
$\bar b^{\,\pi}$ be the $n$-tuple with $c^{\fr{A}}$ in every place $i$ with
$\pi(i)=c$ and the elements of $\bar b$, in order, in the other places; thus,
under an assignment $g$, an atom $P(\bar t)$ with pattern $\pi$ and variables
$\bar x$ is true iff $g(\bar x)^{\pi}\in P^{\fr{A}}$.  We next prove the following:
\begin{itemize}
\item[\rmbr{C}] Let $\fr{A}$ be a structure for the signature of $\chi_1$ such
  that $B:=U^{\fr{A}}$ is non-empty and contains no element denoted by a
  constant, and let $\fr{B}$ be the structure with universe $B$ that
  interprets every letter $P_\pi$ of $\chi^{\circ}$ as
  $\{\bar b\in B^{n-|\dom\pi|}:\bar b^{\,\pi}\in P^{\fr{A}}\}$, so that
  $U^{\fr{B}}=B$.  Then $\fr{A}\models^{g}\psi$ iff
  $\fr{B}\models^{g}\psi^{\circ}$, for every subformula $\psi$ of $\chi_1$ and
  every assignment $g$ with values in $B$; in particular, $\fr{A}\models\chi_1$
  iff $\fr{B}\models\chi^{\circ}$.
\end{itemize}
The proof is by induction on $\psi$.  For an atom $P(\bar t)$ with pattern $\pi$
and variables $\bar x$, both sides say that $g(\bar x)^{\pi}\in P^{\fr{A}}$.
The case of $\to$ is immediate, and, by~(B), the quantified subformulas of
$\chi_1$ have the form $\exists x_i\,(U(x_i)\wedge\psi_1)$, so that their
witnesses, in $\fr{A}$ and in $\fr{B}$, are the elements of $B$ satisfying
$\psi_1$, and $\psi_1^{\circ}$, respectively; IH applies.  The last part holds
as $U^{\fr{B}}=B$.

\smallskip\noindent\textit{Proof of \rmbr{$2$}.}  Let
$\fr{B}\models\chi^{\circ}$, with universe $B$ of $n$ elements; then
$U^{\fr{B}}=B$.  Let $\fr{A}$ be the structure whose universe is $B$ together
with distinct fresh elements $e_c$, for $c\in C$, that interprets every
$c\in C$ as $e_c$, the letter $U$ as $B$, and every other letter $P$ of $\chi$
as the set of tuples $\bar b^{\,\pi}$ with $\bar b\in P_\pi^{\fr{B}}$, for the
patterns $\pi$ of the atoms of $P$ in $\chi_1$.  Then (C) applies to $\fr{A}$,
and $\fr{B}$ is the structure it assigns to $\fr{A}$: for $U$, as
$U^{\fr{B}}=B$; and, for the other letters, the inclusion of $P_\pi^{\fr{B}}$
in $\{\bar b:\bar b^{\,\pi}\in P^{\fr{A}}\}$ holds by definition, and the
converse inclusion because the places of a tuple $\bar b^{\,\pi}$ occupied by
elements $e_c$ determine $\pi$, as these elements are distinct and lie outside
$B$.  Hence, $\fr{A}\models\chi_1$ by~(C), and $\fr{A}\models\chi$ by~(B$'$),
as $U^{\fr{A}}=B$ consists of the elements of $\fr{A}$ not denoted by
constants; and $\fr{A}$ has $n+m$ elements.

Conversely, let $\fr{A}\models\chi$, with universe $A$ of $n$ elements.  As
all elements of $\fr{A}$ may be denoted by constants, we first duplicate
$\fr{A}$, which is why the bound in~\rmbr{$2$} is $2n$ rather than $n$.  Let
$\fr{A}'$ be the $2$-fold Cartesian product of $\fr{A}$
\cite[Section~5.5.1]{PH23}: its universe is $A\times\{0,1\}$, it interprets
every $c\in C$ as $\langle c^{\fr{A}},0\rangle$, and every letter $P$ as the
set of tuples $\langle\langle a_1,i_1\rangle,\ldots,\langle
a_n,i_n\rangle\rangle$ with $\langle a_1,\ldots,a_n\rangle\in P^{\fr{A}}$.
By~\cite[Lemma~5.14]{PH23}, $\fr{A}'$ satisfies the same sentences without
equality as $\fr{A}$, so $\fr{A}'\models\chi$.  The set $B$ of elements of
$\fr{A}'$ not denoted by constants contains every $\langle a,1\rangle$, so it
is non-empty, and $|B|\leqslant2n$.  Interpreting $U$ as $B$, we obtain
$\fr{A}'\models\chi_1$ by~(B$'$), and the structure with universe $B$ assigned
to $\fr{A}'$ by~(C) is a model of $\chi^{\circ}$.  This proves~\rmbr{$2$},
and~\rmbr{$1$} follows.

\section{Proof of Theorem~\ref{thm:rigid}}
\label{sec:thm:rigid}

We use the lemmas of Sections~\ref{sec:transfer} and~\ref{sec:complexity} in
their versions for sentences with constants, established in
Section~\ref{sec:constants}, with $(\Ax^{d}_{\fr{F}})^{C}$ in place of
$\Ax^{d}_{\fr{F}}$; the sentences $\Phi^d_{\fr{F}}(\vp)$, $\Psi^d_{\fr{F}}(\vp)$
and $\Theta^d_{\fr{F}}(\vp)$ then contain the constants of $\vp$.
In~\rmbr{$1$}--\rmbr{$5$}, the lower bounds are those of the corresponding
results for formulas without constants, so only the upper bounds need proof;
and we may assume that $\vp$ is a sentence (in~\rmbr{$3$}, as in the proof of
Theorem~\ref{thm:guarded-complexity}).

(1): By Lemmas~\ref{lem:Phi} and~\ref{lem:renaming}, $\vp$ is refuted in a
$d$-model over $\fr{F}$ if, and only if, $\Phi^d_{\fr{F}}(\vp)$ is
satisfiable; this is a sentence of $(\mathcal{L}_{\mathrm{C}}^{\rc})^{2}$ of
size polynomial in $|\fr{F}|+|\vp|$, and a sentence of
$(\mathcal{L}_{\mathrm{C}}^{\rc})^{1}$ if $\vp$ has one variable and of
$(\mathcal{L}_{=}^{\rc})^{2}$ if $\vp$ has no counting quantifiers.  The upper
bounds follow from Proposition~\ref{prop:classical-const}\rmbr{$1$}.

(2): For $k\leqslant2$, the claim follows from~\rmbr{$1$}, as in the proof of
Theorem~\ref{thm:fluted-complexity}.  For $k\geqslant3$, by
Lemmas~\ref{lem:Psi-size} and~\ref{lem:fluted_renaming}, $\vp$ is refuted in a
$d$-model over $\fr{F}$ if, and only if, the sentence
$\Psi:=\Psi^d_{\fr{F}}(\vp)$ of $(\mathcal{FL}^{\rc})^{k}$, of size polynomial
in $|\fr{F}|+|\vp|$, is satisfiable.  By Lemma~\ref{lem:fluted-const}, this is
the case if, and only if, the sentence $\Psi^{\circ}$ of $\mathcal{FL}^{k}$
without constants is satisfiable; it is computed in time polynomial in
$(c_\vp+1)^k\cdot(|\fr{F}|+|\vp|)$, that is, in time polynomial in
$|\fr{F}|+|\vp|$ for each fixed $k$, and, as $c_\vp,k\leqslant|\vp|$, in time
$2^{O(|\vp|\cdot\log|\vp|)}\cdot(|\fr{F}|+|\vp|)^{O(1)}$ in general.  The upper
bounds follow from Proposition~\ref{prop:FL}, as $\tower$ is closed under
elementary reductions.

(3): By Lemmas~\ref{lem:Theta} and~\ref{lem:guarded-renaming}, $\vp$ is refuted
in a $d$-model over $\fr{F}$ if, and only if, the sentence
$\Theta:=\Theta^d_{\fr{F}}(\vp)$ of $\mathcal{GF}_{=}^{\rc}$, of size
polynomial in $|\fr{F}|+|\vp|$, is satisfiable; by
Lemma~\ref{lem:guarded-syntax}\rmbr{$2$}, for $k\geqslant2$, $\Theta$ has at
most $k$ variables if so does $\vp$.  The $\coTwoEXP$ upper bound follows from
Proposition~\ref{prop:classical-const}\rmbr{$2$}.  For formulas with at most
$k$ variables, we use~\cite[Theorem~2]{tCF05}: by
Lemma~\ref{lem:guarded-syntax}\rmbr{$1$}, every quantifier of $\Theta$ is
guarded by an atom containing the free variables of its matrix, so, once
negations are pushed inwards, $\Theta$ is loosely $\forall$-guarded in the
sense of~\cite{tCF05}, with at most $k$ variables, and satisfiability of such
sentences is in $\EXP$ for every fixed $k$, whatever the number of constants.
The same bounds hold for formulas without equality.

(4): Let $\vp$ be monadic, with letters among $P_1,\ldots,P_s$ and constants
$c_1,\ldots,c_m$.  The sentence $\vp':=\forall\bar y\,\vp(\bar y)$ of
Lemma~\ref{lem:const-param} is monadic, is computable in polynomial time, has
the letters of $\vp$ and $v_\vp+m$ variables, and is $d$-valid on $\fr{F}$ if,
and only if, $\vp$ is (as in Section~\ref{sec:types}, we may assume that
$v_\vp+m\geqslant1$, as otherwise $\vp$ is propositional).  The claims follow
by applying Theorem~\ref{thm:nexptime}, Corollary~\ref{cor:complexity-monadic}
and Proposition~\ref{prop:pspace} to $\vp'$, and
Proposition~\ref{prop:bounded-var}, with $v+m$ in place of $v$, if $\vp$ has at
most $v$ variables and at most $m$ constants.

(5): The sentences $\Phi^d_{\fr{F}}(\vp)$, $\Psi^d_{\fr{F}}(\vp)$,
$\Theta^d_{\fr{F}}(\vp)$ and $\vp'$ are computed in time polynomial in
$|\fr{F}|+|\vp|$, and $\Psi^d_{\fr{F}}(\vp)^{\circ}$ in time polynomial in
$|\fr{F}|+|\vp|$ for each fixed $k$ and in elementary time in general, which
suffices for $\tower$; the claims follow as in Remarks~\ref{rem:frame-input}
and~\ref{rem:mon-frame-input}.

(6): Let $\vp$ be refuted in a $d$-model over $\fr{F}$.  We consider the
fragments in turn; the two-variable fragments with counting quantifiers lack
the finite domain property already without constants
(Remark~\ref{rem:no-FDP-counting}).

\smallskip\noindent\textit{Two variables, without counting quantifiers.}  Let
$\vp\in(\MLangeqrc)^2$.  By Lemmas~\ref{lem:Phi} and~\ref{lem:renaming}, the
sentence $\Phi:=\Phi^d_{\fr{F}}(\vp)$ of $(\mathcal{L}_{=}^{\rc})^2$, of size
polynomial in $|\fr{F}|+|\vp|$, is satisfiable, and a model of $\Phi$ with $n$
elements yields a $d$-model over $\fr{F}$ refuting $\vp$ whose global domain
has at most $n$ elements.  We eliminate the constants of $\Phi$ as
in~\cite[Section~3.1]{PH23}.  Fix a model of $\Phi$, let $\approx$ be the
relation of denoting the same element in it, and let $\Phi_\approx$ be
obtained from $\Phi$ by replacing every constant with a representative of its
$\approx$-class.  Then $\Phi_\approx$ has a model in which distinct constants
denote distinct elements, and every such model yields a model of $\Phi$ over
the same domain~\cite[p.~61]{PH23}; by~\cite[Lemmas~3.4 and~3.6]{PH23},
$\Phi_\approx$ has such a model over a given domain if, and only if, a sentence
$\chi$ of $\mathcal{L}^2_=$ without constants, over unary and binary letters
and computable from $\Phi_\approx$ in polynomial time, has a model over that
domain.  By~\cite[Lemmas~3.3 and~3.8]{PH23}, as in
Appendix~\ref{sec:thm:fdp-two-var}, $\chi$, and hence $\Phi$, has a model with
at most $3\cdot2^{9|\Sigma|}$ elements, where $\Sigma$, the signature of the
normal form of $\chi$, has size polynomial in $|\fr{F}|+|\vp|$.  This gives the
bound $2^{p(|\fr{F}|+|\vp|)}$.

\smallskip\noindent\textit{One variable, with counting quantifiers.}  We argue
as in the proof of Theorem~\ref{thm:fdp-one-var}.  The constants of $\vp$ occur
in $\Phi^d_{\fr{F}}(\vp)$ only in atoms, including the ground atoms $D_w(c)$,
so the sentences $\ex{M}x\,\theta$ for which witnesses are chosen there are the
same.  Adding to the chosen elements the at most $c_\vp$ elements denoted by
constants, we obtain a substructure that interprets the constants and in which
atoms with constants have the same truth values; so the induction goes through
unchanged, and the bound increases by $c_\vp$.

\smallskip\noindent\textit{Fluted.}  A model of $\Psi^d_{\fr{F}}(\vp)^{\circ}$
of the size given by the proof of Theorem~\ref{thm:FDP-fluted} yields, by
Lemma~\ref{lem:fluted-const}\rmbr{$2$}, a model of $\Psi^d_{\fr{F}}(\vp)$ with
$c_\vp$ more elements, and hence, by Lemma~\ref{lem:fluted_renaming}, a
refuting $d$-model whose global domain has at most that many elements; as
$|\Psi^d_{\fr{F}}(\vp)^{\circ}|$ is polynomial in $|\fr{F}|+|\vp|$ only for
fixed $k$, the exponential function in the bound now depends on $k$.

\smallskip\noindent\textit{Guarded.}  We argue as in the proof of
Theorem~\ref{thm:FDP-guarded}, for the sentences $\Theta^d_{\fr{F}}(\vp)$,
which now contain the constants of $\vp$.  The results of~\cite{Fiuk26} used
there admit constants: once constants with the same denotation have been
identified, as in~\cite[Section~2]{Fiuk26}, which changes neither
satisfiability nor the domains of models, the bound
of~\cite[Proposition~4.1]{Fiuk26} applies to the elements not denoted by
constants, and, by~\cite[Claim~2.1]{Fiuk26}, the bound on $|T|$ becomes
$(\varrho_\chi+1)\cdot2^{r_\chi\cdot(\varrho_\chi+m)^{\varrho_\chi}}$ for a
sentence $\chi$ with $m$ constants.  Hence, every satisfiable
$\mathcal{GF}_{=}^{\rc}$-sentence $\chi$ with at most $m$ constants and without
proposition letters has a model with at most
$C\cdot2^{2r_\chi\cdot(2(\varrho_\chi+m))^{\varrho_\chi}}+m$ elements, for an
absolute constant $C$.  As the estimates of $r_\chi$ and $\varrho_\chi$ for the
sentence $\Theta'$ of Appendix~\ref{sec:thm:FDP-guarded} are unchanged,
$\Theta'$ has a model with at most
$C\cdot2^{12|W|\cdot|\vp|\cdot(2(\varrho+c_\vp))^{\varrho}}+c_\vp$ elements,
which gives the stated bound.

\smallskip\noindent\textit{Monadic.}  If $v_\vp=c_\vp=0$, then $\vp$ contains
neither variables nor constants, so its truth at a world does not depend on the
domains; hence, $\vp$ is refuted in the $\mathit{cd}$-model over $\fr{F}$ with
the one-element domain $\{a\}$ at every world, in which every monadic letter is
interpreted as $\varnothing$ and every proposition letter as in a refuting
model; this model has $1\leqslant|\Typ^{d}(\fr{F},s)|$ elements.  Otherwise, by
Lemma~\ref{lem:const-param}, $\vp'=\forall\bar y\,\vp(\bar y)$ is refuted in a
$d$-model over $\fr{F}$; it has counting threshold $M_\vp$ and
$v_\vp+c_\vp\geqslant1$ variables, so $\kappa_{\vp'}=\kappa_\vp+c_\vp$.  By
Theorem~\ref{thm:small} (respectively, Proposition~\ref{prop:fdp-ML}), $\vp'$
is refuted in a $d$-model over $\fr{F}$ with at most
$(\kappa_\vp+c_\vp)\cdot|\Typ^{d}(\fr{F},s)|$ (respectively,
$|\Typ^{d}(\fr{F},s)|$) elements, and, as the proof of
Lemma~\ref{lem:const-param} does not change the global domain, $\vp$ is refuted
in a $d$-model with rigid constants with as many elements.

\section{Proof of Proposition~\ref{prop:const-fixed}}
\label{sec:prop:const-fixed}

(1): This is Theorem~\ref{thm:rigid}\rmbr{$4$}.

(2): For the lower bound, we reduce propositional validity to the set in
question.  Let $\beta(Q_1,\ldots,Q_n)$ be a propositional formula, let
$c_0,c_1,\ldots,c_n$ be constants, and let $\beta^{\ast}$ be obtained from
$\beta$ by replacing each $Q_i$ with $P_1(c_i)$ if $s\geqslant1$, and with
$c_i=c_0$ if $s=0$; thus, $\beta^{\ast}$ has no variables.  If $\beta$ is a
tautology, then so is $\beta^{\ast}$, which is therefore true at every world of
every model.  Conversely, if $\beta$ is false under a truth assignment
$\alpha$, then $\beta^{\ast}$ is false in the $\mathit{cd}$-model over $\fr{F}$
in which every world has the domain $\{a_0,a_1,\ldots,a_n\}$ of pairwise
distinct elements, every letter is interpreted at every world as
$\{a_i:1\leqslant i\leqslant n\mbox{ and }\alpha(Q_i)\mbox{ is true}\}$, $c_0$
as $a_0$, and $c_i$, for $i\geqslant1$, as $a_i$ if $s\geqslant1$, and as
$a_0$ or $a_i$, according as $\alpha(Q_i)$ is true or false, if $s=0$.

For the upper bound, let $\vp$ be a formula of the kind described, with
constants $c_1,\ldots,c_m$; we may assume that $\vp$ is a sentence, as its
universal closure has the same variables, letters and constants.  The sentence
$\vp':=\forall\bar y\,\vp(\bar y)$ of Lemma~\ref{lem:const-param} has
$v':=v_\vp+m$ variables; as $m$ is not bounded,
Proposition~\ref{prop:bounded-var} does not apply to $\vp'$, and we guess,
instead, a type assignment to $y_1,\ldots,y_m$.  As in
Section~\ref{sec:types}, we may assume that $v'\geqslant1$; we put
$N:=|\Typ(\fr{F},s)|$, a constant.  By Lemma~\ref{lem:const-param} and
Proposition~\ref{prop:capped-hist}, $\vp$ is not valid if, and only if,
$\hist,w_0\nsat^{\tau_\varnothing}\vp'$ for some $w_0\in W$ and some
$d$-admissible $\hist\colon\Typ(\fr{F},s)\to\{0,1,\ldots,\kappa_{\vp'}\}$.  As
$\forall y\,\chi$ abbreviates $\neg\ex{1}y\,\neg\chi$ and $\hist$-compatible
extensions have positive multiplicity, $\hist,w\sat^{\tau}\forall y\,\chi$ if,
and only if, $\hist,w\sat^{\tau'}\chi$ for every $\hist$-compatible extension
$\tau'$ of $\tau$ to $y$ at $w$.  Applying this $m$ times, we see that
$\hist,w_0\nsat^{\tau_\varnothing}\vp'$ if, and only if,
$\hist,w_0\nsat^{\tau}\vp(\bar y)$ for some $\hist$-compatible type assignment
$\tau$ on $\{y_1,\ldots,y_m\}$ at $w_0$, as every such $\tau$ arises from
$\tau_\varnothing$ by $m$ successive $\hist$-compatible extensions.

A nondeterministic algorithm, therefore, guesses $w_0$, $\hist$ and $\tau$,
using $O(N\cdot(|\vp|+\log m))$ bits for $\hist$ and
$O(m\cdot((s+1)|W|+\log m))$ bits for $\tau$, which specifies, for each $y_i$,
a type or an earlier variable in the same class.  It checks that $\hist$ is
$d$-admissible and $\tau$ is $\hist$-compatible, and evaluates
$\hist,w_0\sat^{\tau}\vp(\bar y)$ by the recursion of the proof of
Lemma~\ref{lem:cost}.  By~\eqref{eq:restrict}, the type assignments arising in
the recursion agree with $\tau$ on $y_1,\ldots,y_m$ and assign to each of the
at most $v$ other variables either one of the at most $m+v$ existing classes or
a new class with one of the $N$ types; so at most
$|\vp|\cdot|W|\cdot(m+v+N)^{v}$ values are computed, each in time polynomial in
$|\vp|+m$, as in the proof of Lemma~\ref{lem:cost}\rmbr{$1$}.  As $N$, $|W|$ and
$v$ are constants, the algorithm runs in nondeterministic polynomial time, and
so the set in question is in $\coNP$.

%\bibliographystyle{plain}
%\bibliography{foml}

\end{document}